\documentclass[preprint,showpacs,showkeys,preprintnumbers,amsmath,amssymb,floatfix]{revtex4}

\usepackage[dvips]{graphicx,color}
\usepackage{dcolumn}
\usepackage{bm}

\def\pmb#1{\setbox0=\hbox{#1}%
  \kern-.025em\copy0\kern-\wd0 
  \kern.05em\copy0\kern-\wd0
  \kern-.025em\raise.0433em\box0 }

\newcommand{\Figuretable}[1]{%
  \begin{center} --------- {\bf #1} --------- \\ \end{center}} 
\def\lambdabar{\protect\@lambdabar}
\def\@lambdabar{%
\relax
\bgroup
\def\@tempa{\hbox{\raise.73\ht0
\hbox to0pt{\kern.25\wd0\vrule width.5\wd0
height.1pt depth.1pt\hss}\box0}}%
\mathchoice{\setbox0\hbox{$\displaystyle\lambda$}\@tempa}%
{\setbox0\hbox{$\textstyle\lambda$}\@tempa}%
{\setbox0\hbox{$\scriptstyle\lambda$}\@tempa}%
{\setbox0\hbox{$\scriptscriptstyle\lambda$}\@tempa}%
\egroup
}

\begin{document}

\preprint{}

\title{
Isospin Properties of ($K^-$, $N$) Reactions
for the Formation of Deeply-bound Antikaonic Nuclei
}

\author{TAKAHISA KOIKE\footnote
{E-mail:tkoike@riken.jp}
}%
\affiliation{%
Advanced Meson Science Laboratory, 
RIKEN Nishina Center, Wako-shi, Saitama 351-0198, Japan
}

\author{TORU HARADA\footnote
{E-mail:harada@isc.osakac.ac.jp}
}%
\affiliation{%
Research Center for Physics and Mathematics,
Osaka Electro-Communication University, Neyagawa, Osaka, 572-8530, Japan
}

\date{\today}

\begin{abstract}
The formation of deeply-bound antikaonic $K^-/\bar{K}^0$
nuclear states by nuclear ($K^-$, $N$) reactions is 
investigated theoretically within a distorted-wave impulse approximation 
(DWIA), considering the isospin properties of the Fermi-averaged 
$K^-+ N \to N + \bar{K}$ elementary amplitudes.
We calculate the formation cross sections of 
the deeply-bound $\bar{K}$ states by 
the ($K^-$, $N$) reactions on the nuclear targets, $^{12}$C and $^{28}$Si, 
at incident $K^-$ lab momentum $p_{K^-}$ = 1.0 GeV/c 
and $\theta_{\rm lab} = 0^{\circ}$, introducing a complex 
effective nucleon number $N_{\rm eff}$ for unstable bound states in the DWIA.
The results show that the deeply-bound $\bar{K}$ states
can be populated dominantly by the ($K^-$, $n$) reaction 
via the total isoscalar $\Delta T=0$ transition 
owing to the isospin nature 
of the $K^-+ N \to N + \bar{K}$ amplitudes, 
and that the cross sections described by ${\rm Re}N_{\rm eff}$ and 
${\rm Arg}N_{\rm eff}$ enable to deduce the structure of the 
$\bar{K}$ nuclear states;
the calculated inclusive nucleon spectra for a deep 
$\bar{K}$-nucleus potential do not show distinct peak structure
in the bound region. 
The few-body 
$\bar{K}\otimes [NN]$ and $\bar{K}\otimes [NNN]$ states 
formed in ($K^-$, $N$) reactions on $s$-shell nuclear targets, 
$^3$He, $^3$H and  $^4$He, are also discussed.

\end{abstract}
\maketitle

\section{Introduction}
\label{sect:1}

The antikaon-nucleon ($\bar{K}N$) interaction in nuclei
is very important to elucidate the nature of 
high dense nuclear matter \cite{Kaplan86,Brown96,Lee96}.
The attractive $\bar{K}N$ interaction 
in the isospin $I=$ 0 channel is sufficiently strong to produce 
a quasibound state $\Lambda(1405)$ at about 27 MeV below 
the $K^-p$ threshold.
This state is often described as a $\bar{K}N$ bound state
decaying into the main $\pi \Sigma$-channels with a width of 
about 50 MeV \cite{Dalitz67}.
Such a $\bar{K}N$ interaction, together with the moderately 
attractive $\bar{K}N$ $I=$ 1 channel, would give 
a strong attractive potential for the $\bar{K}$ in nuclei, 
leading naively to the existence of deeply-bound 
$\bar{K}$ nuclear states \cite{Dote04}.
Indeed, comprehensive analysis of the shifts and widths of
${K}^-$ atomic X-ray data suggests 
strongly attractive $\bar{K}$ optical potentials, 
of the depth between 150-200 MeV at the nuclear 
center \cite{Friedman93,Friedman94,Batty97}.
Since the $K^-$ atomic state is sensitive 
only to behavior of the $K^-$ wave function 
at the nuclear surface, 
it is ambiguous to extrapolate smoothly to nuclear 
matter density, and it would be inconclusive \cite{Cieply01};
see however \cite{Friedman07}.
Enhancement in $K^-$ production by heavy ion reactions 
suggests a strongly attractive interaction \cite{Li97}
although the recent analysis of details of the reaction 
mechanism leads to attraction of only
$-80$ MeV \cite{Scheinast06}.
The existence of the deeply-bound $\bar{K}$ 
nuclear states would provide constraints on 
kaon condensation in compact stars and neutron 
stars \cite{Prakash98};
the equation of state (EOS) of nuclear matter
is softened due to the appearance of the kaon 
condensation. 

Recent theoretical calculations have found that  
the $\bar{K}$-nucleus attraction 
amounts to 110 MeV at normal nuclear density $\rho_0$,   
being constructed by the chiral SU(3) effective Lagrangian
involving $\bar{K}N$-$\pi Y$ 
coupled channels \cite{Waas97}, 
whereas the $\bar{K}$ potential calculated
self consistently has a relatively 
shallow depth of 40-60 MeV with a broad width 
of about 100 MeV \cite{Waas97,Lutz98,Schaffner00,Ramos00}. 
We belive that more theoretical developments are 
needed in order to describe a quasibound state 
having a broad width in the complicated 
nuclear medium.

On the other hand, 
the production of the deeply-bound $\bar{K}$ nuclear 
states has been investigated theoretically and experimentally 
for the forward ($K^-$, $N$) reaction, 
in order to search directly for signals of 
$\bar{K}$ nuclear bound states, 
as proposed  by Kishimoto \cite{Kishimoto99}.
Ikuta et al.~\cite{Ikuta02} attempted to calculate 
the inclusive ($K^-$, $p$) spectra by the Green's function 
technique \cite{Morimatsu94}, using (energy-dependent) 
$K^-$-nucleus potentials by fits to the $K^-$ atomic 
data \cite{Friedman93,Friedman94,Batty97}.
Kishimoto and his collaborators \cite{Kishimoto05} performed 
experimentally the measurements for searching 
deeply-bound $\bar{K}$ 
nuclei by (in-flight $K^-$,$N$) reactions on $^{12}$C 
and $^{16}$O targets in the PS-E548 experiment at KEK, 
but there was no clear evidence 
in both ($K^-$, $n$) and ($K^-$, $p$) spectra. 
Yamagata et al.~\cite{Yamagata05,Yamagata06} also calculated the 
corresponding cross sections by ($K^-$, $p$) reactions 
on $^{12}$C and $^{16}$O targets, 
using $K^-$ optical potentials based on Chiral 
unitary and phenomenological approaches, and showed
the difficulties to find clear signals for $\bar{K}$ nuclear 
states due to their broad width, even if the bound states exist, 
within Green's function technique \cite{Yamagata06}.

Akaishi and Yamazaki \cite{Akaishi99,Akaishi02} predicted that 
the light $\bar{K}$ nuclear $K^-ppn$/$\bar{K}^0pnn$
system with total isospin $T=$ 0 has a deeply bound state 
with a relatively narrow width, by over $B_{\bar{K}}=$ 108 MeV 
for which the main decay channel $\bar{K}N \to \pi \Sigma$ 
would be kinematically closed. Thus its width becomes 
relatively narrow, about 20 MeV due to the isospin selection. 
Surprisingly, Suzuki et al. \cite{Suzuki04} reported 
experimentally evidence of the tribaryon S$^{0}$(3115) by the 
$^4$He(stopped $K^-$, $p$) reaction, and the signal was 
interpreted as such a state.
Although it was withdrawn \cite{Iwasaki06a}, 
the data analysis from the improved-statistics experiment
E549 is in progress \cite{Sato07}. 
Is there such a narrow deeply-bound $\bar{K}$ state ?

Recently, FINUDA collaboration at DA$\Phi$NE \cite{Agnello05}
reported experimentally the evidence of a deeply-bound $K^-pp$ state 
in invariant mass spectroscopy by 
$K^-pp$ $\rightarrow$ $\Lambda$ + $p$ decay processes from 
$K^-$ absorption on $^6$Li, $^7$Li and  $^{12}$C at rest, 
where the measured energy and width are $B_{\bar{K}}=$ 115 MeV 
and ${\it \varGamma}_{\bar{K}}$ = 67 MeV, respectively.
The three-body $\bar{K}$ bound state with a 
configuration of $[{\bar K}\otimes \{NN \}_{I=1}]_{T=1/2}$ \cite{Nogami63}
is expected to play a fundamental role 
in constructing the various $\bar{K}$ nuclear bound states.
This state is often called ``$K^-pp$'' symbolically.
Several theoretical works 
\cite{Yamazaki02,Shevchenko06,Shevchenko07a,Ivanov05,Ikeda07,Dote07,Nishikawa07,Arai07} 
supported the existence of the $K^-pp$ bound state:
Yamazaki and Akaishi \cite{Yamazaki02} 
predicted the binding energy of $B_{\bar{K}}=$ 48 MeV and the 
width of ${\it \varGamma}_{\bar{K}}$ = 61 MeV for 
the $K^-pp$ bound state by 
a few-body calculation using ATMS method.
Shevchenko et al. \cite{Shevchenko06,Shevchenko07a}
obtained $B_{\bar{K}}=$ 55-70 MeV and 
${\it \varGamma}_{\bar{K}}=$ 95-110 MeV 
for the $K^-pp$ bound state
by a $\bar{K}NN$-$\pi \Sigma N$ coupled-channel Faddeev 
calculation. 
We believe that the search for a deeply-bound $K^-pp$ state 
is one of the most important subjects to verify 
the reliable evidence of the deeply-bound $\bar{K}$ nuclei, 
and that this state might be a doorway for 
access to kaon condensation in neutron stars 
\cite{EXA05}. 
More experimental data are required 
in order to confirm clearly whether deeply-bound 
states exist or not.

Very recently, Iwasaki et al. \cite{Iwasaki06} proposed 
a new experiment searching the deeply-bound 
$K^-pp$ state at J-PARC, 
by the missing-mass spectrum of the $^3$He(in-flight $K^-$,$n$) 
reaction, together with the invariant-mass spectra 
detecting all particles via decay processes (J-PARC E15).
In a previous paper \cite{Koike07}, 
we demonstrated the inclusive 
and semi-exclusive spectra of the $^3$He(in-flight $K^-$, $n$) reaction 
at $p_{K^-}$ = 1.0 GeV/c and $\theta_{\rm lab} = 0^{\circ}$ 
within the distorted wave impulse approximation (DWIA),
in order to see theoretically the expected spectra 
of the $^3$He(in-flight $K^-$,$n$) reaction for preparing the 
forthcoming J-PARC E15 experiment \cite{Iwasaki06}. 
Yamazaki and Akaishi \cite{Yamazaki07} also proposed the 
$K^-pp$ formation 
via $p + p \to K^+ + ``\Lambda(1405)p" \to K^+ + K^-pp$, 
assuming a $\Lambda(1405)p$ doorway.
Such a $K^-pp$ experiment with a $\sim$4 GeV proton beam on 
a proton target is planned by FOPI collaboration 
at GSI \cite{Suzuki05}.

In this paper, we investigate theoretically the (in-flight $K^-$, $N$) 
reaction for the formation of the deeply-bound $\bar{K}$ states 
on nuclei, within a distorted-wave impulse
approximation (DWIA). 
One of our main purpose is to understand the formation mechanism 
for the deeply-bound $\bar{K}$ states by the nuclear ($K^-$, $n$) 
and ($K^-$, $p$) reactions, which would provide the selectivity 
of the isospin excitation in the $\bar{K}$ nuclear bound states. 
We discuss in some detail the isospin properties of the cross section 
by the nuclear ($K^-$, $n$) or ($K^-$, $p$) reaction on a $^{12}$C target, 
together with the isospin-dependence of the 
$\bar{K}$-nucleus potentials which are suggested from 
the recent analyses of $K^-/K^+$ production
in heavy ion reactions at KaoS \cite{Li97,Scheinast06}, and 
of $^{12}$C(in-flight $K^-$, $N$) 
reaction experiments (PS-E548) at KEK \cite{Kishimoto05}.

The outline of this paper is as follows.
In Section \ref{sect:2}, we present the amplitudes and the 
cross sections of the forward $K^- + N \to N + \bar{K}$ reaction, 
on the basis of the $K^-N$ elastic scattering amplitudes 
analyzed by Gopal et al.~\cite{Gopal77}. The isospin properties 
of the forward $K^- + N \to N + \bar{K}$ reactions are shown, 
together with Fermi-averaged $K^- + N \to N + \bar{K}$ 
amplitudes which take into account the Fermi motion of 
a nucleon in the nucleus. 
In Section \ref{sect:3}, we discuss the formation of the 
deeply-bound $\bar{K}$ nuclear state by nuclear ($K^-$, $N$) 
reactions within the DWIA, 
introducing a {\it complex} effective nucleon number for 
the {\it unstable} bound states.
We stress the importance of the kinematical effects 
in terms of the momentum transfer 
to a residual $\bar{K}$ nucleus for the ($K^-$, $N$) reaction, 
which differs considerably from hypernuclear production reactions 
by the nuclear ($\pi^+$, $K^+$) and ($K^-$, $K^+$) ones. 
We discuss the properties of the formation cross sections of the 
deeply-bound $K^-$ states by the nuclear ($K^-$, $N$) reactions 
on the closed-shells $^{12}$C and $^{28}$Si targets 
at the incident ${K}^-$ lab momentum $p_{K^-}=$ 1.0 GeV/c, 
and the isospin selection by the ($K^-$, $n$) and ($K^-$, $p$) 
reactions. 
In Section \ref{sect:4}, we concentrate on isospin dependence 
of the relative cross sections for the deeply-bound
three-body $\bar{K}\otimes [NN]$ and 
four-body $\bar{K}\otimes [NNN]$ states
in order to study the formation of light 
deeply-bound $\bar{K}$ states.
Summary and conclusion are given in Section \ref{sect:6}.
A derivation of kinematical factors in the DWIA is
given in in Appendices \ref{sect:a1} and \ref{sect:a2},
together with the relation between 
the integrated cross section and the inclusive spectrum.
Spin-isospin states for $s$-shell $\bar{K}$-nuclear systems 
are given in Appendix \ref{sect:a3}.

\section{The ($K^-$, $N$) reactions}
\label{sect:2}

\subsection{The elementary reactions $K^- + N \to \bar{K} +N $ 
and $K^- + N \to N + \bar{K}$}
\label{sect:2-1}

Kishimoto \cite{Kishimoto99} suggested that the forward 
($K^-$, $N$) reaction on a nuclear target 
could result in a discovery of
$\bar{K}$ nuclear states. This has triggered experimental and 
theoretical studies of the ($K^-$, $N$) reaction for searching 
deeply-bound $\bar{K}$ nuclear states. 
Because the forward $K^- + N \to N + K^-$ scattering amplitudes 
$(\theta_{\rm c.m.}=0^\circ)$ in the c.m.~frame are equivalent 
to the backward $K^- + N \to K^- + N$ scattering ones 
$(\theta_{\rm c.m.}=180^\circ)$, 
we considered the $K^- + N \to \bar{K} + N$ elastic and 
charge-exchange reactions,
in order to understand the nature of the 
$K^- + N \to N + \bar{K}$ reaction which is needed to 
calculate the production cross section by the nuclear ($K^-$, $N$) 
reaction within the DWIA.

\begin{figure}[htb]
	\begin{minipage}{.48\linewidth}
	\includegraphics[width=\linewidth]{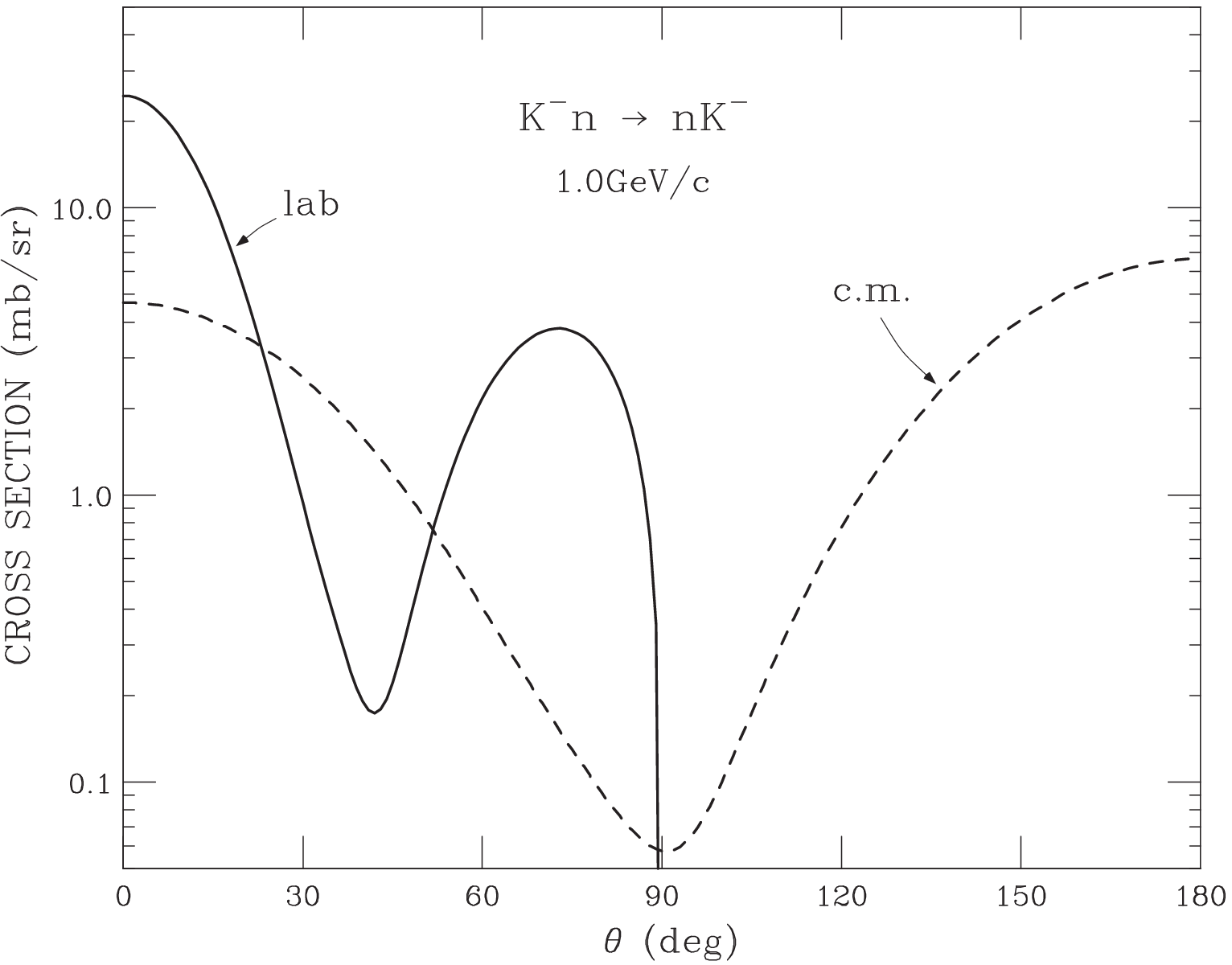}
	\vspace{5mm}
	\end{minipage}
	\begin{minipage}{.48\linewidth}
	\includegraphics[width=\linewidth]{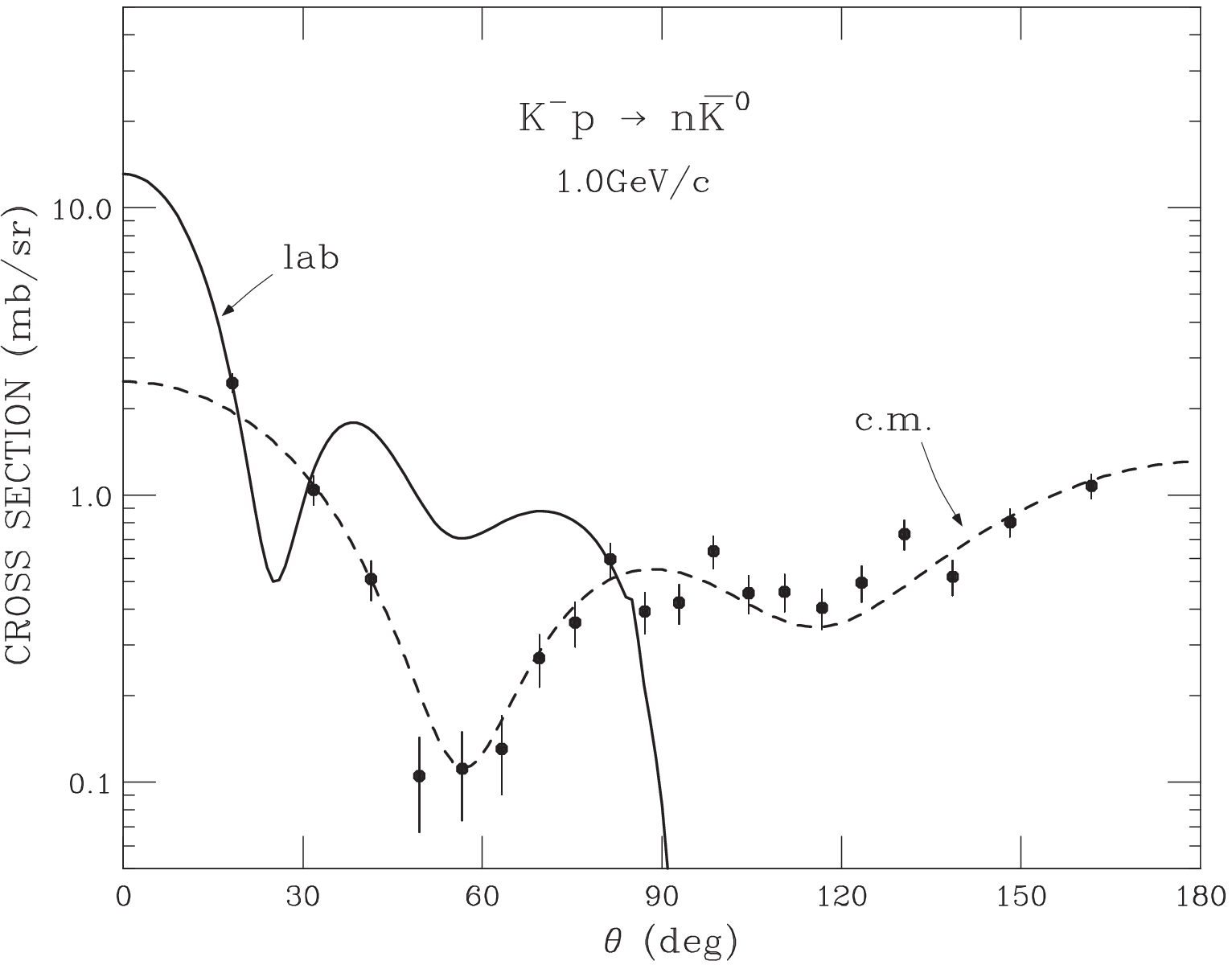}
	\vspace{5mm}
	\end{minipage}
	\begin{minipage}{.48\linewidth}
	\includegraphics[width=\linewidth]{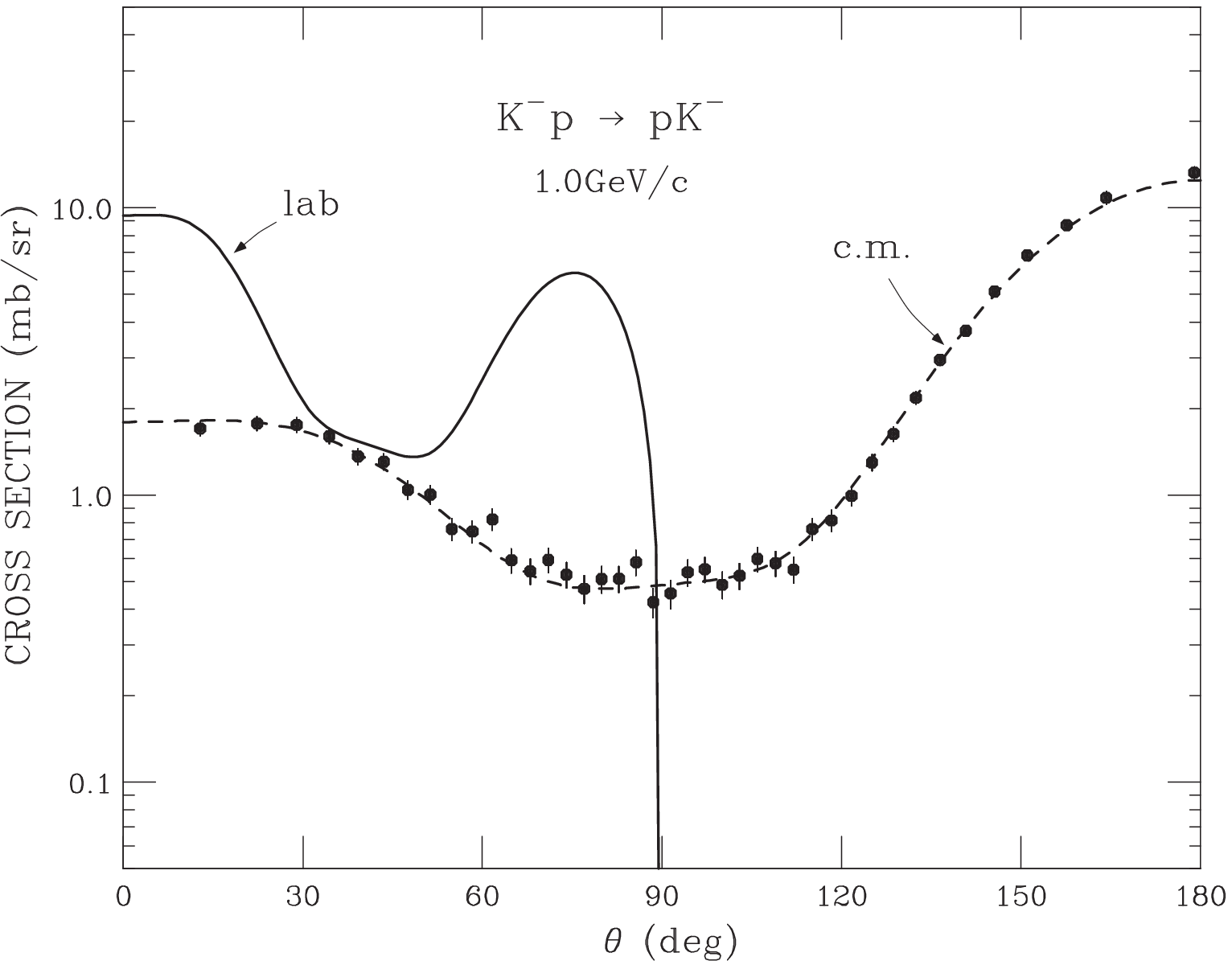}
	\end{minipage}
        \vspace{0.5cm}
	\caption{\label{fig:2}
   The differential cross sections of 
   $K^- + n \to n + K^-$, $K^- + p \to n + \bar{K}^0$ and 
   $K^- + p \to p + K^-$ reactions in free space 
   at the incident $K^-$ lab momentum $p_{K^-}=$ 1.0 GeV/c, as a function 
   of the scattering angle $\theta$. 
   The dashed and solid curves denote the c.m. and lab cross sections, 
   respectively, which are obtained by Gopal et al. \cite{Gopal77}.
   The experimental data are taken from Conforto et al.~\cite{Conforto76}.
}
\end{figure}

The c.m.~backward $K^- + N \to \bar{K} + N$ scattering amplitudes
are equivalent to the c.m.~forward $K^- + N \to N + \bar{K}$ ones  
in free space \cite{Kishimoto99}.
Thus, the c.m.~$K^- + N \to N + \bar{K}$ amplitudes are given as 
\begin{eqnarray}
& f_{K^- n \to n K^-}(\theta_{\rm c.m.}) &  
= f_{K^- n \to K^- n}(\pi-\theta_{\rm c.m.}), \nonumber \\
&f_{K^- p \to n \bar{K}^0}(\theta_{\rm c.m.}) & 
= f_{K^- p \to \bar{K}^0 n}(\pi-\theta_{\rm c.m.}), \label{eqn:e04}\\
& f_{K^- p \to p K^-}(\theta_{\rm c.m.}) &= 
f_{K^- p \to K^- p}(\pi-\theta_{\rm c.m.}), \nonumber
\end{eqnarray}
where the labels by $K^- n \to n K^-$, 
$K^- p \to n \bar{K}^0$ and $K^- p \to p K^-$ 
denote the c.m.~backward cross sections for 
$K^- + n \to K^- + n$, $K^- + p \to \bar{K}^0 + n$
and $K^- + p \to K^- + p$ reactions, respectively; 
$\theta_{\rm c.m.}$ is the nucleon angle relative to $K^-$ beam 
direction in the c.m.~frame. 
The c.m.~differential cross sections for 
the $K^- + N \to N + \bar{K}$ reactions can be obtained by
\begin{equation}
\left( {d \sigma \over d \Omega}(\theta_{\rm c.m.})
\right)_{\rm c.m.}^{K^-N \to N \bar{K}}
= |f_{K^-N \to N \bar{K}}(\theta_{\rm c.m.})|^2,
\label{eqn:e07}
\end{equation}
for these channels. 
In Fig.~\ref{fig:2}, we show the c.m.~differential cross sections
for the $K^- + N \to N + \bar{K}$ reactions at the incident $K^-$ lab 
momentum $p_{K^-}=$ 1.0 GeV/c, as a function of the scattering angle 
$\theta$, together with the lab differential cross sections.
The dashed and solid curves are the angular distributions 
of $(d\sigma/d\Omega)^{K^-N \to N \bar{K}}_{\rm c.m.}$ and 
$(d\sigma/d\Omega)^{K^-N \to N \bar{K}}_{\rm lab}$, respectively,
which are obtained by Gopal et al. \cite{Gopal77}. 
The experimental data are taken from Conforto et al. \cite{Conforto76}.
These values are connected by the relation \cite{Dover83} 
\begin{eqnarray}
{(d\sigma/d\Omega)^{K^-N \to N \bar{K}}_{\rm lab}
\over (d\sigma/d\Omega)^{K^-N \to N \bar{K}}_{\rm c.m.}} &=&
{{m_{N} p_{K^-} p_{N}} \over {p_{K^-{\rm c.m.}} p_{N {\rm c.m.}}}}
\left(E_{K^-}+m_N-E_{N}{p_{K^-} \over p_{N}}\cos{\theta_{\rm lab}}
\right)^{-1},
\label{eqn:e08}
\end{eqnarray}
where \{$p_{K^-}$, $E_{K^-}$\} and \{$p_{N}$, $E_{N}$\} are 
lab momenta and total energies for the incoming $K^-$ and the outgoing nucleon, 
respectively, and the quantities labeled by ${\rm c.m.}$ subscript are 
in the c.m. frame. $m_N$ is the mass of a nucleon.
The lab scattering angle $\theta_{\rm lab}$ satisfies
\begin{equation}
\cos{\theta_{\rm lab}}={
E_{K^-}E_{N}-E_{K^-{\rm c.m.}}E_{N{\rm c.m.}}+
p_{K^-{\rm c.m.}}p_{N{\rm c.m.}}\cos{\theta_{\rm c.m.}}
\over p_{K^-}p_{N}},
\label{eqn:e09}
\end{equation}
where there is a kinematical focussing \cite{Dover83} of 
the lab cross section to a cone for the maximum angle 
$\theta_{\rm lab}^{\rm max}={\pi/2}$, because 
$m_{\bar{K}}/m_{N} < 1$ in the $K^- + N \to N + \bar{K}$ reactions.

\Figuretable{FIG. 1}

\begin{figure}[htb]
  \begin{center}
  \includegraphics[width=0.6\textwidth]{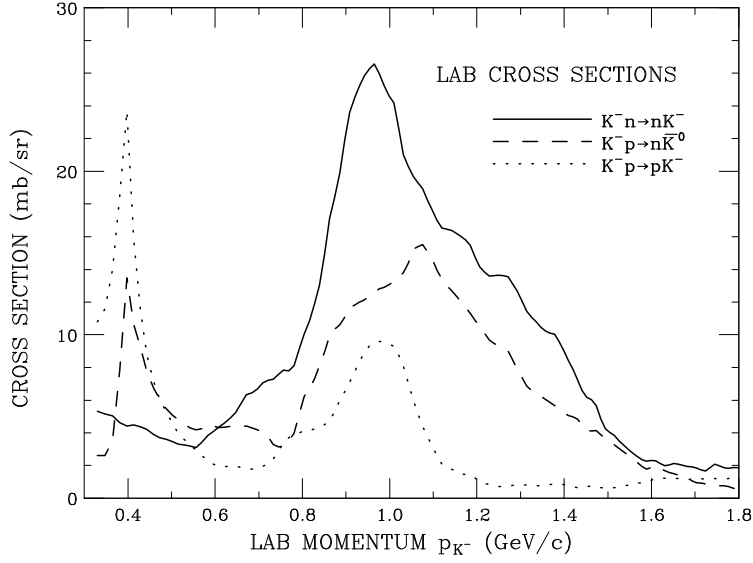}
  \end{center}
  \caption{\label{fig:4}
   The lab differential cross sections for the free space 
   $K^-+ n \to n + K^-$, $K^- + p \to n + \bar{K}^0$ and 
   $K^-+ p \to p + K^-$ reactions at the detected nucleon angle 
   $\theta_{\rm lab}=0^\circ$, as a function of the $K^-$ lab momentum $p_{K^-}$.
   The curves are constructed from the 
   $K^-+ N \to \bar{K} + N$ scattering 
   amplitudes analyzed by Gopal et al.~\cite{Gopal77}. 
  }
\end{figure}

In Fig.~\ref{fig:4}, we show the lab differential cross sections for 
$K^- + n \to n + K^-$,  $K^- + p \to n + \bar{K}^0$ and 
$K^- + p \to p + K^-$ 
reactions at $\theta_{\rm lab}=0^\circ$, as a function of 
the incident $K^-$ lab momentum. 
These values are maximum near $p_{K^-}$= 1.0 GeV/c because of the existence 
of several $\Sigma^*$ and $\Lambda^*$ resonances in the 0.9-1.2 GeV/c region, 
e.g., $S_{11}(1750)$, $D_{15}(1775)$, $S_{01}(1800)$, $F_{05}(1820)$, etc. 
At $p_{K^-}$= 1.0 GeV/c, thus,  
\begin{eqnarray}
\left( {d\sigma \over d\Omega}(0^\circ)\right)^{K^-N \to N \bar{K}}_{\rm lab}
=& 24.5 \, \mbox{mb/sr}, \quad \mbox{for } & K^- + n \to n + K^-, \nonumber\\
=& 13.1 \, \mbox{mb/sr}, \quad \mbox{for } & K^- + p \to n + \bar{K}^0, \label{eqn:e10} \\
=&  9.4 \, \mbox{mb/sr}, \quad \mbox{for } & K^- + p \to p + K^-. \nonumber
\end{eqnarray}

\Figuretable{FIG. 2}

\subsection{Isospin properties for the $K^- + N \to N + \bar{K}$ reaction}
\label{sect:2-2}

Now we introduce the $s$-channel isospin transition 
amplitudes $f_I$ for the ($K^-$,~$N$) reactions, 
in order to clarify the isospin properties of these reactions.
The amplitudes 
for $K^- + n \to n + K^-$, $K^- + p \to n + \bar{K}^0$ 
and $K^- + p \to p + K^-$ processes have the relations 
\begin{eqnarray}
& f_{K^- n \to n K^-} & = f_1, \label{eqn:e11}\\
& f_{K^- p \to n \bar{K}^0} & = (f_1+f_0)/2, \label{eqn:e12}\\
& f_{K^- p \to p K^-} &= (f_1-f_0)/2, \label{eqn:e13}
\end{eqnarray}
where $f_0$ and $f_1$ denote the $s$-channel isospin 0 and 1 amplitudes 
for the ($K^-$,~$N$) reactions, respectively. 
One has to notice that the definitions of 
Eqs.~(\ref{eqn:e12}) and (\ref{eqn:e13}) 
are interchanged for the ($K^-$,~$\bar{K}$) reactions, 
as given by Eqs.~(9) and (10) of Ref.~\cite{Kishimoto99}.

Considering the isospins
${\bm i}_1 + {\bm i_2} \to {\bm i}_1' + {\bm i}_2'$ 
for the $K^- + N \to N + \bar{K}$ reactions 
in the isospin algebra \cite{Dover83}, 
we can represent the $t$-channel forward two-body 
reaction amplitude 
$f^{(t)}=\langle (i_2 i_2') t |f|(i_1 i_1') t \rangle$ 
by expanding the $s$-channel amplitude  
$f_I=\langle (i_1' i_2') I |f|(i_1 i_2) I \rangle$, 
so as to derive the isospin decomposition of the nuclear reactions:
\begin{equation}
f^{(t)} = \sum_{I} (-)^{i_1+i_2+I }
(2I+1)
    \left\{ \begin{array}{ccc}
                  i_1'  &  i_2'    &  I       \\
                  i_2 &  i_1   & t \\
    \end{array}  \right\} 
f_I.
\label{eqn:e14}
\end{equation}
For the ($K^-$, $N$) reactions, thus, we obtain
\begin{eqnarray}
& f^{(0)} & 
=(f_0+3f_1)/2
=f_{K^- n \to n K^-}+f_{K^- p \to n \bar{K}^0}, \label{eqn:e15}\\
& f^{(1)} & 
=(f_1-f_0)/2
=f_{K^- n \to n K^-}-f_{K^- p \to n \bar{K}^0}
=f_{K^- p \to p K^-}, \label{eqn:f1_t} \label{eqn:e16}
\end{eqnarray}
where 
$f^{(0)}$ and $f^{(1)}$ denote (unnormalized) $t$-channel isospin 0 and 1 
amplitudes for the $K^-+N \to N + \bar{K}$ reactions, 
respectively.
Note that these amplitudes correspond to $u$-channel 
isospin 0 and 1 amplitudes for 
the $K^-+N \to \bar{K}+N$ reactions.
The amplitudes in Eq.~(\ref{eqn:e16}) are related by 
isospin to the amplitudes of the neutron and proton reactions,  
\begin{eqnarray}
& f_{K^- n \to n K^-} & 
=f_{K^- p \to n \bar{K}^0}+f_{K^- p \to p K^-},
\label{eqn:e17}
\end{eqnarray}
which is easily verified by a combination of 
Eqs.~(\ref{eqn:e11})-(\ref{eqn:e13}).
The isospin relation between the amplitudes for ($K^-$,~$N$)
reactions is summarized in Table~\ref{tab:table1}.

\Figuretable{TABLE I}

\subsection{Fermi-averaging in nuclear medium}
\label{sect:2-3}

When we calculate the nuclear ($K^-$, $N$) cross sections 
with the elementary $K^- + N \to N + \bar{K}$ amplitudes 
$f_{K^-N \to N \bar{K}}$, 
it is important to take into account the Fermi motion 
of a struck nucleon moving with a Fermi-momentum 
$p_F \simeq$ 270 MeV/c in nuclear medium 
\cite{Allardyce73,Rosenthal80}, 
because a momentum transfer 
$|\, q \,|\simeq$ 200 MeV/c in
the forward $K^- + N \to N + \bar{K}$ reaction 
at $p_{K^-}=$ 1.0 GeV/c is scarcely smaller than the Fermi-momentum. 
Dover and Walker \cite{Dover79} have pointed out that 
Fermi-averaging for
$K^- + N \to \bar{K} + N$ reaction acts considerably on 
the ($K^-$, $\bar{K}$) inelastic cross sections;
particularly, the effect appears near a narrow resonance, 
e.g., $\Lambda$(1520) at $p_{K^-}\sim$ 390 MeV/c, 
because its width is smaller than the Fermi-motion energy of 
a struck nucleon.
In the case of the nuclear ($K^-$, $N$) reaction, 
let us obtain the Fermi-averaged amplitudes 
which play an important role of calculating 
the cross section with the DWIA.

\begin{figure}[htb]
  \begin{center}
  \includegraphics[width=0.6\textwidth]{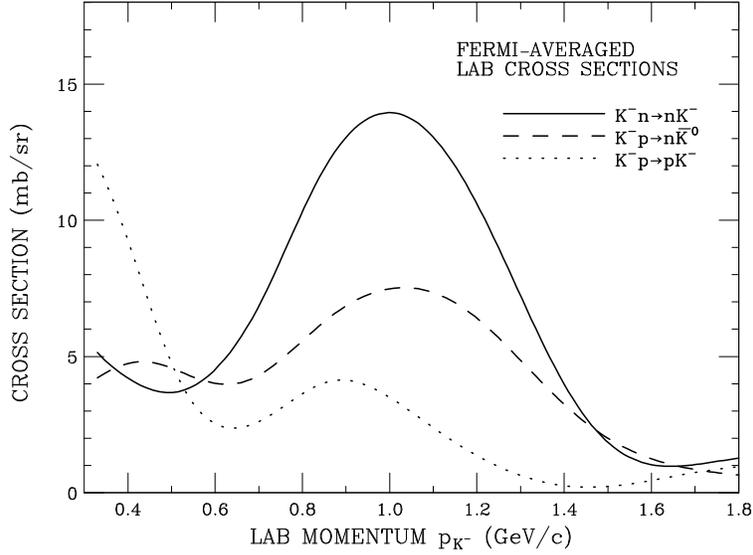}
  \caption{\label{fig:5}
  The absolute values of the Fermi-averaged cross sections at 
  $\theta_{\rm lab}= 0^\circ$ 
  for the $K^- + n \to n + K^-$, $K^- + p \to n  + \bar{K}^0$ and 
  $K^-+ p \to p + K^-$ reactions, 
  as a function of the $K^-$ lab momentum $p_{K^-}$.
  The free space amplitudes of Gopal et al.~\cite{Gopal77} were used.
  }
  \end{center}
\end{figure}

According to the procedure by Rosental and Tabakin \cite{Rosenthal80}, 
we perform the Fermi-averaging of the 
$K^- + N \to N + \bar{K}$ scattering $T$-matrix in the lab frame.
Using the $K^-N$ amplitudes by Gopal et al. \cite{Gopal77},
we obtain the Fermi-averaged $T$-matrix for the lab momenta 
of 0.3-2.0 GeV/c at $\theta_{\rm lab}=0^\circ$,
which is given by
\begin{eqnarray}
 \bar{t}_{K^-N \to N\bar{K}} (0^\circ) 
&= & {1 \over 2}\int_{-1}^{1}dx \int_0^\infty p^2dp \, \rho(p) 
      t_{K^-N \to N\bar{K}}(E_{K^-N}), 
\label{eqn:e18}
\end{eqnarray}
as a function of the effective lab energy 
$E_{K^-N} = \sqrt{{\bm p}_{K^-}^2 + m^2_{K^-}} 
+\sqrt{{\bm p}^2 + m^2_{N}}$, where 
${\bm p}$ and $m_N$ are the momentum and the mass of 
a struck nucleon in the nucleus, respectively, 
and $x = \hat{\bm p}_{K^-} \cdot \hat{\bm p}$.
Here we use a lab momentum distribution
$\rho(p)$ of the struck nucleon with a sum of 
squares of harmonic oscillator wave 
functions \cite{Dover79}
\begin{eqnarray}
\rho(p)
&= & \rho_0\left(N_{s}+ {2 \over 3}N_{p}(bp)^2\right)
\exp{(-(bp)^2)}, 
\label{eqn:e19}
\end{eqnarray}
where 
$N_s$ and $N_p$ are the numbers of $s$- and $p$-shell 
nucleon, respectively, and $\rho_0=(b/\sqrt{\pi})^{3}$ with 
the size parameter $b=$ 1.64 fm for $^{12}$C.
The resultant $T$-matrix is not so sensitive to  
a choice of the form of $\rho(p)$ constrained by 
a comparable value of $ \langle p^2 \rangle^{1/2}$, 
as shown in Ref.~\cite{Dover79}.
Thus, the  Fermi-averaged forward amplitude in nuclear medium 
is given as 
\begin{eqnarray}
\bar{f}_{K^-N \to N\bar{K}}(0^\circ)
&=& - {1 \over 2 \pi}\left({p_N \over p_{K^-}}\right)
\left[{E_{K^-}E_NE_{\bar{K}} 
       \over E_{K^-}+m_N-E_N(p_{K^-}/p_N)}
\right]^{1 \over 2}
\bar{t}_{K^-N \to N\bar{K}}(0^\circ),
\label{eqn:e18a}
\end{eqnarray}
where a participant nucleon involving a Fermi-averaged 
$T$-matrix is considered approximately to be at rest in the target.
In Fig.~\ref{fig:5}, we show the Fermi-averaged forward cross 
section for $^{12}$C in the lab frame, 
\begin{equation}
\left\langle{d\sigma \over d\Omega}(0^{\circ}) 
\right\rangle_{\rm lab}^{K^- N \to N \bar{K}}
   =  | \bar{f}_{K^-N \to N\bar{K}}(0^\circ) |^2
\label{eqn:e20}
\end{equation}
for $K^- + n \to n + K^-$, $K^- + p \to n + \bar{K}^0$ 
and $K^- + p \to p + K^-$ channels.
The nucleon Fermi-motion appreciably moderates 
the effects of resonances. 
The shape of the cross sections near 1.0 GeV/c becomes sizably 
broader, and a narrow $\Lambda$(1520) resonance 
affects more strongly the cross sections near $0.4$ GeV/c. 
Even after the Fermi-averaging, 
there remains rather strongly energy-dependence for the 
$K^- + N \to N + \bar{K}$ reactions, and 
also their charge (isospin) state-dependence, 
as suggested by Rosental and Tabakin \cite{Rosenthal80}.
The absolute values of the Fermi-averaged forward 
cross sections at $p_{K^-}$= 1.0 GeV/c are calculated by
\begin{eqnarray}
\left\langle{d\sigma \over d\Omega}(0^{\circ}) 
\right\rangle_{\rm lab}^{K^- N \to N \bar{K}}
=& 13.9 \, \mbox{mb/sr}, \quad \mbox{for } & K^- + n \to n + K^-, \nonumber\\
=&  7.5 \, \mbox{mb/sr}, \quad \mbox{for } & K^- + p \to n + \bar{K}^0, 
\label{eqn:e21}\\
=&  3.5 \, \mbox{mb/sr}, \quad \mbox{for } & K^- + p \to p + K^-. \nonumber
\end{eqnarray}
We find that these values for 
$K^- + n \to n + K^-$, $K^- + p \to n + \bar{K}^0$ and 
$K^- + p \to p + K^-$
are reduced by about 0.57, 0.57 and 0.37, respectively, in comparison 
with those for free space, Eq.~(\ref{eqn:e10}).
If we replace the $T$-matrix $t_{K^-N \to N\bar{K}}$ in Eq.~(\ref{eqn:e18})
by $|t_{K^-N \to N\bar{K}}|$ neglecting its phase, 
we find the values of 18.3 mb/sr, 11.3 mb/sr and 4.8 mb/sr 
for the corresponding Fermi-averaged cross sections 
at $p_{K^-}$= 1.0 GeV/c, 
which would be equivalent to the Fermi-averaging for the differential 
cross sections instead of the $T$-matrices. 
This means that the resultant cross sections are further more reduced 
by the phase at this momentum region.
When we do the Fermi-averaging for $^4$He ($^3$He) in the $s$-shell 
nuclei, we use $b=$ 1.21 fm (1.38 fm) with $N_p=$ 0 
in the momentum distribution $\rho(p)$ in Eq.~(\ref{eqn:e19}).
We confirm that the calculated cross sections are not so changed 
by choosing the targets, e.g., $^{12}$C or $^4$He, 
as discussed by Dover et al.~\cite{Dover89}.

\Figuretable{FIG. 3}

\begin{figure}[htb]
  \begin{center}
  \includegraphics[width=0.7\textwidth]{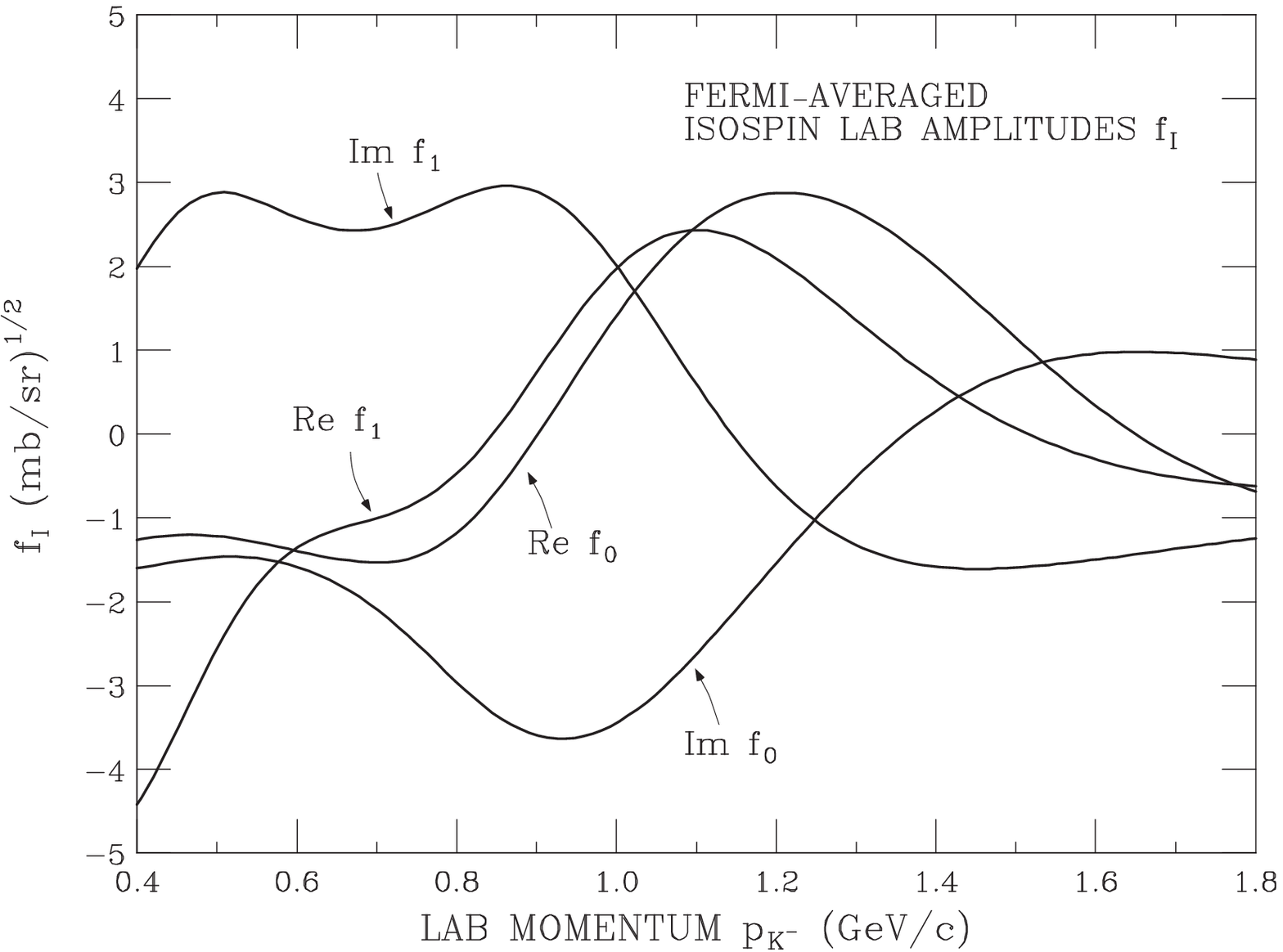}
  \caption{\label{fig:6}
  Fermi-averaged $K^- + N \to N + \bar{K}$ amplitudes for $f_I$ 
  for $s$-channel isospin $I=0$, $1$, as a function of the incident 
  $K^-$ lab momentum at $\theta_{\rm lab}= 0^\circ$. 
  The amplitudes are calculated by averaging the $T$-matrices for $^{12}$C 
  with the harmonic oscillator momentum distribution.
  The free-space amplitudes of Gopal et al.~\cite{Gopal77} were used.  
  }
  \end{center}
\bigskip
  \begin{center}
  \includegraphics[width=0.7\textwidth]{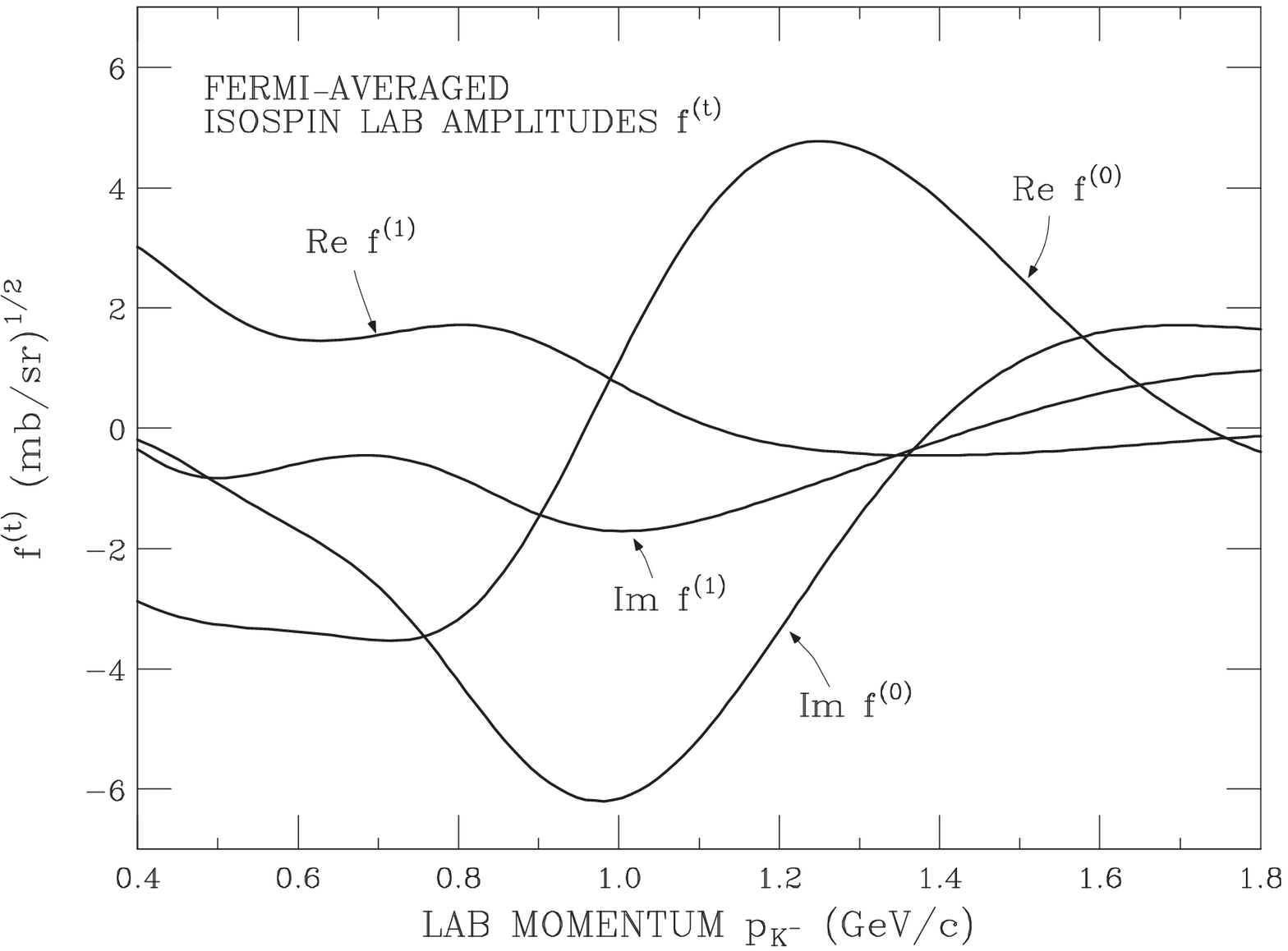}
  \end{center}
  \caption{\label{fig:7}
  Fermi-averaged $K^- + N \to N + \bar{K}$ amplitudes for $f^{(t)}$ 
  for $t$-channel
  isospin $t=0, 1$, as a function of $K^-$ lab momentum at 
  $\theta_{\rm lab}= 0^\circ$. 
  The free space amplitudes of Gopal et al.~\cite{Gopal77} were used.
  }
\end{figure}

In Figs.~\ref{fig:6} and \ref{fig:7}, we show the 
Fermi-averaged lab amplitudes for $s$-channel isospin 
$\bar{f}_I$ and  for $t$-channel isospin 
$\bar{f}^{(t)}$, respectively.   
We shall employ these Fermi-averaged amplitudes 
$\bar{f}_{K^-N \to N\bar{K}}$ in the following 
calculations.

\Figuretable{FIG. 4}
\Figuretable{FIG. 5}

\section{Formation of the Deeply-bound antikaonic nuclei}
\label{sect:3}

\subsection{Distorted-wave impulse approximation (DWIA)}
\label{sect:3-1}

\begin{figure}[htb]
  \begin{center}
  \includegraphics[width=0.9\textwidth]{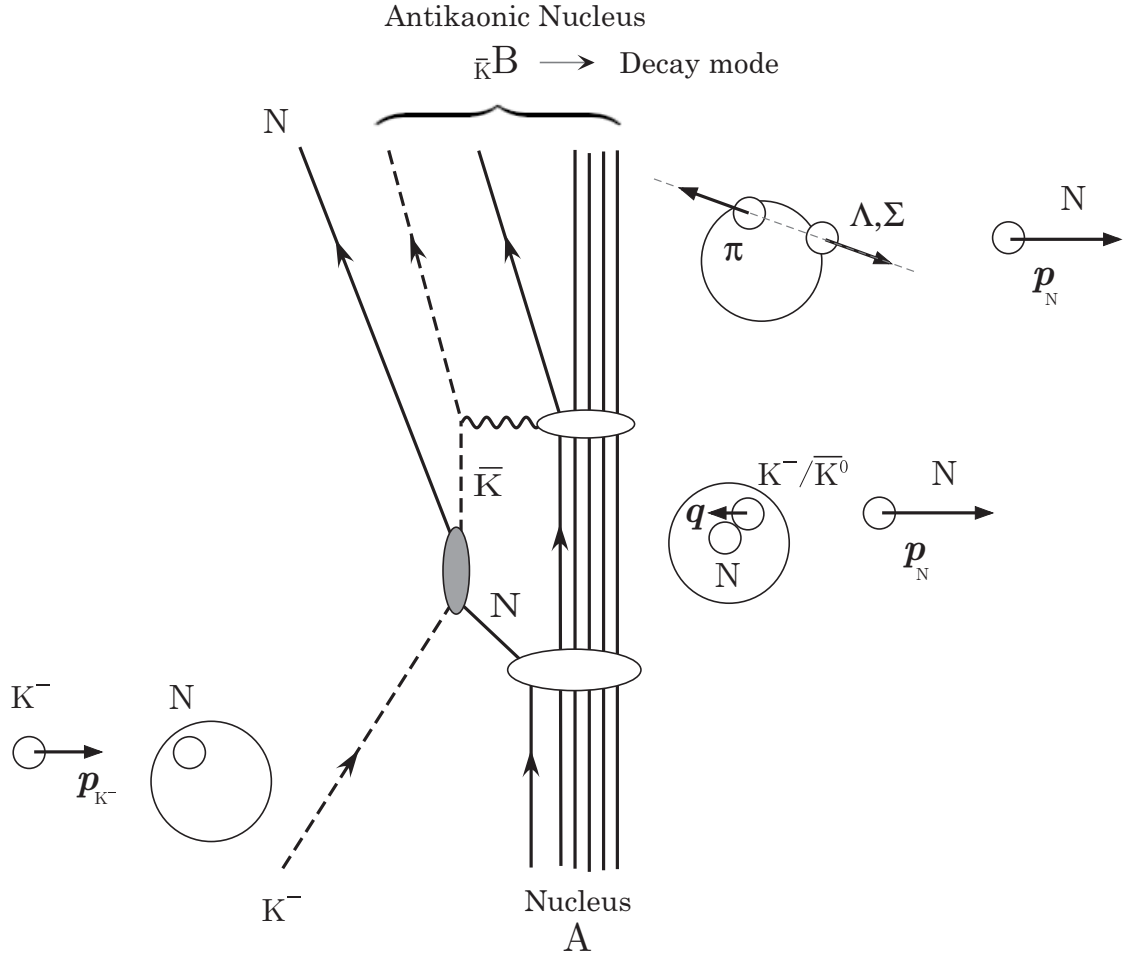}
  \caption{\label{fig:8}
  Diagram of the impulse approximation for the ($K^-$,$N$) reaction 
  on the nuclear target A.
  }
  \end{center}
\end{figure}

The formation of the deeply-bound $\bar{K}$ nuclear 
states for the nuclear ($\bar{K}$, $N$) reaction
has been investigated experimentally 
\cite{Kishimoto05,Suzuki04,Agnello05} and 
theoretically 
\cite{Kishimoto99,Cieply01,Ikuta02,Akaishi02,Yamagata05,Yamagata06,Koike07}.
Iwasaki et al., \cite{Iwasaki06a}  have proposed the experiment searching 
the deeply-bound $K^-pp$ states by 
$^3$He(in-flight $K^-$, $N$) reaction at $p_{K^-}=$ 1.0 GeV/c 
and $\theta_{\rm lab}=$ 0$^\circ$.
In order to clarify these reaction mechanism, 
let us consider the nuclear ($K^-$, $N$) reaction 
for the formation of deeply-bound $\bar{K}$ nuclear 
states within the distorted-wave impulse approximation (DWIA)
\cite{Hufner74,Bouyssy77,Auerbach83,Dover83}.
In Fig.~\ref{fig:8}, we illustrate a diagram of the 
nuclear $A$($K^-$, $N$)${_{\bar{K}}{B}}$ reaction with 
the impulse approximation; the final states 
in ${_{\bar{K}}{B}}$ will be unstable 
due to the strong-interaction decay processes by
the one-body $\bar{K}$ absorption 
$K^-/\bar{K}^0 + ``N" \to \pi + \Sigma/\Lambda$ 
and the multi-nucleonic $\bar{K}$ absorption, e.g., 
$K^-/\bar{K}^0 + ``NN" \to\Sigma/\Lambda + N$,  
in the nucleus.
In this paper, we focus on the formation stage of 
the deeply-bound $\bar{K}$ nuclear states via 
$K^- + N \to N + \bar{K}$ processes on the nuclear targets.
In a previous paper \cite{Koike07}, 
we studied the inclusive and semi-exclusive spectra of 
the $^3$He(in-flight $K^-$, $n$) reaction 
at $p_{K^-}$ = 1.0 GeV/c and $\theta_{\rm lab} = 0^{\circ}$ 
for searching the deeply-bound $K^-pp$ states,
and discussed briefly a kinematical factor for 
the $K^-pp$ states and the isospin excitation of their cross 
sections in the DWIA.
Such formation reactions have an unique kinematics 
similar to the formation of $\pi$, $\rho$ and $\omega$-nuclear 
bound states \cite{Marco01}, 
rather than hypernuclear production by
($K^-$,$\pi$) \cite{Hufner74,Bouyssy77,Auerbach83,Harada98}, 
($\pi$,$K^+$) \cite{Dover80,Motoba88,Hausmann89,Harada05}
and ($K^-$, $K^+$) \cite{Dover83,Tadokoro95}
reactions, as we will mention below.

\Figuretable{FIG. 6}

The double-differential cross section for the formation 
of deeply-bound $\bar{K}$ nuclear states 
by the ($K^-$,$N$) reaction
at a nucleon direction angle $\theta_{\rm lab}$ is
given within the DWIA \cite{Morimatsu94,Koike07,Dover83,
Auerbach83,Harada98},
assuming a zero-range interaction for elementary 
$K^- + N \to N + \bar{K}$ reactions, as
\begin{eqnarray}
   \left({{d^2\sigma} 
    \over {d\Omega_{N} dE_{N}} }\right)_{\rm lab}
 &  = & {\beta}
   {1 \over {2J_A+1}} \sum_{M_A}  \sum_{m_s,B} \vert\langle \Psi_B \vert
   {\hat{F}} \vert \Psi_A \rangle\vert^{2} \delta (\omega+E_{B}-E_{A})
\label{eqn:e22}
\end{eqnarray}
with
\begin{equation}
{\hat{F}} 
 =  \int d{\bm r} \> \chi_{N,m_s}^{(-) \ast}({\bm p}_{N},{\bm r},\sigma)
                     \chi_{K^-}^{(+)}({\bm p}_{K^-},{\bm r})  
                     \sum_{j=1}^{A} 
                     {\bar{f}}_{K^-N \to N \bar{K}} \hat{\cal O}_{N_j} 
                      \delta ({\bm r}-{\bm r}_{j}) ,
\label{eqn:e23}
\end{equation}
where $\vert \Psi_B \rangle$ and $\vert \Psi_A \rangle$ are final states of 
the $\bar{K}$ nuclear states with total spin $J_B$ 
and initial states of the target nucleus
with total spin $J_A$, respectively. 
$\chi_{N,m_s}^{(-)}({\bm p}_N,{\bm r},\sigma)$ and 
$\chi_{K^-}^{(+)}({\bm p}_{K^-},{\bm r})$ denote distorted waves for
an outgoing spin-(${1 \over 2}$,$m_s$) nucleon with the detected lab 
momentum ${\bm p}_N$
and for an incoming $K^-$ with the incident lab momentum 
${\bm p}_{K^-}$, respectively. 
The operator ${\hat{\cal O}}_{N_j}$ changes a $j$-th nucleon into 
the $\bar{K}$ in the nucleus; 
${\bar{f}}_{K^-N \to N \bar{K}}$ is the Fermi-averaged lab amplitude for 
the $K^- + N \to N + \bar{K}$ reaction in nuclear medium, 
and $\beta$ is a kinematical factor, as mentioned in 
Appendix~\ref{sect:a1}. 
Here we consider only the non-spin-flip processes, 
because we are interested in the cross section at the forward direction.
In our previous paper \cite{Koike07}, the formation cross 
sections for $K^-pp$ bound states were evaluated 
by the Green's function technique \cite{Morimatsu94}.
This technique can describe unstable hadron nuclear systems 
such as the $\Sigma$ and $\Xi$ nuclear bound states very well, 
so as to compare theoretical spectra with experimental 
ones \cite{Tadokoro95,Harada98,Harada05}.
In $\bar{K}$ nuclear physics, the spectral calculations have 
been performed by several authors 
\cite{Ikuta02,Kishimoto05,Yamagata06,Koike07}. 
Using Green's function ${G}(\omega)$ for the $\bar{K}$-nucleus system, 
a sum of the final states of Eq.~(\ref{eqn:e22}) is written as
\begin{equation}
 \sum_{B}  \vert \Psi_B \rangle \delta (\omega+E_{B}-E_{A})
    \langle \Psi_B \vert 
  = (-){1 \over \pi} {\rm Im}{G}(\omega),
\label{eqn:e24}
\end{equation}
where $\omega$ is the energy transfer to the residual $B$ nucleus.
Then the inclusive spectrum of the double-differential cross section 
in Eq.~(\ref{eqn:e22}) is 
rewritten as
\begin{eqnarray}
   \left({{d^2\sigma} 
    \over {d\Omega_{N} dE_{N}} }\right)_{\rm lab}
 &  = & {\beta}(-){1 \over \pi}{\rm Im} \left[ \sum_{\alpha' \alpha} 
 \int\,d{\bm r}' d{\bm r}
  F_{\alpha'}^\dagger({\bm r}') G_{\alpha' \alpha}(\omega;{\bm r}',{\bm r})
  F_{\alpha}({\bm r})\right],
 \label{eqn:e25}
\end{eqnarray}
where ${\bm r}$ is the relative coordinate between the $\bar{K}$ and 
the core-nucleus. 
$F_{\alpha}({\bm r}) $ presents the $\bar{K}$-nucleus doorway states 
excited initially as
\begin{equation}
F_{\alpha}({\bm r}) = 
\chi^{(-)*}_{N,m_s}({\bm p}_{N},{M_C \over M_B}{\bm r},\sigma)
\chi^{(+)}_{K^-}({\bm p}_{K^-},{M_C \over M_A}{\bm r})
\bar{f}_{K^-N \to N\bar{K}}
\langle \alpha | \hat{\psi}_N ({\bm r}) | \Psi_{A} \rangle,
\label{eqn:e26}
\end{equation}
where
$\langle \alpha \, | \hat{\psi}_N ({\bm r}) |\Psi_{A} \rangle$ 
is a hole-state wave function for a struck nucleon in the 
target, and $\alpha$ denotes the complete set of eigenstates 
for the system.
The factors of $M_C/M_B$ and $M_C/M_A$ 
take into account the recoil effects, where 
$M_A$, $M_B$ and $M_C$ are masses
of the target, the $\bar{K}$-nucleus and the core-nucleus, 
respectively.

\subsection{Kinematics}
\label{sect:3-2}

Now let us consider the kinematics for the ($K^-$, $N$) reaction on a
nuclear target.  
The momentum-energy transfers to the $\bar{K}$ nuclear bound states 
are given by
\begin{eqnarray}
  {\bm q} &=& {\bm p}_{K^-}- {\bm p}_N, \label{eqn:e27}\\
  \omega &=& E_{K^-}-E_N= E_B-E_A\simeq m_{\bar{K}} - m_N -B_{\bar{K}} -
 \varepsilon_N + T_{\rm recoil},
\label{eqn:e28}
\end{eqnarray}
where 
$m_{\bar{K}}$ and $m_N$ are masses of an $\bar{K}$ and a nucleon, 
respectively, and $B_{\bar{K}}$ is an $\bar{K}$ binding energy 
measured from the $K^-$+core-nucleus threshold.
$\varepsilon_N$ is a single-particle energy of a nucleon-hole 
state in the target, and $T_{\rm recoil}$ is a recoil energy 
to the $\bar{K}$ nuclear bound state.
The momentum transfer ${\bm q}$ into the residual nucleus $B$ 
in the lab frame is very important for characterizing a nuclear reaction
\begin{equation}
  {a} + {A} \to {b} + {B},
\label{eqn:e29}
\end{equation}
where $a$, $b$, $A$ and $B$ are the incident, the detected, the target 
and the residual particles, respectively.
Dalitz and Gal \cite{Dalitz78} discussed the kinematics 
of the exothermic reaction 
($m_{a}+M_{A} > m_{b}+M_{B}$)
such as $K^- + N \to \pi + \Lambda$
for $\Lambda$-hypernuclear production with 
``substitutional'' states due to the recoilless 
condition with $q \simeq$ 0 MeV/c.   
Dover and his collaborators \cite{Dover80,Dover83} studied 
for the endothermic reaction 
($m_{a}+M_{A} < m_{b}+M_{B}$)
such as $\pi + N \to K^+ + \Lambda$ for $\Lambda$-hypernuclear production 
and $K^- + N \to K^+ + \Xi^-$ for $\Xi$-hypernuclear production, 
leading to ``spin-stretched'' states due to a high momentum 
transfer $q \simeq$ 400-500 MeV/c. 
In the case of the ($K^-$, $N$) reaction, 
the momentum transfer at $p_{K^-}=$ 1.0 GeV/c and 
$\theta_{\rm lab}=$ 0$^\circ$ becomes $q \simeq$ 200-400 MeV/c, 
depending on $B_{\bar{K}}$ in the $\bar{K}$ nuclear bound 
state \cite{Kishimoto99,Cieply01,Yamagata05}.
This situation seems to be similar to hypernuclear production by
($\pi$, $K^+$) reactions \cite{Cieply01}, but $M_{A} > M_{B}$ is 
different from $M_{A} < M_{B}$ occurred in 
hypernuclear production.
Such a kinematical condition is often found in 
the ($\gamma$, $p$) or ($\pi$, $p$) reaction 
for searching the $\pi$, $\omega$, $\rho$-nuclear bound 
states \cite{Marco01}. 
The kinematics for the ($K^-$, $N$) reaction differs completely from 
that for any hypernuclear production reactions,
as we will discuss below;
the momentum transfer $q(0^\circ)$ is negative, 
$q(0^\circ)= p_{K^-} - p_{N} < 0$, 
because the detected nucleon momentum $p_N$ becomes 1.2-1.4 GeV/c 
at the incident $K^-$ momentum $p_{K^-}$= 1.0 GeV/c \cite{Koike07}.
The negative value of $q(0^\circ)$ means that the residual 
$\bar{K}$ recoils backward relative to the incident 
particle $K^-$ in the lab frame, as illustrated in Fig.~\ref{fig:8}.

\begin{figure}[htb]
  \begin{center}
  \includegraphics[width=0.8\textwidth]{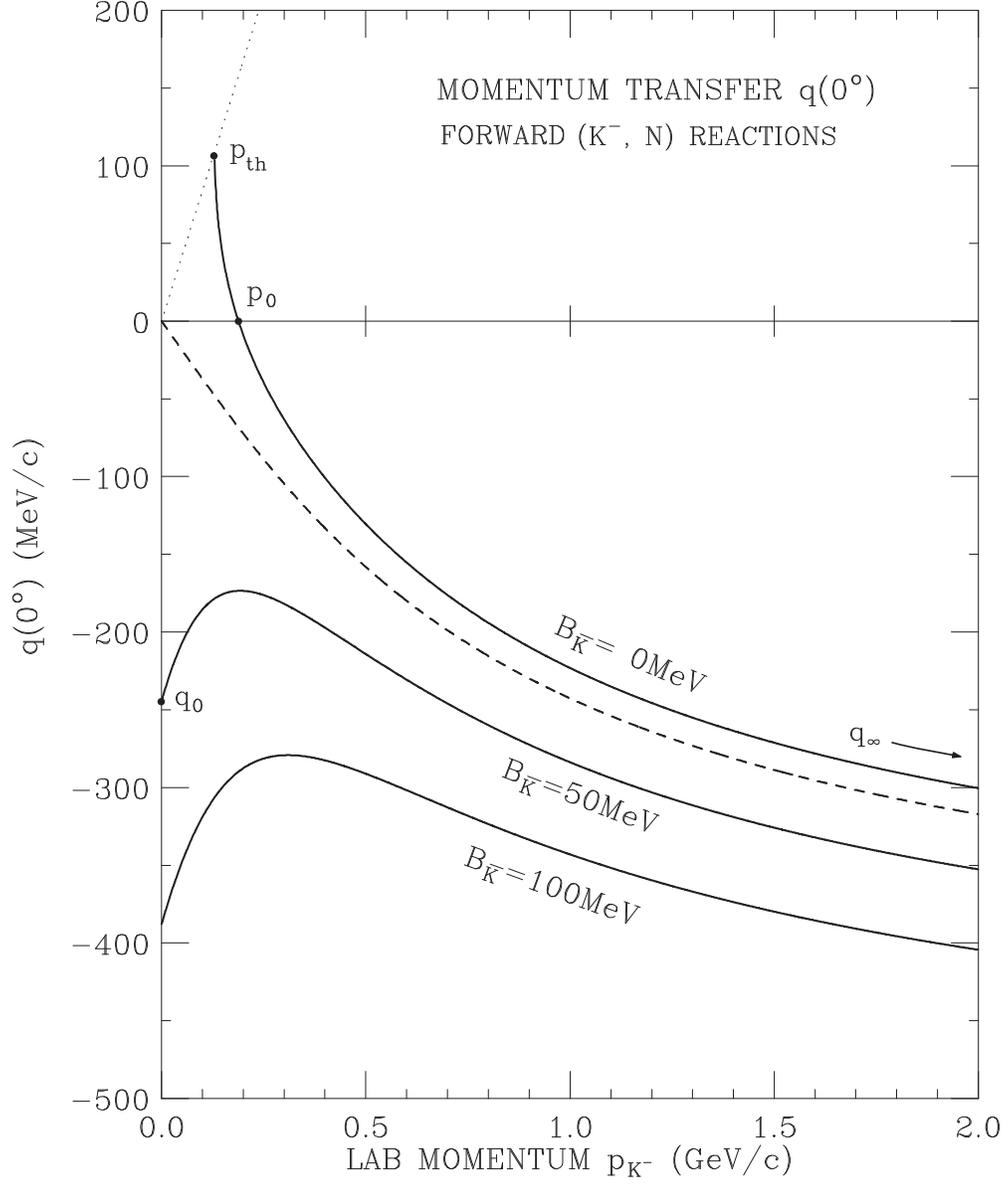}
  \caption{\label{fig:9}
   Momentum transfer $q(0^\circ)$ to the $\bar{K}$ nuclear states 
   for the forward $K^- + N \to N + \bar{K}$ reaction on the target, 
   as a function of the $K^-$ lab momentum. Here the $^{12}$C target 
   is used. The $\bar{K}$ binding energies are taken to be 
   $B_{\bar{K}}=$ 0, 50, 100 MeV. 
   The negative value of $q(0^\circ)$ means that the residual 
   $\bar{K}$ recoils backward relative to the forward incident $K^-$. 
   See also in the text. 
  }
  \end{center}
\end{figure}

Now we study the ($K^-$, $N$) reactions on the $^{12}$C 
target, following the helpful arguments 
by Dalitz and Gal \cite{Dalitz78}. 
In Fig.~\ref{fig:9}, we show the momentum transfer of 
$q(0^\circ)= p_{K^-} - p_{N}$ in the ($K^-$, $N$) reaction 
on $^{12}$C at $\theta_{\rm lab}=0^\circ$, 
as a function of the incident $K^-$ lab momentum $p_{K^-}$. 
Similar figures drawing the momentum transfer would be 
found in several papers \cite{Kishimoto99,Yamagata05}, 
but Fig.~\ref{fig:9} represents explicitly a sign of $q(0^\circ)$;  
the solid curve denotes $q(0^\circ)$ for 
$\bar{K}$ binding energies with $B_{\bar{K}}=$ 0, 50 and 100 MeV 
in the nucleus.
The dashed curve corresponds to the recoil $q(0^\circ)$ 
in the case of $B_{\bar{K}}=-\varepsilon_N=$ 16.0 MeV, which 
gives a ``nonthermic'' condition 
as $m_{K^-}+M_{A}=m_{N}+M_{B}$.

In the endothermic case of $m_{K^-}+M_{A} < m_{N} + M_{B}$, 
there is a threshold lab momentum 
$p_{\rm th}$, which is 129.4 MeV/c when we choose  
$B_{\bar{K}}=$ 0 MeV. The dotted line in Fig.~\ref{fig:9}
denotes the relation 
$q_{\rm th}=(M_{B}-m_N)/(M_{B}+m_N)p_{\rm th}$ 
at the threshold in the endothermic condition. 
Moreover, the magic momentum $p_0$ can be found at 188.1 MeV/c, 
fulfilling the recoilless condition 
$\sqrt{p_0^2+m_{K^-}^2}+M_{A}=\sqrt{p_0^2+m_{N}^2}+M_{B}$; 
its value is obtained by the relation
\begin{equation}
p_0=\sqrt{\prod{(M_{B}-M_{A} \pm m_{K^-} \pm m_N)}}
\bigg/ 2(M_{A}-M_{B}), 
\label{eqn:e30}
\end{equation}
where $\prod(a \pm b \pm c)=(a+b+c)(a+b-c)(a-b+c)(a-b-c)$, as 
given in Ref.~\cite{Dalitz78}.
In the exothermic case of $m_{K^-}+M_{A} > m_{N} + M_{B}$, 
the recoil momentum $q_0$ at $p_{K^-}=$ 0 GeV/c is 
given by
\begin{equation}
q_0=\sqrt{\prod{(m_{K^-} + M_A \pm m_N \pm M_{B})}}
\bigg/ 2(m_{K^-}+M_A), 
\label{eqn:e31}
\end{equation}
under the condition 
$m_{K^-}+M_A=\sqrt{q_0^2+m_{N}^2}+\sqrt{q_0^2+M_{B}^2}$. 
Thus, the values of $q_0$ are $-$244.2 and $-$387.8 MeV/c for 
$B_{\bar{K}}=$ 50 and 100 MeV, respectively.
Moreover, there is a point of inflection 
in $q(0^\circ)$, which is given by the condition 
${{\partial q(0^\circ)}/{\partial p_{K^-}}}=0$.
This point gives a minimum of $|q(0^\circ)|$, 
$q_{\rm min}$, of which values are $-$173.4 MeV/c
at $p_{K^-}=$ 192.6 MeV/c and $-$279.2 MeV/c at $p_{K^-}=$ 310.0 MeV/c 
for $B_{\bar{K}}=$ 50 and 100 MeV, respectively.
When we consider $p_{K^-} \to +\infty$, 
we confirm that $q(0^\circ)$ goes into the 
limit $q_{\infty} = (M_{B}^2-M_{A}^2)/2M_{A}$, 
leading to $-$420.4, $-$468.4 and $-$516.1 MeV/c for $B_{\bar{K}}=$ 0, 
50 and 100 MeV, respectively.

\Figuretable{FIG. 7}

The kinematical factor $\beta$ \cite{Tadokoro95} used in Eq.~(\ref{eqn:e22}) 
expresses the translation from the two-body $\bar{K}$-nucleon lab system 
to the many-body $\bar{K}$-nucleus lab system \cite{Dover83}.
It is defined by  
\begin{equation}
 \beta = \biggl(1+ {E^{(0)}_N \over E^{(0)}_{\bar{K}}}
        {{p^{(0)}_N - p_{K^-} \cos\theta_{\rm lab}} 
        \over p^{(0)}_N} \biggr){p_N E_N \over p^{(0)}_N E^{(0)}_N},
\label{eqn:e32}
\end{equation}
where $p_{K^-}$ and $p_N$ ($E_{\bar{K}}$ and $E_N$) are 
momenta of the incident $K^-$ and the detected $N$ 
(energies of $\bar{K}$ and $N$)
in the many-body $K^- + {A} \to N + {_{\bar{K}}{B}}$ 
reaction, respectively, 
and the quantities bearing an $(0)$ superscript are in 
the two-body $K^- + N \to N + \bar{K}$ reaction. 
For the forward direction ($\theta=0^\circ$), 
the kinematical factor $\beta$ in Eq.~(\ref{eqn:e32}) is reduced to 
\begin{equation}
 \beta(0^\circ) =
\biggl(1- {v^{(0)}_{\bar{K}} \over v^{(0)}_N} \biggr)
{p_N E_N \over p^{(0)}_N E^{(0)}_N},
\label{eqn:e33}
\end{equation}
where $v^{(0)}_N=p^{(0)}_N/E^{(0)}_N$ is a velocity of  
the detected nucleon, and 
$v^{(0)}_{\bar{K}}=q(0^\circ)/E^{(0)}_{\bar{K}}$ that of 
the residual $\bar{K}$.
The recoilless reaction under $q(0^\circ)\simeq$ 0 MeV/c 
appears in the case of $B_{\bar{K}}=$ 0 MeV and $p_{K^-}\simeq$ 0.18 GeV/c, 
leading to $\beta(0^\circ) \simeq $ 1. 
We stress that when we choose $p_{K^-} \geq$ $\sim$0.5 GeV/c, 
the value of $\beta(0^\circ)$ is larger than 1, because the momentum 
transfer $q(0^\circ) < 0$ 
and also $v^{(0)}_{\bar{K}} < 0$.
On the other hand, the hypernuclear production by ($\pi^+$,~$K^+$) and 
($K^-$,~$K^+$) reactions shows that the value of $\beta(0^\circ)$ 
is smaller than 1, 
because $q(0^\circ)= p_{\pi^+} - p_{K^+} >0$ for the $\Lambda$-hypernuclei 
\cite{Dover80} and
$q(0^\circ)= p_{K^-}- p_{K^+} >0$ for the $\Xi$-hypernuclei \cite{Dover83}. 
One should recognize the different nature of the nuclear ($K^-$,~$N$) reaction 
from the well-known ($\pi^+$,~$K^+$) and ($K^-$,~$K^+$) reactions.  

\begin{figure}[htb]
  \begin{center}
  \includegraphics[width=0.8\textwidth]{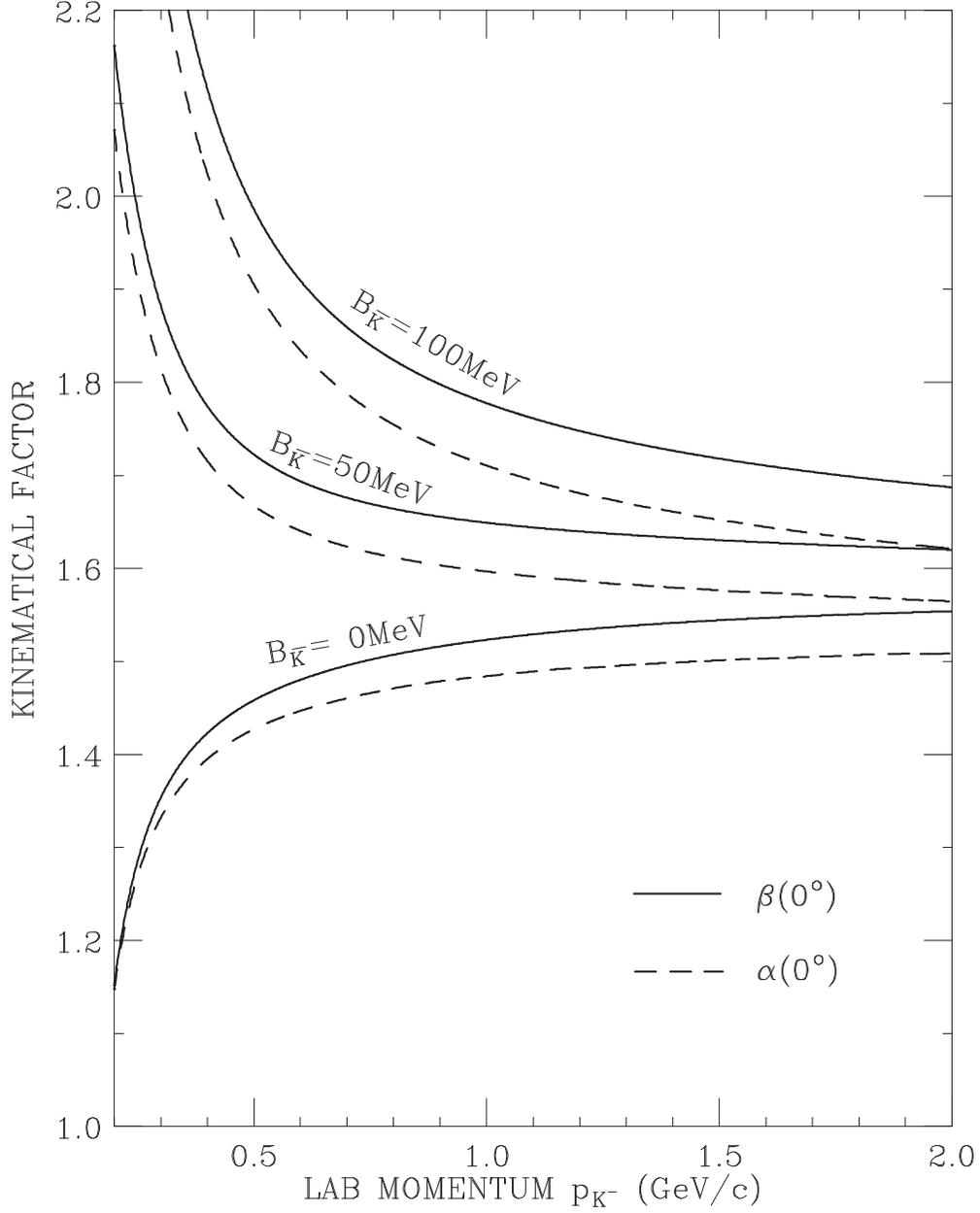}
  \end{center}
  \caption{\label{fig:10}
  The kinematical factor $\beta(0^\circ)$ and $\alpha(0^\circ)$,
  which are given by Eq.~(\ref{eqn:e33}) and Eq.~(\ref{eqn:b12}), respectively,  
  transforming from the two-body $K^-$+$N$ system to the many-body 
  $K^-$+nucleus system, as a function of the $K^-$ lab momentum $p_{K^-}$. 
  The $\bar{K}$ binding energies are taken to be 
  $B_{\bar{K}}=$ 0, 50, 100 MeV. 
  The $^{12}$C target was used. 
  }
\end{figure}

In Fig.~\ref{fig:10}, we show a kinematical factor 
$\beta(0^\circ)$ in the ($K^-$, $N$) reaction 
on the $^{12}$C target at $\theta_{\rm lab}=0^\circ$, 
as a function of the incident $K^-$ lab momentum, 
together with a kinematical factor $\alpha(0^\circ)$ 
given by Eq.~(\ref{eqn:b12}) in Appendix~\ref{sect:a2}, 
which is often used in DWIA calculations~\cite{Dover83,Cieply01}.
The calculated values also depend on the $\bar{K}$ binding 
energies of $B_{\bar{K}}$; 
we find $\beta(0^\circ)=$ 1.52, 1.62 and 1.75 for 
$B_{\bar{K}}=$ 0, 50 and 100 MeV 
at $p_{K^-}=$ 1.0 GeV/c, respectively, 
and these effects are not negligible. 
Ciepl$\acute{\rm y}$ et al.~\cite{Cieply01} calculated the integrated 
cross section for the $\bar{K}$ nuclear 
($1s$)$_{\bar{K}}$ bound state by ($K^-$, $p$) 
reactions on $^{12}$C within the DWIA.  
When we assume $B_{K^-}$= 122 MeV, which corresponds to the same binding 
energy given in Ref.~\cite{Cieply01}, for the $(1s)_{\bar{K}}$ bound state 
in the $K^-$-$^{11}$B nucleus, 
we find $\alpha(0^\circ)=$ 1.76, rather than 0.69 
obtained in Ref.~\cite{Cieply01} which seems to arise as an error 
by the opposite sign of $q(0^\circ)$. 
Obviously, the kinematical factor is needed to 
reproduce the absolute value of 
the cross section of the ($K^-$, $N$) reaction.

\Figuretable{FIG. 8}

\subsection{Eikonal distortion}
\label{sect:3-3}

Full distorted-waves of the nucleon- and $K^-$-nucleus are 
important to reproduce absolute values of the production cross sections.
Since the ($K^-$,~$N$) reaction requires a large momentum transfer and
a high angular-momentum, we simplify the computational procedure
by using the eikonal approximation to the distorted waves of the 
nucleon- and $K^-$-nucleus states 
\cite{Dover80,Hufner74,Bouyssy77,Hausmann89,Motoba88}:
\begin{eqnarray}   
\chi^{(-)*}_{N,m_s}({\bm p}_{N},{\bm r},\sigma)
&=&
\exp{\left( - i {\bm p}_{N}\cdot{\bm r} 
- \frac{i}{v_n} 
\int^{+\infty}_{z} U_N(\mbox{\boldmath{$b$}},z')dz'
\right)}\chi_{{1 \over 2},m_s}^\dagger(\sigma), \nonumber
\label{eqn:e34}
\\
\chi^{(+)}_{K^-}({\bm p}_{K^-},{\bm r})
&=&
\exp{\left( + i{\bm p}_{K^-}\cdot{\bm r}
- \frac{i}{v_{K^-}} 
  \int^z_{-\infty} U_{K^-}({\bm b},z')dz'
\right)}
\label{eqn:e35}
\end{eqnarray}
with an impact parameter coordinate ${\bm b}$ and 
the optical potential for $\lambda=$ $N$ or $K^-$,
\begin{eqnarray}   
U_{\lambda}(r) = -i \, {v_{\lambda} \over 2} \, 
\bar{\sigma}^{\rm tot}_{\lambda N} \, (1 - i \alpha_{\lambda N}) \, \rho(r), 
\label{eqn:e36}
\end{eqnarray}
where $\rho(r)$ is a nuclear density distribution, and 
$\bar{\sigma}^{\rm tot}_{\lambda N}$ and $\alpha_{\lambda N}$ 
are an isospin-averaged total cross section
and a ratio of the real to imaginary parts of the forward amplitude
for the $\lambda + N$ scattering, respectively. 
$\chi_{{1 \over 2},m_s}(\sigma)$ is a spin-(${1 \over 2}$,$m_s$) state 
of the nucleon. 
Reducing the r.h.s. in Eq.~(\ref{eqn:e35}) 
by partial waves expansion, 
the distorted-waves at $\theta_{\rm lab}$ in Eq.~(\ref{eqn:e23}) 
can be expressed as
\begin{eqnarray}
&&\chi^{(-)*}_{N,m_s}({\bm p}_{N},{\bm r},\sigma)
\chi^{(+)}_{K^-}({\bm p}_{K^-},{\bm r}) \nonumber\\
&&=\sum_{J=L \pm {1 \over 2},M'} \sqrt{4 \pi (2L+1)} i^L 
\tilde{j}_{LM}(p_N, p_{K^-}, \theta_{\rm lab}; r)
[Y_{LM}(\hat{\bm r}) \otimes \chi_{{1 \over 2},m_s}^\dagger(\sigma)]_{JM'}, 
\label{eqn:e37}
\end{eqnarray}
where $\tilde{j}_{LM}(p_N, p_{K^-}, \theta_{\rm lab}; r)$
is a radial part of the distorted-wave for angular-momentum transfer $L$, 
depending on the momentum transfer $q$.
If the distortion is switched off, i.e., 
$\bar{\sigma}^{\rm tot}_{N N}=
\bar{\sigma}^{\rm tot}_{K^- N}=0$, 
it becomes ${j}_{L}(qr)\delta_{M0}$ for the plane-wave approximation,
where ${j}_{L}(x)$ is the spherical Bessel function. 

\subsection{Isospin states}
\label{sect:3-4}

Now we discuss isospin properties of the residual $\bar{K}$-nuclear 
bound states in the forward ($K^-$, $p$) 
and ($K^-$, $n$) reactions on a nuclear target. 
Dover and his collaborators examined that the isospin properties 
in ($K^-$, $\pi^\mp$) and ($\pi^\pm$, $K^+$) reactions for 
$\Sigma$-hypernuclear production \cite{Dover89}, 
and also ($K^-$, $K^+$) reactions for 
$\Xi$-hypernuclear production \cite{Dover83}. 
We attempt to apply their treatment into the $\bar{K}$ nuclear bound 
states, and to discuss the ($K^-$, $n$) and ($K^-$, $p$) reactions on 
the nuclear target.
The isospin vectors relevant 
in the nuclear $A$($K^-$, $N$)$B$ reaction, 
and their $z$-components, satisfy
\begin{equation}
{\bm {1 \over 2}} + {\bm T}_{A} = {\bm {1 \over 2}} + {\bm T}_{B}, 
\qquad i_{K^-} + \tau_{A} = i_{N} + \tau_{B},
\label{eqn:e41}
\end{equation}
respectively.
For instance, we obtain the $\bar{K}$ nuclear states 
by the ($K^-$, $p$) reaction 
on the $^{12}$C target with $J_A^{\pi}=$ 0$^+$ and $T_A=$ 0 as 
\begin{equation}
|^{11}_{\bar{K}}{\rm Be}\rangle_{T_B=1,\tau_B=-1}  
= |K^- \rangle | ^{11}{\rm B} \rangle
\label{eqn:e38}
\end{equation} 
for $T_B=$ 1 states, and by the ($K^-$, $n$) reactions, 
\begin{eqnarray}
|^{11}_{\bar{K}}{\rm B}\rangle_{T_B=1,\tau_B=0} 
&=& \sqrt{1 \over 2} |\bar{K}^0 \rangle | ^{11}{\rm B}\rangle
  + \sqrt{1 \over 2} |K^- \rangle | ^{11}{\rm C} \rangle, \nonumber 
\label{eqn:e39}\\
|^{11}_{\bar{K}}{\rm B}\rangle_{T_B=0,\tau_B=0} 
&=& \sqrt{1 \over 2} |\bar{K}^0 \rangle | ^{11}{\rm B}\rangle
  - \sqrt{1 \over 2} |K^- \rangle | ^{11}{\rm C} \rangle,
\label{eqn:e40}
\end{eqnarray}
for both $T_B=$ 1 and $T_B=$ 0 states. 

Using the Fermi-averaged elementary amplitudes 
$\bar{f}_{K^-N \to N \bar{K}}$ which give the isospin transition 
with $\Delta T=$ 1 and $\Delta T=$ 0,  
we obtain the explicit isospin dependence of the nuclear amplitudes
for the nuclear ($K^-$, $N$) reactions
involving an isospin transition 
($T_A$, $\tau_A$) $\to$ ($T_B$, $\tau_B$) \cite{Dover83};
\begin{eqnarray}
F^{(K^-,\,p)}
& = & \sqrt{3(2T_A+1)(2T_B+1)}(-)^{T_B-\tau_B}
    \left( \begin{array}{ccc}
                  T_B  &  1  &  T_A \\
             -\tau_B   &  -1 &  \tau_A    \\
    \end{array}  \right)  \nonumber\\
&& \times (-)^{T_B+T_C-{1 \over 2}}
    \left\{ \begin{array}{ccc}
                  {1 \over 2}  &  T_B  &  T_C \\
                  T_A   & {1 \over 2} &  1    \\
    \end{array}  \right\} \bar{f}^{(1)}, \label{eqn:e42}\\
F^{(K^-,\,n)} 
& = & \delta_{T_A,T_B}{1 \over 2}\bar{f}^{(0)}
-\sqrt{{ 3(2T_A+1)(2T_B+1) \over 2}}(-)^{T_B-\tau_B}
    \left( \begin{array}{ccc}
                  T_B  &  1  &  T_A \\
             -\tau_B   &  0 &  \tau_A    \\
    \end{array}  \right)  \nonumber\\
&& \times (-)^{T_B+T_C-{1 \over 2}}
    \left\{ \begin{array}{ccc}
                  {1 \over 2}  &  T_B  &  T_C \\
                  T_A   & {1 \over 2} &  1    \\
    \end{array}  \right\} \bar{f}^{(1)}, \label{eqn:e43}
\end{eqnarray}
where $\bar{f}^{(0)}$ and $\bar{f}^{(1)}$ are 
the Fermi-averaged $t$-channel 0 and 1 
amplitudes, respectively, and $T_C$ denotes the 
isospin of the core-nucleus. 
These expression indicates clearly that 
the nuclear ($K^-$, $p$) reaction has $\Delta T=$ 1, 
while the nuclear ($K^-$, $n$) reaction $\Delta T=$ 1, 0.

When we assume closed-shells targets with $J_A^{\pi}=$ 0$^+$ 
and $T_A=$ 0 such as $^4$He, $^{12}$C and $^{28}$Si, 
we find that only $T_B=1$ levels are excited 
by the ($K^-$, $p$) reaction 
due to the restriction to $T_A=0$, as seen in Eq.(\ref{eqn:e42}).
Thus, we obtain
\begin{equation}
F^{(K^-,\, p)}_{T_A=0, T_B=1}
= - {1 \over \sqrt{2}}\bar{f}^{(1)}.
\label{eqn:e44}
\end{equation}
For the ($K^-$, $n$) reaction, both $T_B=0, 1$ types of levels can 
be excited in the $\bar{K}$ nuclear states; 
\begin{equation}
F^{(K^-,\, n)}_{T_A=0, T_B=0}= {1 \over 2}\bar{f}^{(0)},
\quad
F^{(K^-,\, n)}_{T_A=0, T_B=1}= {1 \over 2}\bar{f}^{(1)}.
\label{eqn:e45}
\end{equation}
In order to clarify the isospin properties of the lab cross section 
with the final $T_B$ states in the $\bar{K}$ nuclear bound states, 
therefore, we evaluate the ratios of the cross sections 
\begin{equation}
{ \sigma(K^-, n)_{T_A=0, T_B=1} \over \sigma(K^-, p)_{T_A=0, T_B=1}}
= {1 \over 2}, 
\label{eqn:e46}
\end{equation}
and for the summed cross section
\begin{eqnarray}
&&{ \sum_B \sigma(K^-,\, n)_{T_A=0, T_B=0} \over 
\sum_B \sigma(K^-,\, p)_{T_A=0, T_B=1}}
= {1 \over 2}
  {|\bar{f}^{(0)}|^2 
   \over |\bar{f}^{(1)}|^2} \nonumber\\
&&= {1 \over 2}
  {|f_{K^-n \to nK^-}+f_{K^-p \to n\bar{K}^0}|^2 \over |f_{K^-p \to pK^-}|^2}
= {1 \over 2}
  \left|{2
  {f_{K^-n \to nK^-} \over f_{K^-p \to pK^-}}} -1 \right|^2.
\label{eqn:e47}
\end{eqnarray}
Here we used the last equality in  Eq.(\ref{eqn:e17}). 
In the case that the forward $K^- + n \to n + K^-$ cross section 
is much larger than the forward $K^- + p \to p + K^-$ cross section
at 1.0 GeV/c, the ratio (\ref{eqn:e47})
shows approximately  
\begin{eqnarray}
{\sum_B \sigma(K^-, n)_{T_A=0, T_B=0} \over 
\sum_B \sigma(K^-, p)_{T_A=0, T_B=1}}
& \simeq & 2 
{\langle d \sigma/d\Omega \rangle_{0^\circ}^{K^-n \to n {K}^-} \over 
 \langle d \sigma/d\Omega \rangle_{0^\circ}^{K^-p \to p {K}^-}} 
= 2 \times {13.9 \over 5.2} \simeq 5.3. 
\label{eqn:e48}
\end{eqnarray}
Comparing Eq.(\ref{eqn:e46}) with Eq.(\ref{eqn:e48}), we obtain roughly the ratio
\begin{eqnarray}
{\sum_B \sigma(K^-, n)_{T_A=0, T_B=0} \over 
 \sum_B \sigma(K^-, n)_{T_A=0, T_B=1}}
 \simeq 10,
\label{eqn:e49}
\end{eqnarray}
which shows a strong preference for the excitation of $T_B=0$ states 
by the ($K^-$, $n$) reaction.
This fact indicates that the ($K^-$, $n$) reaction on the closed-shells 
targets $J_A^{\pi}=$ 0$^+$, $T_A=$ 0 
can populate dominantly isospin $T_B=0$ states.
We obtain the isoscalar $\Delta T=$ 0 and 
isovector $\Delta T=$ 1 transition cross sections in the lab frame,
\begin{eqnarray}
\sigma(\Delta T= 0)
&=& {1 \over 2}|\bar{f}^{(0)}|^2 
=\left|{1 \over \sqrt{2}} \bar{f}_{K^-n \to nK^-}+
 {1 \over \sqrt{2}} \bar{f}_{K^-p \to n\bar{K}^0} \right|^2, 
\label{eqn:e50} \nonumber \\
\sigma(\Delta T= 1)
&=&
{1 \over 2}|\bar{f}^{(1)}|^2
=\left|{1 \over \sqrt{2}} \bar{f}_{K^-n \to nK^-}-
 {1 \over \sqrt{2}} \bar{f}_{K^-p \to n\bar{K}^0} \right|^2, 
\label{eqn:e51}
\end{eqnarray}
for the nuclear ($K^-$, $n$) reaction on the closed-shell 
targets. 
These transition cross sections in Eq.~(\ref{eqn:e51}) are 
rewritten as 
\begin{eqnarray}
\sigma(\Delta T= 0)
&=& {1 \over 2}\sigma(\bar{K}^0)+{1 \over 2}\sigma(K^-)
+ \sqrt{\sigma(\bar{K}^0)\sigma(K^-)}
\cos{\varphi_R},
\nonumber
\label{eqn:e51a} \nonumber\\
\sigma(\Delta T= 1)
&=& {1 \over 2}\sigma(\bar{K}^0)+{1 \over 2}\sigma(K^-)
- \sqrt{\sigma(\bar{K}^0)\sigma(K^-)}
\cos{\varphi_R},
\label{eqn:e51a} 
\end{eqnarray}
where $\sigma(\bar{K}^0)$ and $\sigma(K^-)$ are the Fermi-averaged 
cross sections of the $\bar{K}^0$ and the $K^-$ production, respectively,
in the ($K^-$, $n$) reaction. $\varphi_R$ denotes 
the relative phase between 
$\bar{f}_{K^-n \to nK^-}$ and $\bar{f}_{K^-p \to n\bar{K}^0}$.

\begin{figure}[thb]
  \begin{center}
  \includegraphics[width=0.79\textwidth]{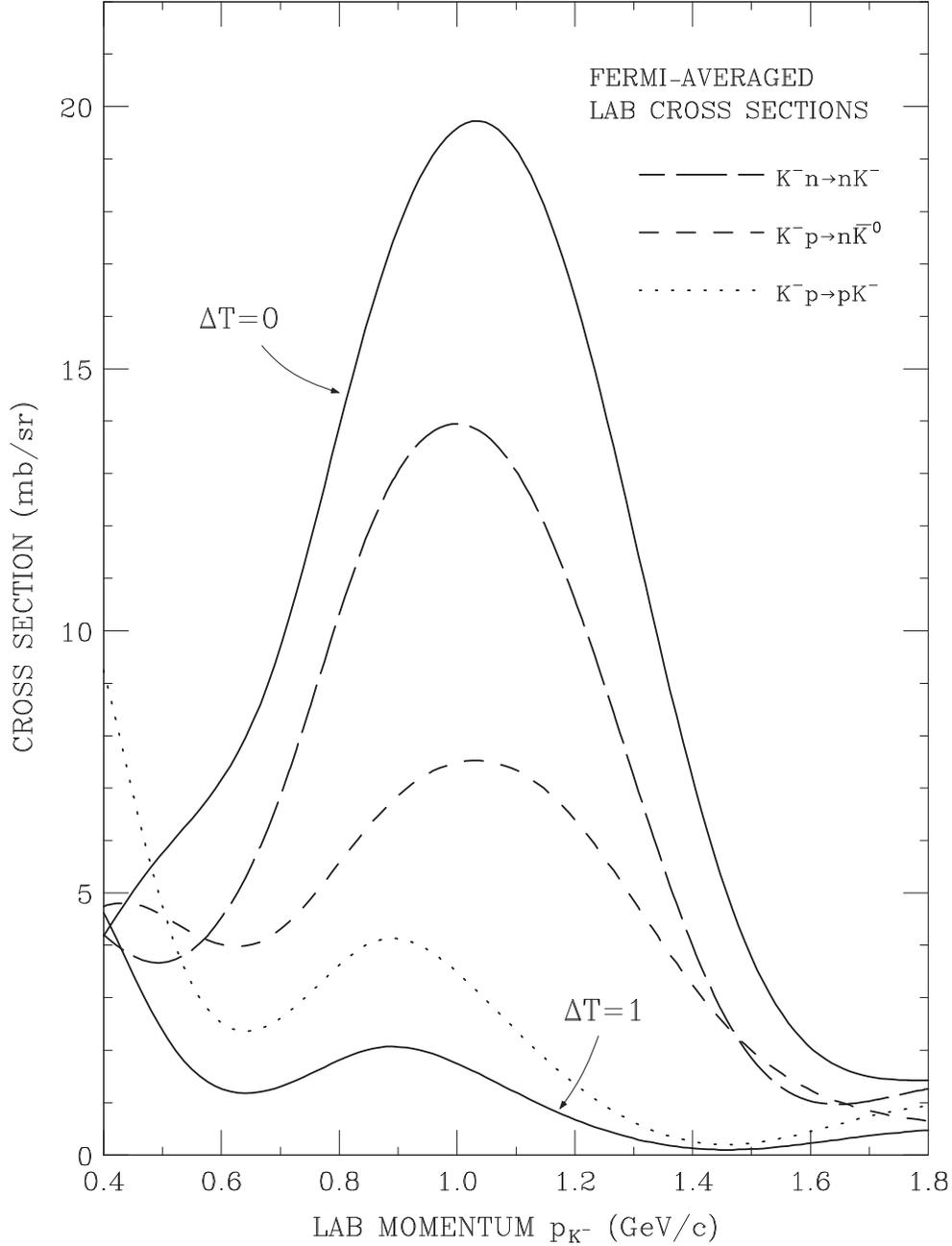}
  \end{center}
  \caption{\label{fig:11}
  The absolute values of the Fermi-averaged lab cross sections at 
  $\theta_{\rm lab}=0^\circ$ for the isospin $\Delta T=0$, $1$ 
  states for the ($K^-$, $n$) reaction 
  on an isospin-0 target such as $^{12}$C, 
  as a function of the $K^-$ lab momentum $p_{K^-}$. 
  The Fermi-averaged lab cross sections for $K^- + n \to n + K^-$, 
  and $K^- + p \to n + \bar{K}^0$, and 
  $K^- + p \to p + K^-$ processes are also shown.  
  The free space amplitudes of Gopal et al.~\cite{Gopal77} were used.
  }
\end{figure}

In Fig.~\ref{fig:11}, we show the Fermi-averaged transition 
cross sections of $\sigma(\Delta T=~0)$ and $\sigma(\Delta T=~1)$, 
together with $\sigma(\bar{K}^0)$ and $\sigma(K^-)$, 
and also that for $K^- + p \to p + K^-$.
We find that the magnitude of $\Delta T= 0$ is quite larger
than that of $\Delta T= 1$ over a wide momentum range, 
since the interference terms between 
$K^-+n \to n+K^-$ and $K^-+p \to n+\bar{K}^0$ channels 
in the r.h.s. of Eq.~(\ref{eqn:e51a})
are positive in the momentum region of $p_{K^-}=$ 0.6-2.0 GeV/c 
where the relative phase has 14.4$^\circ$ $\leq \varphi_R \leq$ 64.1$^\circ$, 
leading to $\sigma(\Delta T= 0) > \sigma(\Delta T= 1)$.
Indeed, the value of $\sigma(\Delta T= 0)$ at $p_{K^-}=$ 1.0 GeV/c
amounts to 19.6 mb/sr, while the value of 
$\sigma(\Delta T= 1)$ is 1.7 mb/sr, which corresponds 
to a half of the cross section of the $K^- + p \to p + K^-$ reaction. 
The interference effects are significant 
and have to be considered in calculations of cross sections 
for nuclear targets.

\Figuretable{FIG. 9}

\begin{figure}[thb]
  \begin{center}
  \includegraphics[width=0.79\textwidth]{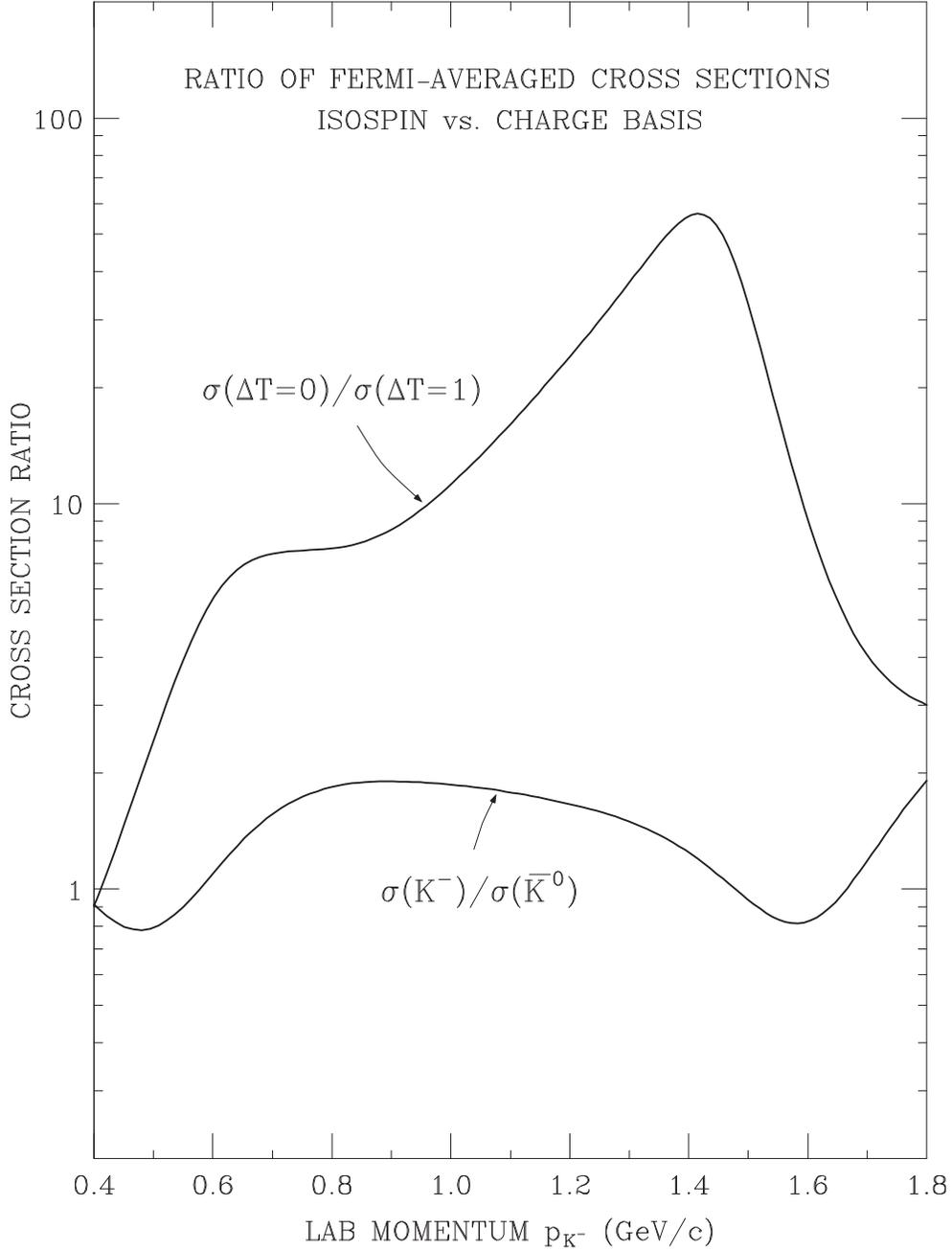}
  \caption{\label{fig:12}
  Ratio of the Fermi-averaged cross section 
  $\sigma$($\Delta T$=0)/$\sigma$($\Delta T$=1)
  in isospin basis or $\sigma(K^-)/\sigma(\bar{K}^0)$ in the charge basis, 
  as a function of $K^-$ lab momentum $p_{K^-}$. 
  The free space amplitudes of Gopal et al. \cite{Gopal77} were used.
  These ratios refer to relative ($K^-$, $n$) cross sections for an isospin-0
  target, leading to final $\bar{K}$ nuclear configuration having a good 
  isospin, or, a specific charge state.  
  }
  \end{center}
\end{figure}

In Fig.~\ref{fig:12}, we show the ratios of the Fermi-averaged cross 
sections $\sigma(\Delta T=0)/\sigma(\Delta T=1)$ in isospin base
and $\sigma(K^-)/\sigma(\bar{K}^0)$ in charge base
in the nuclear ($K^-$, $n$) reactions, as a function of $p_{K^-}$. 
We find that the excitation of the isospin states with $\Delta T=0$ 
dominates the 0.6-1.6 momentum region.
Particularly, the ratio of $\sigma(\Delta T=0)/\sigma(\Delta T=1)$ is 
quite large near 1.4 GeV/c, because the cross section of 
$\sigma(\Delta T=1)$ is reduced, as seen in Fig.~\ref{fig:11}. 
On the other hand, the cross sections of charged $K^-$ and 
$\bar{K}^0$ states in this region 
are not so different from each other; 
$\sigma(K^-)/\sigma(\bar{K}^0)\simeq$ 1-2.
We recognize that 
the contributions of both $K^-$ and $\bar{K}^0$
are necessary to explain the cross section in the 
nuclear ($K^-$, $n$) reaction.

\Figuretable{FIG. 10}

\subsection{$\bar{K}$-nucleus potentials 
and $\bar{K}$ nuclear bound states}
\label{sect:3-5}

The $\bar{K}$-nucleus potentials have been obtained by analyzing 
strong-interaction shifts and widths of $K^-$ atomic X-ray data 
of 65 data points \cite{Batty97}.
However, it is known that it is difficult to clarify the geometry and 
strengths inside the nucleus from the $K^-$ atomic data, 
because the calculated values of the shifts and widths are sensitive 
to almost only a tail part of the potential 
outside the nuclear surface \cite{Batty97}.
Hence, several types of the $\bar{K}$-nucleus potential have 
been proposed to reproduce the $K^-$ atomic data, 
taking theoretical constraints into consideration:
(a) the $t_{\rm eff}\rho$-type potential \cite{Batty81} determined earlier 
by an effective $\bar{K}N$ scattering length, e.g., 
${\bar a}= 0.62 +i 0.92$ fm which corresponds to
$U_{\bar{K}}(0)=$ $-89 -i133$ MeV at the nuclear center in $^{11}$B;
(b) the density-dependent (DD) potential \cite{Batty97} determined 
by fitting its phenomenological parameters 
based by the $\bar{K}N$ scattering length;
(c) the Chiral-motivated potential \cite{Kaiser95,Waas97,Oset98} constrained 
microscopically by the $\bar{K}N$-$\pi Y$ 
coupled-channels describing the $\Lambda$(1405) in the nuclear medium, 
leading to a strongly attraction of ${\rm Re}U_{\bar{K}}\simeq$ $-$110 MeV. 
But a version of 
the potentials calculated {\it self consistently} in the in-medium propagator 
yields ${\rm Re}U_{\bar{K}}\simeq$ $-$40-($-$60) MeV 
\cite{Lutz98,Schaffner00,Ramos00,Hirenzaki00,Baca00}; 
(d) the relativistic mean-field (RMF) potential calculated for
the L-SH or NL-SH model with the $\alpha_\sigma$ and 
$\alpha_\omega$ parameters \cite{Mares06}.
One of the current problems in $\bar{K}$-nuclear physics is 
whether the depth of the $\bar{K}$-nucleus potential is 
deep, ${\rm Re}U_{\bar{K}} \simeq$ $-$100-($-$200) MeV, 
or shallow, ${\rm Re}U_{\bar{K}} \simeq$ $-$40-($-$60) MeV.

In this paper, we employ the {\it deep} $\bar{K}$-nucleus DD 
potential \cite{Batty97} involving an isovector term,
in order to see the isospin dependence of the 
deeply-bound $\bar{K}$-nuclear states which are expected to be 
observed as relative narrow signals 
if these bound states exist below the $\pi\Sigma$-threshold.
The $\bar{K}$-nucleus DD potential \cite{Batty97} is given as 
\begin{equation}
  {2 \mu}U_{\bar{K}}(r)
  = - {4 \pi}\left( 1 + { \mu \over m_{\rm N} } \right)
  \left\{ \left[ b_0 + B_0 ({\rho(r) \over \rho(0)})^\alpha \right] \rho(r) 
  + \left[ b_1 + B_1 ({\rho(r) \over \rho(0)})^\alpha \right] 
  \delta \rho(r)\right\}, 
\label{eqn:e56}
\end{equation}
where the point nucleon distribution 
$\rho(r)=\rho_p(r) + \rho_n(r)$ and 
the isovector one $\delta \rho(r)=\rho_n(r) - \rho_p(r)$, 
and phenomenological parameters
$b_0=-0.15+i\,0.62$ fm, $b_1=-0.20+i\,0.15$ fm, $B_0=1.58+i\,0.02$ fm, 
$B_1= 0.0$ fm and $\alpha=$ 0.17. 
It is also rewritten in a general form by  
\begin{equation}
U_{\bar{K}}(r)=U_0^{\bar{K}}(r) + 
U_1^{\bar{K}}(r)({\bf T}_C \cdot{\bf t}_{\bar{K}})/A_{\rm core}, 
\label{eqn:e56a}
\end{equation}
where ${\bf T}_C$ is the isospin operator of the 
core nucleus with its $z$-projection $\tau_C= (Z-N)/2$, 
and ${\bf t}_{\bar{K}}$ is the $\bar{K}$ isospin operator 
with $\tau_{\bar{K}}=$ $\pm 1/2$ for the ($\bar{K}^0, K^-$) 
isodoublet;
$U_0^{\bar{K}}$ and $U_1^{\bar{K}}$ are 
isoscalar and isovector (Lane term) parts of the potential, respectively.
This real parts of the potential 
has the attraction corresponding to 
${\rm Re}U_{\bar{K}} \simeq$ $-$150-($-$200) MeV 
at nuclear matter density.

The study of the ($K^-$, $n$) reaction on the closed shells targets
yields information concerning the potential properties of the isoscalar and 
isovector parts in the $\bar{K}$-nucleus potential,
because the reaction can populate both isospin $T_B=$ 0 and $T_B=$ 1 states. 
The configurations of the potentials for isospin $T_B=$ 0, 1 
states represent 
\begin{eqnarray}
U_{\bar{K}}^{T_B=0}
&=& U_0^{\bar{K}} - U_1^{\bar{K}}/{4A_{\rm core}}, 
\nonumber \\
U_{\bar{K}}^{T_B=1}
&=& U_0^{\bar{K}} + U_1^{\bar{K}}/{4A_{\rm core}}, 
\label{eqn:e57}
\end{eqnarray}
whereas the ($K^-$, $p$) reaction gives information on only
$U_{\bar{K}}^{T_B=1}$ for $T_B=$ 1 states.

\Figuretable{FIG. 11}

\begin{figure}[htb]
  \begin{center}
  \includegraphics[width=0.65\textwidth]{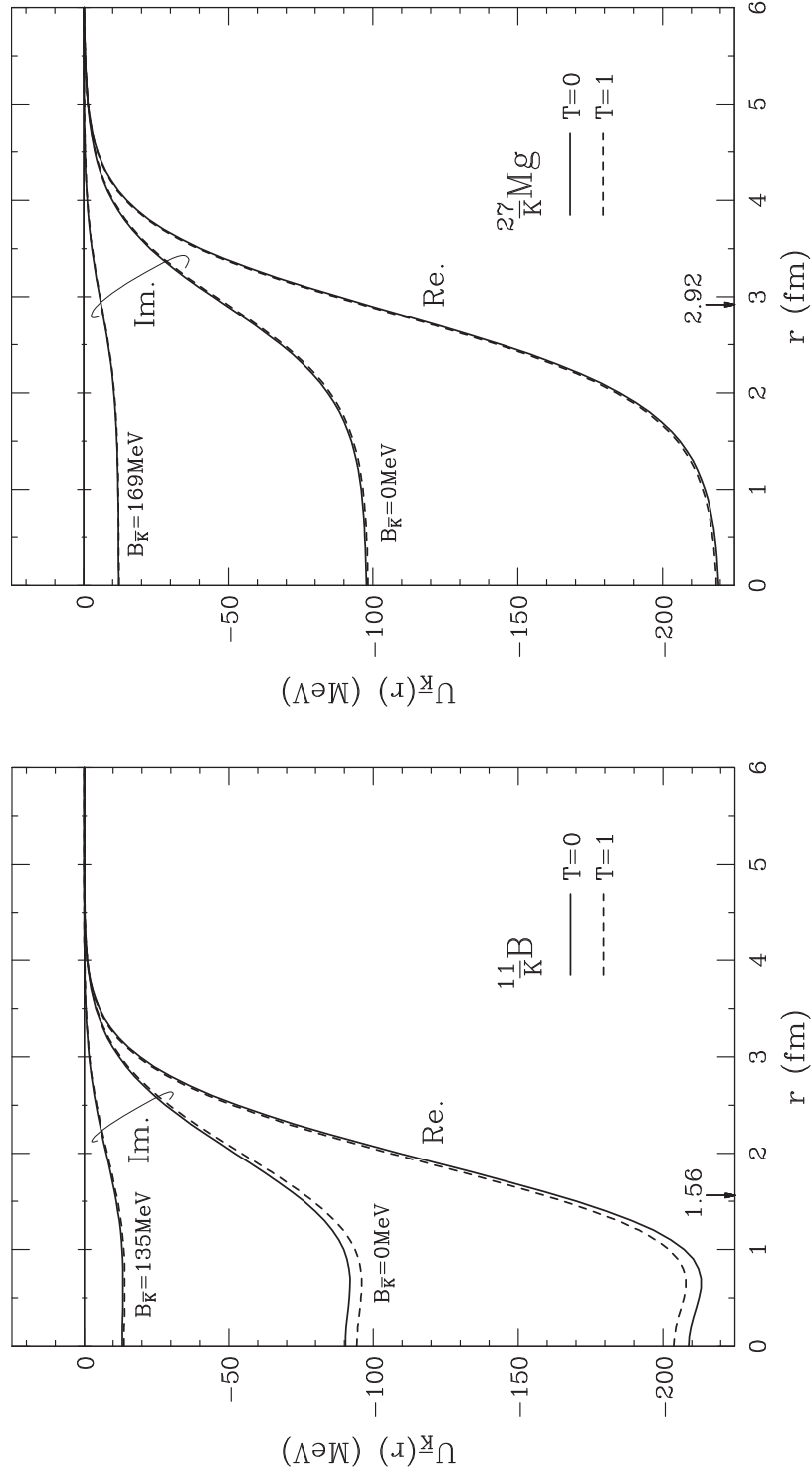}
  \caption{\label{fig:13}
  Real and imaginary parts of the $\bar{K}$-nucleus DD potentials 
  for (left) $^{11}_{\bar{K}}$B and 
  (right) $^{27}_{\bar{K}}$Mg.
  The solid and dashed curves denote the potentials for total isospin 
  of $T=$ 0 and 1 states, respectively. The imaginary parts of the 
  potentials draw for $B_{\bar{K}}=$ 0 MeV and 135 MeV to see the effects 
  of the phase space factors given by Mare$\breve{\rm s}$ \cite{Mares06}. 
  }
  \end{center}
\end{figure}

In Fig.~\ref{fig:13}, we illustrate the real and imaginary parts of 
the $\bar{K}$-nucleus DD potentials for isospin $T=$ 0, 1 
states in $A=$ 11 and 27
which are populated via nuclear ($K^-$, $N$) reactions on 
the $^{12}$C and $^{28}$Si targets, respectively.
We find that ${\rm Re}\, U_0^{\bar{K}}(0) \simeq$ $-$210-($-$220) 
MeV and ${\rm Re}\, U_1^{\bar{K}}(0) \simeq$ 120 MeV 
at the nuclear center.
It is also shown that the value of 
${\rm Re}\, U_1^{\bar{K}}(0) \sim$ 152-210 MeV
is obtained in a Breuckner calculation with 
$\bar{K}N$ interaction by fits to the low-energy 
$\bar{K}N$ scattering data \cite{Akaishi02}.
However, the isospin dependence on the $U_1^{\bar{K}}$ 
Lane term has to be reduced by $A^{-1}_{\rm core}$.
Recently, Kishimoto et al. \cite{Kishimoto05} have suggested  
${\rm Re}\, U_{\bar{K}}^{T_B=0} \sim$ $-$190 MeV
and 
${\rm Re}\, U_{\bar{K}}^{T_B=1} \sim$ $-$160 MeV 
for $^{11}_{\bar{K}}$B by the DWIA analysis of 
the $^{12}$C(in-flight $K^-$, $N$) data (PS-E548) 
at KEK. This means ${\rm Re}\, U_1^{\bar{K}} \sim$ 660 MeV.
Consequently, this value seems to be too large in terms of 
the isospin dependence of the $\bar{K}$-nucleus potential. 

The search for deeply-bound $\bar{K}$ states
is one of the most important subjects in $\bar{K}$ 
nuclear physics.
However, signals of the states lying above 
the $\pi\Sigma$-threshold might be unclear in the 
($K^-$, $p$) spectrum due to large decay widths of 
the main $\bar{K}N \to \pi\Sigma$ channels. 
If the deeply-bound $\bar{K}$ state exists below 
the $\pi \Sigma$-threshold, 
as discussed by Yamazaki and Akaishi \cite{Yamazaki02} 
and by Mare$\breve{\rm s}$ et al.~\cite{Mares06},
one expects that a clear signal of the bound state with the 
relative narrow width $\varGamma_{\bar{K}} \simeq$ 20-50 MeV 
can be observed.
In order to take account of these effects, 
we introduce the imaginary parts of the $\bar{K}$-nucleus 
potentials including a phase space factor $f(E)$; 
\begin{equation}
{\rm Im}\, U_{\bar{K}}(r) 
\to {\rm Im}\, U_{\bar{K}}(r) \times f(E),
\label{eqn:e56b}
\end{equation}
as attempted in Refs.~\cite{Ikuta02,Yamagata06}, 
Here we used the phase space factor $f(E)$ by
Mare$\breve{\rm s}$ et al.~\cite{Mares06}.

We calculate the $\bar{K}$ nuclear {\it unstable} bound states 
by solving the Schr\"odinger equation 
instead of the Klein-Gordon equation for simplicity.
Since the Hamiltonian $H=T+U_{\bar{K}}$ is not Hermitian, 
we have 
\begin{eqnarray}
&H \, {\varphi}_{n \ell} &= E_{n \ell} \, {\varphi}_{n \ell}, \nonumber\\
&H^\dagger \, \tilde{\varphi}_{n \ell} &= E^*_{n \ell} \, \tilde{\varphi}_{n \ell}, 
\label{eqn:e57c}
\end{eqnarray}
where $E_{n \ell}$ is a complex eigenvalues for the $\bar{K}$ unstable 
bound states.
Thus, the $\bar{K}$ nuclear binding energies and widths
can be evaluated as 
\begin{eqnarray}
E_{n \ell}={(k^{(pole)}_{n \ell})^2/2 \mu}
=-B_{n \ell}-i{\varGamma_{n \ell}/2},
\label{eqn:e57a}
\end{eqnarray}
where $k^{(pole)}_{n \ell}$ denotes a pole position of the bound state 
in the complex momentum plane,
and $\mu$ is the reduce mass of the $\bar{K}$-nucleus system. 
${\varphi}_{n \ell}$ is a wave function for the eigenstate labeled by $E_{n \ell}$, 
and $\tilde{\varphi}_{n \ell}$ is the wave function given by a biorthogonal set;
its conjugate state becomes $(\tilde{\varphi}_{n \ell})^*={\varphi}_{n \ell}$.  
It is noticed that such radial wave functions must be normalized 
by so-called $c$-products \cite{Berggren68},  
\begin{equation}
\int_0^{\infty} r^2d{r} 
(\tilde{\varphi}_{n \ell}(r))^*{\varphi}_{n \ell}(r)=
\int_0^{\infty} r^2d{r} ({\varphi}_{n \ell}(r))^2=1, 
\label{eqn:e56c}
\end{equation}
under the boundary condition for decaying states \cite{Kapur38}. 
Therefore, we can obtain the completeness relation for Green's function as
\begin{eqnarray}
G_\ell(\omega;{r}',{r})
&=&
\sum_n {\varphi_{n \ell}(r')(\tilde{\varphi}_{n \ell}(r))^* \over 
\omega-E_{n \ell}+i\epsilon} + 
{2 \over \pi}\int_0^\infty dk {k^2 S_{\ell}(k) 
u_{\ell}(k,r')(\tilde{u}_{\ell}(k,r))^* \over \omega-E_k+i\epsilon},
\label{eqn:e57d}
\end{eqnarray}
where the summation over $n$ includes all the 
pole of the $S$-matrix in the complex $k$-plane.

In Table~\ref{tab:table2}, we show the numerical results of the 
$K^-$ nuclear binding energies and widths for 
$^{11}_{K^-}$Be=[$K^-$-${^{11}{\rm B}}$] and 
$^{27}_{K^-}$Mg=[$K^-$-${^{27}{\rm Al}}$],
considering the effects of the Coulomb potentials and 
of the imaginary parts of the DD potentials.
Note that the relativistic effects for the $\bar{K}$ binding 
energy and the core-polarization 
(shrinkage effects for the core-nucleus)  
are not taken into account. 
In order to see the effects of the widths,  we listed up 
the calculated results without the imaginary parts of the DD potential 
(Real only).
The wave function for this state resides in the nuclear inside, 
e.g., the rms radius $\sqrt{\langle r^2 \rangle}$= 1.3 fm 
for $(1s)_{K^-}$ in $^{11}_{K^-}$Be, in comparison to the rms 
radius of the $^{11}$B core-nucleus, 2.28 fm.
When the imaginary part multiplied by the phase space 
factor \cite{Mares06} is switched on (Full), 
we find that the $(1s)_{K^-}$ bound state in $^{11}_{K^-}$Be 
have a relative narrow width of $\varGamma_{\over{K}} \simeq$ 25 MeV 
because its pole arises at $k^{(pole)}_{1s}=-0.09 +i 1.82$ fm$^{-1}$ 
below the $\pi\Sigma$ decay threshold. 
For $^{27}_{K^-}$Mg, we find that the $(1s)_{K^-}$ state has a narrow width 
of $\varGamma_{\over{K}} \simeq$ 22 MeV and a rms radius of 
$\sqrt{\langle r^2 \rangle}$= 1.6 fm, 
and $(1p)_{K^-}$ state also $\varGamma_{\over{K}} \simeq$ 25 MeV 
and $\sqrt{\langle r^2 \rangle}$= 2.0 fm.
The $K^-$ is located at the nuclear inside
in terms of the rms radius of 2.93 fm for $^{27}$Al.
As several $K^-$ nuclear bound states with broad widths of 
$\varGamma_{\over{K}} \simeq$ 100 MeV exist,  
the shape of the peaks in the spectrum do not 
necessarily correspond to that of a standard Breit-Wigner 
resonance located at $B_{n \ell}$ \cite{Morimatsu94}, 
as shown in Appendix~\ref{sect:a2}.
The observed peaks in the energy spectrum might be often 
seen near the points of $B_{n \ell}$ calculated by 
the real parts of the optical (complex) potentials.

\Figuretable{TABLE II}

\subsection{Integrated cross sections and ($K^-$, $N$) spectra}
\label{sect:3-6}

Using the inclusive spectrum in the forward $A$($K^-$, $N$) reaction 
in Eq.~(\ref{eqn:e22}), 
we can evaluate the integrated cross section of the $\bar{K}$ 
nuclear unstable bound state by the energy integration
\begin{eqnarray}
\biggl({d\sigma \over d{\Omega_N}}\biggr)
&=& \int {dE_N}\, 
\biggl({d^2\sigma \over dE_N d{\Omega_N}}\biggr),  
\label{eqn:e52}
\end{eqnarray}
even if such a bound state has a large width and also exists
near the $\bar{K}$ emitted threshold. 
Here we consider especially that the $\bar{K}$ nuclear states
have a relative narrow width, 
because the $\bar{K}N \to \pi\Sigma$ 
decay channel is closed if the deeply-bound $\bar{K}$ 
states have $B_{\bar{K}} > $ $\sim$100 MeV. 
Since these states are located far from the $K^-$ threshold, 
we will attempt to calculate the integrated cross sections,
adapting the effective number technique into the 
DWIA \cite{Dover83,Kishimoto99,Cieply01,Yamagata05}, 
as mentioned in Appendix~\ref{sect:a2}. 
Considering that the $\bar{K}$ nuclear state with 
total spin $J_B$ has a $n \ell_{\bar{K}}$ orbit for 
$\bar{K}$ and a $j_N^{-1}$ proton-hole state, 
we obtain the forward ($K^-$, $N$) cross section for the bound state 
with $(j^{-1}_N,n\ell_{\bar{K}})J_B$ in the DWIA; 
\begin{eqnarray}
\biggl({d\sigma \over d{\Omega_N}}(0^\circ) 
\biggr)^{(j^{-1}_N n \ell_{\bar{K}})J_B}_{\rm lab}
& = & 
\alpha(0^\circ) \left\langle{d\sigma \over d\Omega}(0^{\circ}) 
\right\rangle_{\rm lab}^{K^- N \to N \bar{K}}
{\rm Re}N^{(j^{-1}_N n \ell_{\bar{K}})J_B}_{\rm eff}(0^\circ).
\label{eqn:e53}
\end{eqnarray}
Here we introduced precisely the {\it complex} effective nucleon number
$N^{(j^{-1}_N n \ell_{\bar{K}})J_B}_{\rm eff}(0^\circ)$, 
which is defined as 
\begin{eqnarray}
N^{(j^{-1}_N n \ell_{\bar{K}})J_B}_{\rm eff}(0^\circ)
&=& (2J_B+1)(2j_N+1)(2 \ell_{\bar{K}}+1)
    \left( \begin{array}{ccc}
     \ell_{\bar{K}} & j_N & J_B  \\
     0  & -{1 \over 2} & {1 \over 2}
    \end{array} \right)^2
    F({q})F^\dagger({q}).
\label{eqn:e54}
\end{eqnarray}
The form factor $F(q)$ denotes
\begin{eqnarray}
F(q)&=&
\int_0^{\infty} r^2d{r} 
(\tilde{\varphi}_{n \ell_{\bar{K}}}(r))^*
\tilde{j}_{L0}(p_N, p_{K^-} ,{0^\circ}; r)\varphi^{(N)}_{j_N}(r), 
\label{eqn:e55}
\end{eqnarray}
where $L=J_B\pm{1 \over 2}$ and $\ell_{\bar{K}}+L+\ell_N$ must 
be even due to the non-spin-flip processes. 
$\varphi^{(N)}_{j_N}$ denotes a single-particle wave function
for the nucleon, and 
$\tilde{\varphi}_{n \ell_{\bar{K}}}$ is a biorthogonal one 
for the $\bar{K}$, as given by Eq.~(\ref{eqn:e57c}). 
One should notice that
$N^{(j^{-1}_N n \ell_{\bar{K}})J_B}_{\rm eff}(0^\circ)$ 
must be a complex number due to the unstable bound states with the
non-Hermite Hamiltonian.
When the potential has no imaginary part, 
$N^{(j^{-1}_N n \ell_{\bar{K}})J_B}_{\rm eff}(0^\circ)$
is reduced to a real number.
The recoil effects are taken into account 
in the distorted waves of 
$\tilde{j}_{L0}(p_N, p_{K^-}, {0^\circ};r)$
by the factors of ${M_C/M_B}$ and ${M_C/M_A}$, 
as seen in Eq.~(\ref{eqn:e26}). 
The kinematical factor $\alpha(0^\circ)$ is often used in 
the effective number technique within 
the DWIA \cite{Dover80,Dover83,Motoba88,Hausmann89,Tadokoro95,Cieply01}.
The relation between the kinematical factors $\beta$ and $\alpha$ 
is shown in Appendix~\ref{sect:a2}.

The production cross sections for 
the deeply-bound $\bar{K}$ nuclear states by the ($K^-$, $p$) 
reaction were performed theoretically by 
several authors \cite{Kishimoto99,Cieply01,Yamagata05}.  
Moreover, Kishimoto and his collaborators \cite{Kishimoto05}
have analyzed the inclusive spectra in the ($K^-$, $p$) and 
($K^-$, $n$) reactions on the nuclear $^{12}$C and $^{16}$O targets. 
Therefore, a comparison between our results and other ones 
would be warrant, in order to recognize the nature of the 
nuclear ($K^-$, $N$) reactions. 
Let us attempt to calculate the production cross sections 
by the ($K^-$, $p$) reaction on the $^{12}$C and $^{28}$Si targets.
For the $^{12}$C target, we use single-particle wave functions 
for a proton,
which are calculated with a Woods-Saxon potential \cite{Bohr69}:
\begin{equation}
U_N(r)= {V_0^N}f(r)+V^N_{ls} ({\bm l }\cdot{\bm s})
r_0^2 {1 \over r}{d \over dr}f(r)
\label{eqn:e55a}
\end{equation}
with $f(r)=[1 + \exp{((r-R)/a)}]^{-1}$, 
where $V^N_{ls}$= $-0.44 V^N_0$, $a$=0.67 fm, 
$r_0$=1.27 fm and $R=r_0A^{1/3}=$2.91 fm.
We choose the strength of $V^N_{0}$=$-$64.8 MeV, fitting to 
the charge radius of 2.46 fm \cite{Vries86}.
A Coulomb potential with a uniform sphere of the radius $R$ 
is included.
We input the proton hole-energy of
$\varepsilon_N=$ $-$16.0 MeV for a $1p_{3 \over 2}$ hole state, 
and the energy and width of $\varepsilon_N=$ $-$36.0 MeV and 
$\varGamma=$ 10 MeV for a $1s_{1 \over 2}$ state.
For the $^{28}$Si target, 
we use also single-particle wave functions for a proton, 
which are calculated with the WS potential with 
$R=r_0A^{1/3}=$ 3.86 fm.
We choose the strength of $V^N_0$=$-$59.7 MeV, fitting to the charge 
radius of 3.09 fm \cite{Vries86}, and input  
$\varepsilon_N=$ $-$11.6 MeV for a $1d_{5 \over 2}$ proton-hole state, 
and ($\varepsilon_N$, $\varGamma$)=($-16$ MeV, $4$ MeV), ($-23$ MeV, $6$ MeV)
and ($-41$ MeV, $10$ MeV) for $1p_{1 \over 2}$, $1p_{3 \over 2}$ 
and $1s_{1 \over 2}$ proton-hole states, respectively \cite{Jacob66}.

\subsubsection{$^{12}{\rm C}(K^-,\, p)$ reactions}

\begin{figure}[htb]
  \begin{center}
  \includegraphics[width=0.5\textwidth, angle=90]{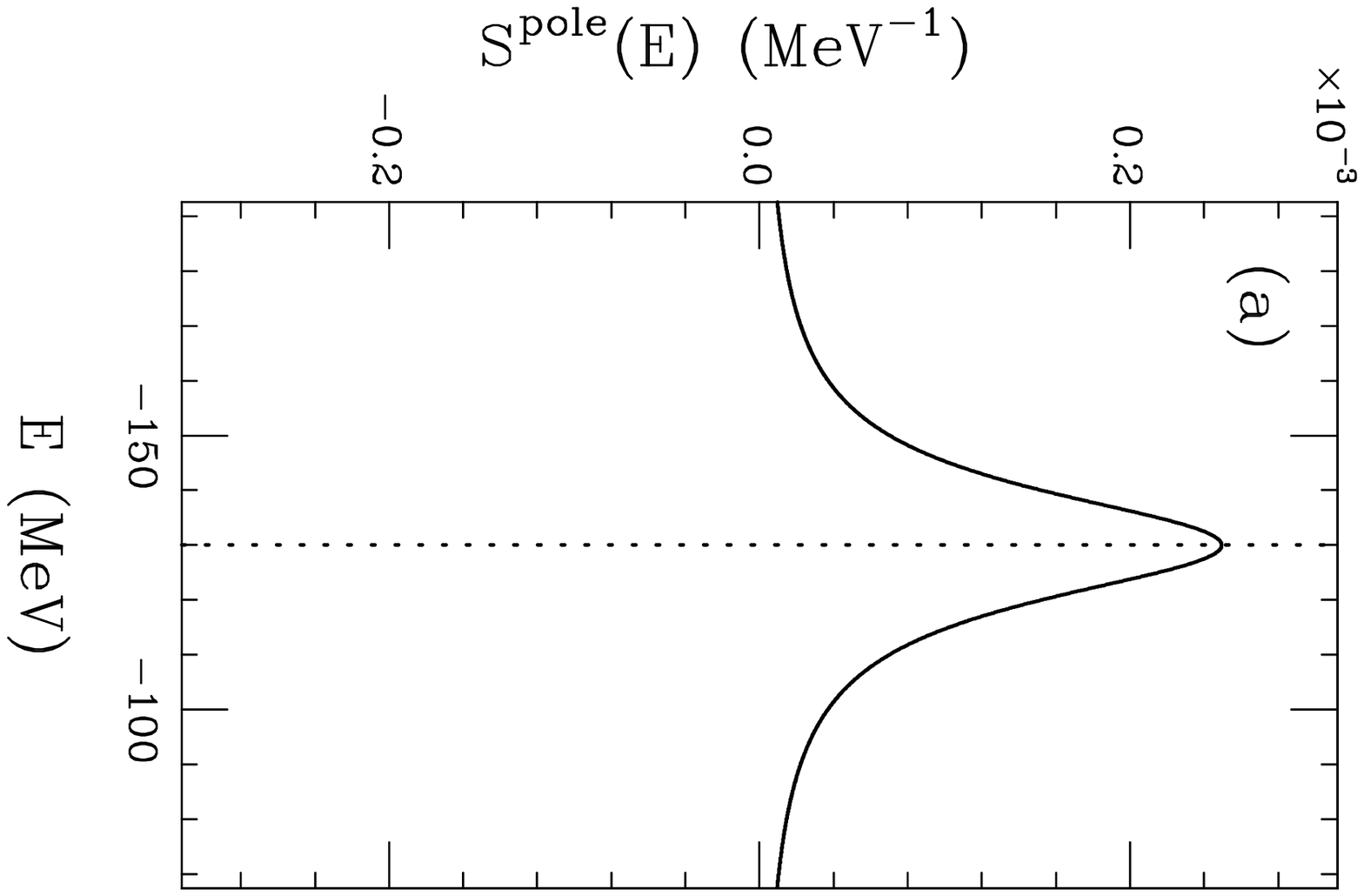}
  \includegraphics[width=0.5\textwidth, angle=90]{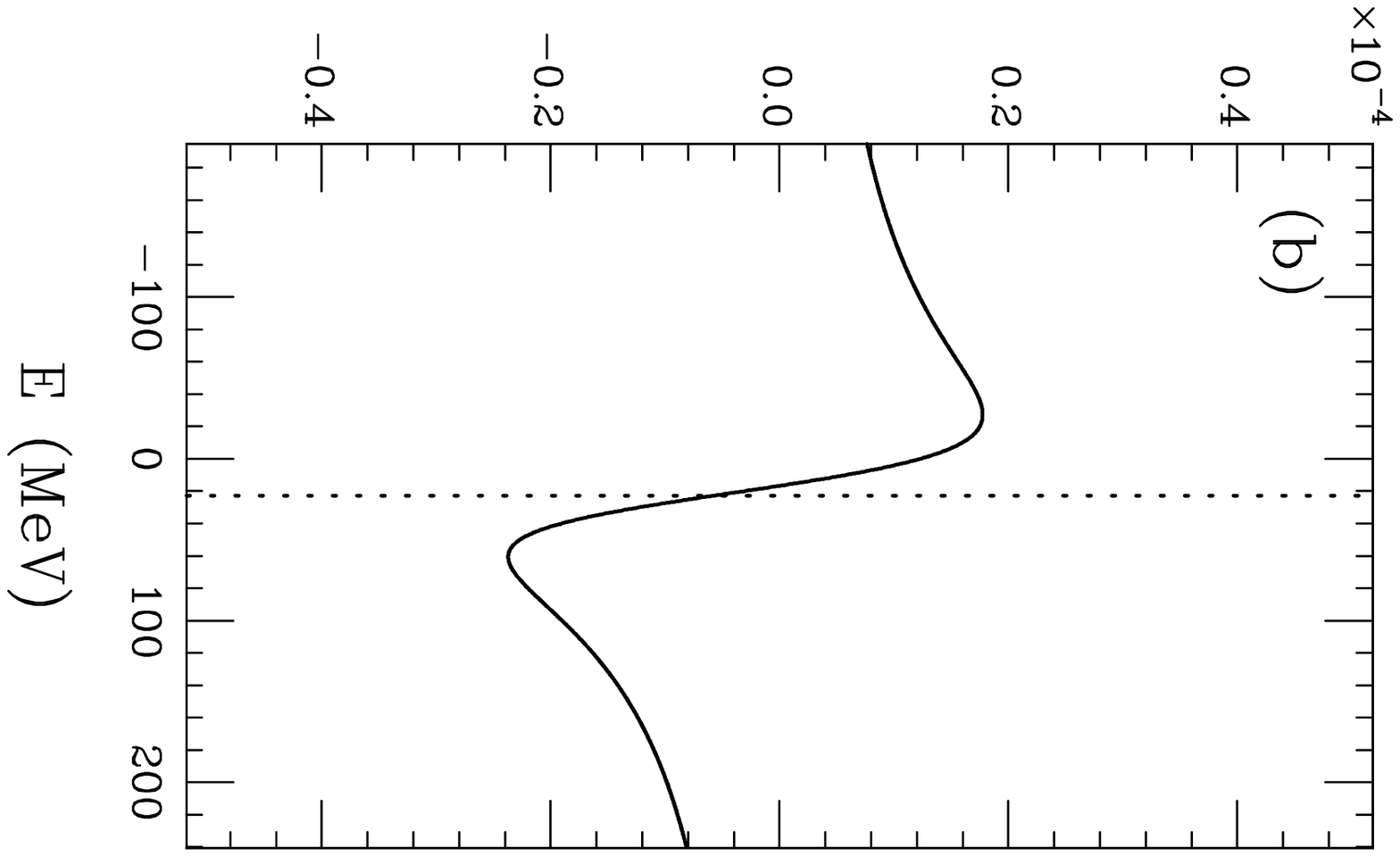}
  \includegraphics[width=0.5\textwidth, angle=90]{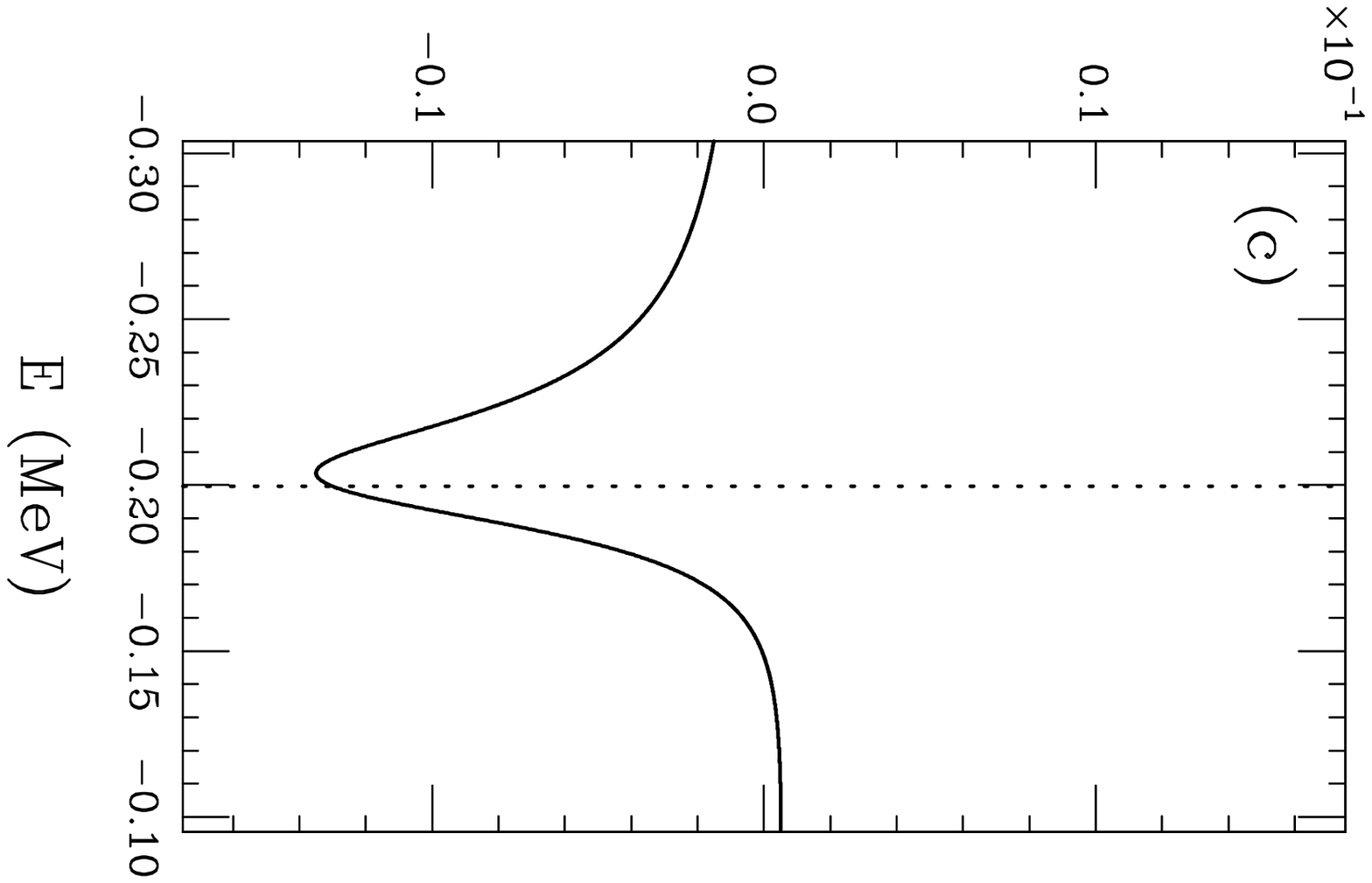}
  \caption{\label{SpecShape}
  Shape of the strength function $S^{\rm (pole)}(E)$ for the unstable
  nuclear and atomic states in the $K^- \otimes {^{11}{\rm B}}$ system
  as a function of $E_{K^-}$;
(a) ${\rm Arg}N_{\rm eff} =$ 0.08$^\circ$ at $B_{K^-}=$ $-$130 MeV and 
$\varGamma_{K^-}=$ 25 MeV for the transition $1p_{3 \over 2} \to 
(1s)_{K^-}$, $\Delta L=$ 1, 
(b) ${\rm Arg}N_{\rm eff} =$ 98.3$^\circ$ at $B_{K^-}=$ +23 MeV and 
$\varGamma_{K^-}=$ 87 MeV for the transition $1s_{1 \over 2} \to 
(2s)_{K^-}$, $\Delta L=$ 0,  and
(c) ${\rm Arg}N_{\rm eff} =$ $-$157.9$^\circ$ at $B_{K^-}=$ $-$199 keV and 
$\varGamma_{K^-}=$ 41.6 keV for the transition $1p_{3 \over 2} \to 
(1s)_{\rm atom}$, $\Delta L=$ 1.
The dotted line denotes the position of $B_{K^-}$ for each case.
}
  \end{center}
\end{figure}

In Table~\ref{tab:table3}, we show the numerical 
results of the integrated lab cross sections of 
$(j_N^{-1},n\ell_{K^-})J^{\pi}$ states, which are given 
in Eq.~(\ref{eqn:e53}),
for the forward $^{12}$C($K^-$, $p$) reactions at the incident 
$K^-$ lab momentum $p_{K^-}=$ 1.0 GeV/c. 
This ($K^-$, $p$) reactions can populate only the $T=$ 1 states in 
$^{11}_{K^-}$Be which consists of the $K^-$ and the $^{11}$B nucleus.
It is noticed that ``spin-stretched'' states are excited selectively 
due to a high momentum transfer $q(0^\circ)=$ 
$-380$-$(-210)$ MeV/c, 
where the negative momentum transfer means that 
the residual $K^-$ recoils backward relative to the incident 
particle $K^-$.
We list up the value of the real part of the effective nucleon numbers
and its argument, ${\rm Re}N_{\rm eff}$ and ${\rm Arg}N_{\rm eff}$, 
in order to see the shape of the inclusive spectrum 
as a function of $E_{K^-}$.
As seen in Table~\ref{tab:table3}, a negative value of the cross section 
in the transition $1s_{1 \over 2} \to (2s)_{K^-}$ means that 
the shapes of the states are ``upside-down peaks"
rather than Breit-Wigner peaks
because the arguments satisfy ${\rm Arg}N_{\rm eff} >$ 90$^\circ$ 
with substantial inelasticity and background \cite{Taylor72} 
(See Fig. \ref{SpecShape}).
As discussed by Morimatsu and Yazaki \cite{Morimatsu94}, 
such peak structures are often predicted in the spectra of 
$\Sigma^-$ atomic states, and also those of the $K^-$ atomic 
states \cite{Yamagata07a}. 
For $[K^- \otimes {^{11}{\rm B}}]$ atomic states, indeed, we found 
${\rm Arg}N_{\rm eff} =$ $-$157.9$^\circ$ 
at $B_{K^-}=$ 199 keV and $\varGamma_{K^-}=$ 41.6 keV 
for the transition $1p_{3 \over 2} \to (1s)_{\rm atom}$, $\Delta L=$ 1, 
and ${\rm Arg}N_{\rm eff} =$ 90.4$^\circ$ 
at $B_{K^-}=$ 81.0 keV and $\varGamma_{K^-}=$ 0.76 keV 
for the transition $1p_{3 \over 2} \to (1p)_{\rm atom}$, $\Delta L=$ 2.

Yamagata et al.~\cite{Yamagata07a} have discussed 
in detail some interesting shapes
of the inclusive ($K^-$, $p$) spectra for $K^-$ atomic states.  
In Fig.~\ref{SpecShape}, thus, we display the shape of the strength function
$S^{\rm (pole)}(E)$ given by Eq. (\ref{eqn:b3}) 
in the $K^- \otimes {^{11}{\rm B}}$ 
system, as a function of $E_{K^-}$;
(a) ${\rm Arg}N_{\rm eff} =$ 0.08$^\circ$ at $B_{K^-}=$ $-$130 MeV,  
$\varGamma_{K^-}=$ 25 MeV for the nuclear transition $1p_{3 \over 2} \to 
(1s)_{K^-}$, $\Delta L=$ 1, 
(b) ${\rm Arg}N_{\rm eff} =$ 98.3$^\circ$ at $B_{K^-}=$ +23 MeV, 
$\varGamma_{K^-}=$ 87 MeV for $1s_{1 \over 2} \to 
(2s)_{K^-}$, $\Delta L=$ 0,  and
(c) ${\rm Arg}N_{\rm eff} =$ $-$157.9$^\circ$ at $B_{K^-}=$ $-$199 keV,  
$\varGamma_{K^-}=$ 41.6 keV for the atomic transition $1p_{3 \over 2} \to 
(1s)_{\rm atom}$, $\Delta L=$ 1.
These spectra are plotted by each suitable energy scale in order 
to compare their shapes. The spectrum (a) has a standard Breit-Wigner 
peak, whereas the spectra (b) and (c) are sizably modified by the 
background; the shape in (c) just indicates the upside-down peak. 
These shapes also depend strongly on the momentum transfer 
by choosing the incident momentum $p_{K^-}$.
Thus we find that the calculated values of ${\rm Re}N_{\rm eff}$ and 
${\rm Arg}N_{\rm eff}$ give valuable information concerning 
the structure of the $\bar{K}$ unstable bound states formed 
by this reaction condition.

\Figuretable{TABLE III}
\Figuretable{FIG. 12}

One of the most important subjects is to clarify the possibility 
of the detected signals of the deeply-bound $K^-$ nuclear states; 
A deeply-bound $K^-$ state with $(1p^{-1}_{3 \over 2},1s_{K^-}){3 \over 2}^+$ 
configuration is located below the $\pi\Sigma$ threshold, 
having the relative narrow widths $\sim$25 MeV, 
for the $K^-$ nuclear $1s$ bound state via the $(1p)_p \to (1s)_{K^-}$ 
transition in $^{12}$C 
at the incident $K^-$ lab momentum $p_{K^-}=$ 1.0 GeV/c. 
Here we used the Fermi-averaged elementary cross section of 
$\langle d\sigma(0^\circ)/d\Omega \rangle^{K^-p \to p K^-}_{\rm lab}$= 3.5 mb/sr
in Eq.~(\ref{eqn:e21}),
which is slightly smaller than 5.2 mb/sr used in Ref.~\cite{Cieply01}.
For distortion parameters in the DWIA, we choose 
$\bar{\sigma}^{\rm tot}_{NN}
=\bar{\sigma}^{\rm tot}_{K^-N}=$ 40 mb and 
${\alpha}_{NN}={\alpha}_{K^-N}=$ 0, 
followed by precedent pioneering 
works \cite{Kishimoto99,Cieply01,Yamagata05,Yamagata06}.

The calculated effective number for $(1p^{-1}_{3 \over 2},1s_{K^-}){3 \over 2}^+$
is
${\rm Re}N^{(1p_{3/2}^{-1}1s_{K^-}){3 \over 2}^+}_{\rm eff}=$ 0.98$\times$10$^{-2}$,
which is in agreement with the result of 0.013 calculated 
in Ref.~\cite{Cieply01}.
Thus, the integrated lab cross section for $(1p_{3 \over 2})_p \to (1s)_{K^-}$
in Eq.~(\ref{eqn:e53}) is obtained by
\begin{equation}
\left({d\sigma \over d\Omega_N}(0^\circ)\right)_{{\rm lab}}
^{(1p^{-1}_{3/2}1s_{K^-}){3 \over 2}^+}
= 1.78 \times 3.5 \, {\rm (mb/sr)}\, \times (0.98\times10^{-2}) 
= 61 \,\, (\mu{\rm b/sr}).
\label{eqn:e55b}
\end{equation}
In Table~\ref{tab:table4}, we compare this value  
of $(d\sigma(0^\circ)/d\Omega)_{\rm lab}$ with those seen 
in other works on the $^{12}$C target \cite{Cieply01,Yamagata05,Kishimoto99}
under similar conditions of the momentum transfer $q$,
together with the case of the $^{28}$Si target:
Our results seem to be in good agreement with 
those calculated by Yamagata et al. \cite{Yamagata05},
who make a 
choice of $(d\sigma(0^\circ)/d\Omega)_{\rm lab}=$ 8.8 mb/sr for 
the input $K^- + p \to p + K^-$ cross section,
whereas the kinematical factor $\alpha(0^\circ)$ 
and recoil effects were not included in their calculation. 
The difference from the results by Ciepl$\acute{\rm y}$ et 
al. \cite{Cieply01} would come from $\alpha(0^\circ)=$ 0.69
which was erroneously employed 
due to missing a negative momentum transfer, in addition to
a use of harmonic-oscillator wave functions for proton-hole and 
$K^-$ bound states, and 
the input $(d\sigma(0^\circ)/d\Omega)_{\rm lab}=$ 5.2 mb/sr. 

\Figuretable{TABLE IV}

\begin{figure}[htb]
  \begin{center}
  \includegraphics[width=0.9\textwidth]{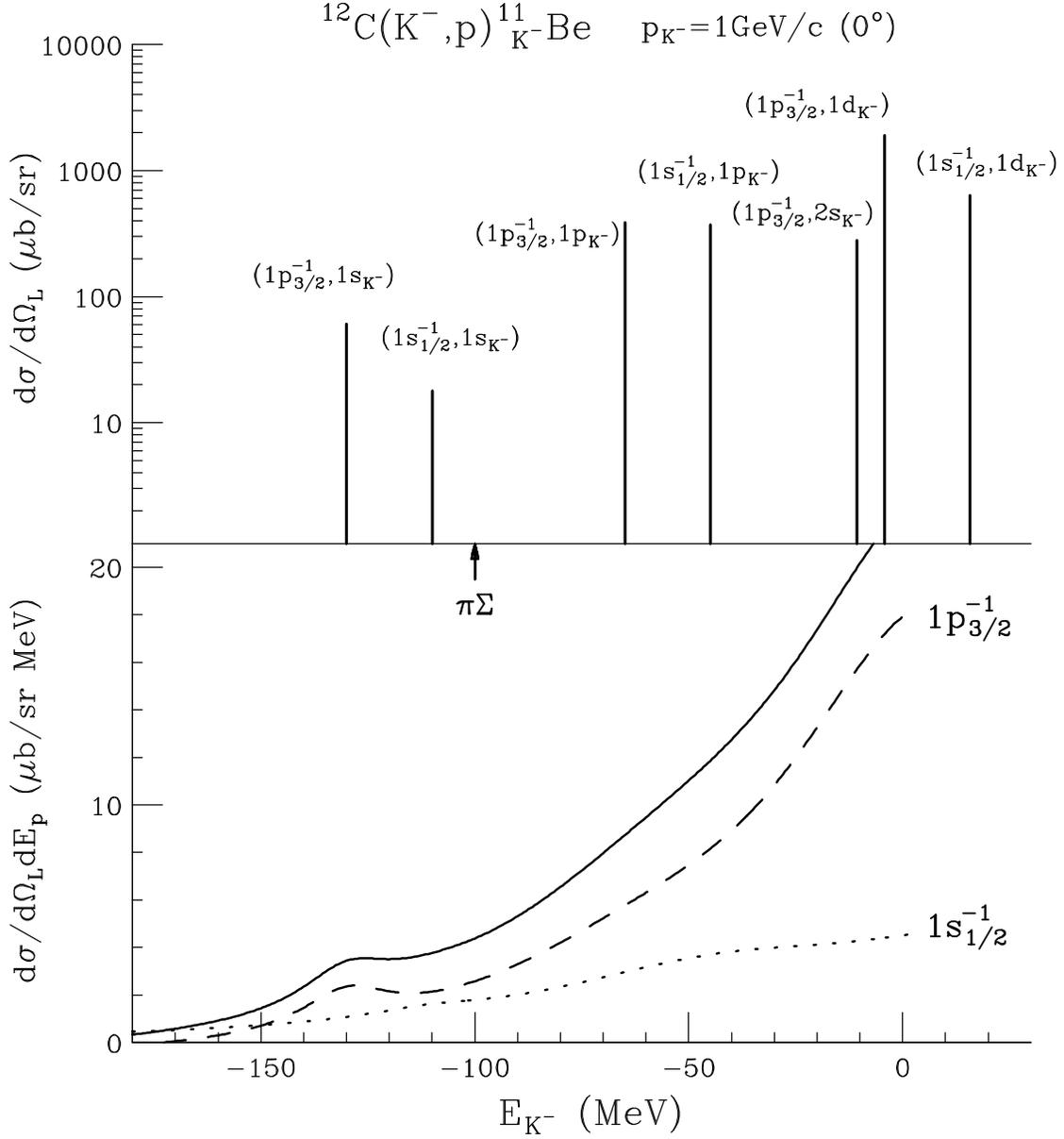}
  \caption{\label{fig:14}
  The integrated cross sections (top) and the corresponding proton spectrum (bottom) 
  for $^{11}_{K^-}$Be from $^{12}$C($K^-$, $p$) reactions 
  at $p_{K^-}=$ 1.0 GeV/c and $\theta_{\rm lab}=$ 0$^\circ$.  
  The values of the integrated cross section are estimated, omitting the imaginary 
  part of the $\bar{K}$-nucleus DD potential \cite{Mares06}, whereas 
  the spectrum is obtained by using the $\bar{K}$-nucleus DD potential
  with the imaginary part multiplying the phase space factor 
  \cite{Mares06}. The arrow denotes the $\pi\Sigma$ decay threshold of about 
  $E_{K^-} \sim$ 100 MeV.
  Note that the background and the continuum states for $K^-$ are not included. 
  }
  \end{center}
\end{figure}

\begin{figure}[htb]
  \begin{center}
  \includegraphics[width=0.9\textwidth]{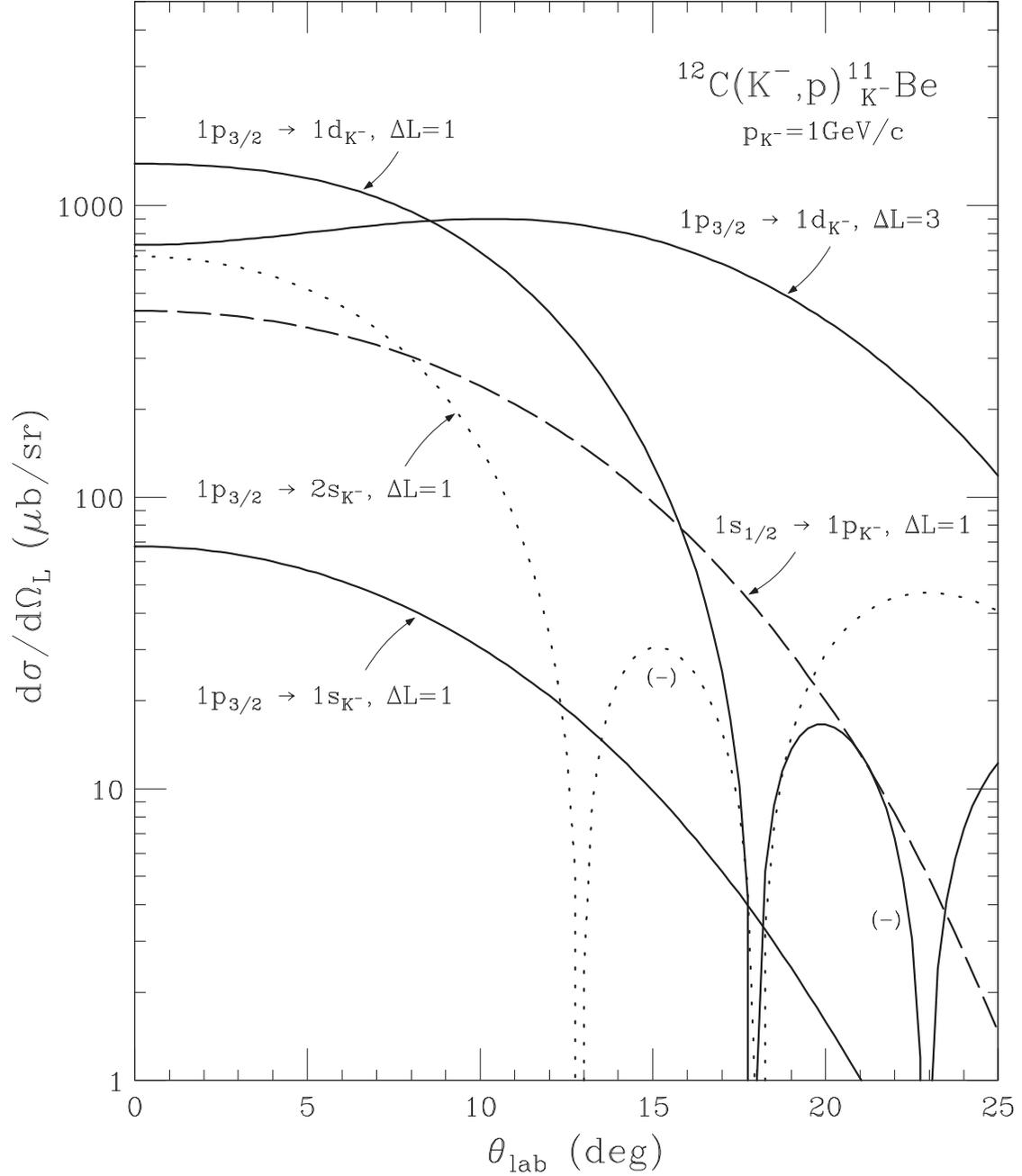}
  \caption{\label{fig:15}
  Angular distribution of the lab cross sections for $^{11}_{{K^-}}$Be 
  by ($K^-$, $p$) reactions on a $^{12}$C at $p_{K^-}=$ 1.0 GeV/c. 
  The solid and dotted curves denote for the transitions 
  $1p_{3 \over 2} \to (1s)_{K^-}$, 
  $(1d)_{K^-}$ with $\Delta L=$ 1 or 3, and $1p_{3 \over 2} \to (2s)_{K^-}$ 
  with $\Delta L=$ 1, respectively. The dashed curve denotes for 
  the transitions $1s_{1 \over 2} \to (1p)_{K^-}$ with $\Delta L=$ 1.
  The labeled $(-)$ denotes the negative value for the cross section 
  which means that the shape of the states grows into an upside-down peak 
  with the background.  
  }
  \end{center}
\end{figure}

\begin{figure}[htb]
  \begin{center}
  \includegraphics[width=0.9\textwidth]{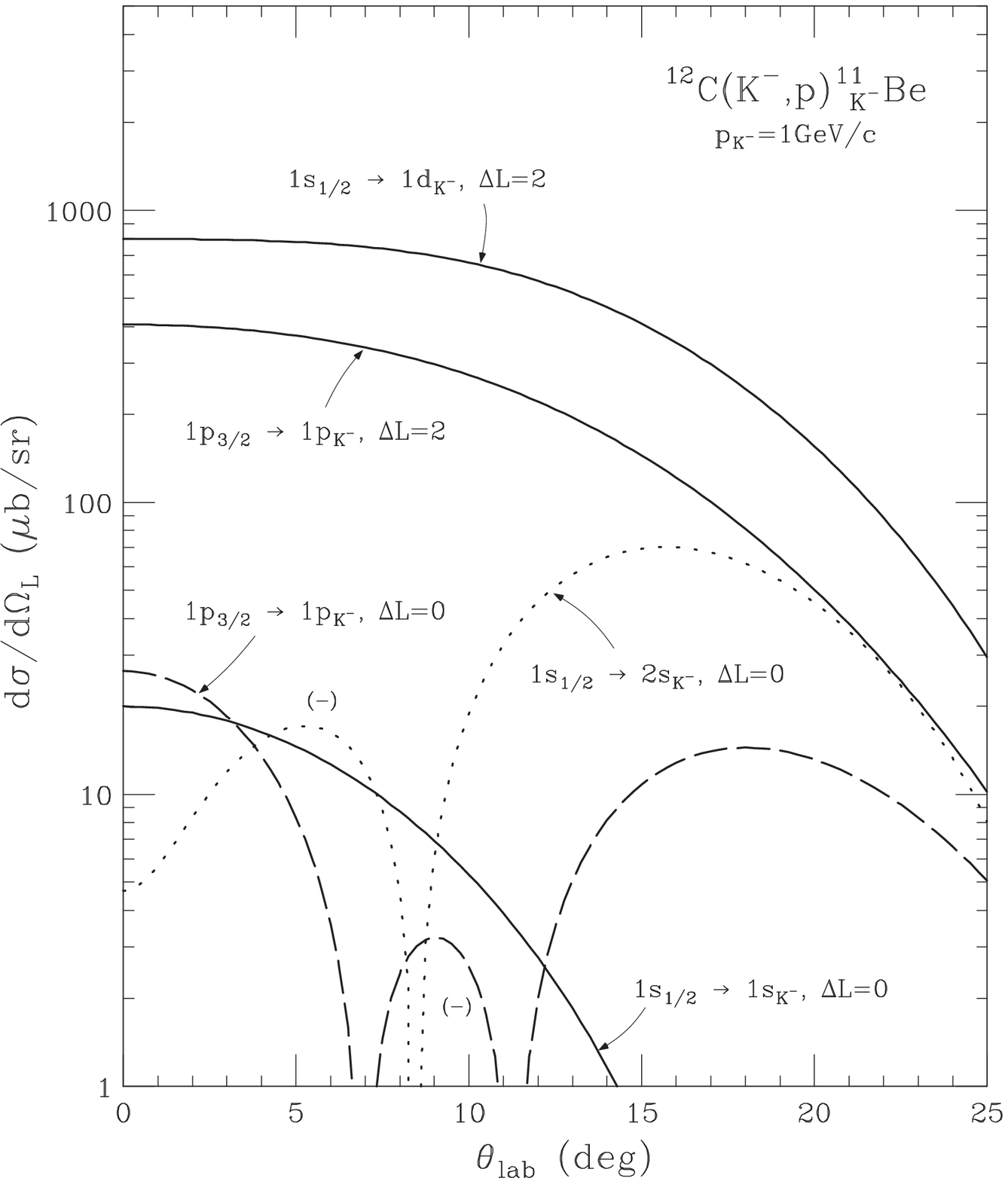}
  \caption{\label{fig:16}
  Angular distribution of the lab cross sections for $^{11}_{{K^-}}$Be 
  by $^{12}$C($K^-$, $p$) reactions at $p_{K^-}=$ 1.0 GeV/c. 
  The solid curves denote for the 
  transitions $1s_{1 \over 2} \to (1d)_{K^-}$ 
  and $1p_{3 \over 2} \to (1p)_{K^-}$ with $\Delta L=$ 2, 
  and the transition $1s_{1 \over 2} \to 1s_{K^-}$ with $\Delta L=$ 0. 
  The dotted and dashed curves denote for the transitions 
  $1s_{1 \over 2} \to (2s)_{K^-}$ with $\Delta L=$ 0 and  
  $1p_{3 \over 2} \to (1p)_{K^-}$ with $\Delta L=$ 0, respectively. 
  See the caption in Fig.~\ref{fig:15}. 
  }
  \end{center}
\end{figure}

In Fig.~\ref{fig:14}, we show the inclusive spectrum 
from $^{12}$C($K^-$, $p$) reactions at $p_{K^-}=$ 1.0 GeV/c, 
in the $^{11}_{K^-}$Be bound region obtained by our results 
listed up in Table~\ref{tab:table2}. 
Note that the contributions of the background of the $K^-$ absorption 
processes and the continuum states for $K^-$ are not included. 
In order to recognize the population by the ($K^-$, $p$) reaction,
we also illustrate the integrated cross sections which 
are calculated by omitting the imaginary parts of the DD potentials 
and the width of the $1s_{1 \over 2}$ proton-hole state.
The resultant spectra confirm that it is difficult to observe 
clear signals of the deeply-bound $K^-$ states 
even if the relative narrow state such as 
$(1p_{3 \over 2}^{-1},1s_{K^-})$ exists, 
as discussed in Refs.~\cite{Yamagata06} and \cite{Kishimoto05};
these cross sections are relatively reduced by the distortion 
of the incoming $K^-$ and outgoing proton waves, i.e., 
the distortion factor becomes $D_{\rm dis} \simeq 
{\rm Re}N^{\rm DW}_{\rm eff}/{\rm Re}N^{\rm PW}_{\rm eff}
=$ 0.98$\times$10$^{-2}$/15.6$\times$10$^{-2}$= 0.063 for the 
$(1p_{3 \over 2})_p \to ({1s})_{K^-}$ transition, 
where the superscripts DW and PW denote the distorted-wave and 
plane-wave approximations, respectively. 
In Figs.~\ref{fig:15} and \ref{fig:16}, 
we display the angular distributions of the formation of 
$\Delta L=$ 1, 3 and $\Delta L=$ 0, 2 in $^{12}_{K^-}$Be 
via the ($K^-$, $p$) reaction at 1.0 GeV/c. 
The curves labeled by $(-)$ indicate the negative value for the cross section 
which means that the shape of these spectra 
for the corresponding bound states behaves such as upside-down peaks 
with substantial background, rather than Breit-Wigner peaks, 
as seen in Fig.~\ref{SpecShape}.

\Figuretable{FIG. 13}
\Figuretable{FIG. 14}
\Figuretable{FIG. 15}

\subsubsection{$^{12}{\rm C}(K^-,\, n)$ reactions}

\begin{figure}[htb]
  \begin{center}
  \includegraphics[width=0.9\textwidth]{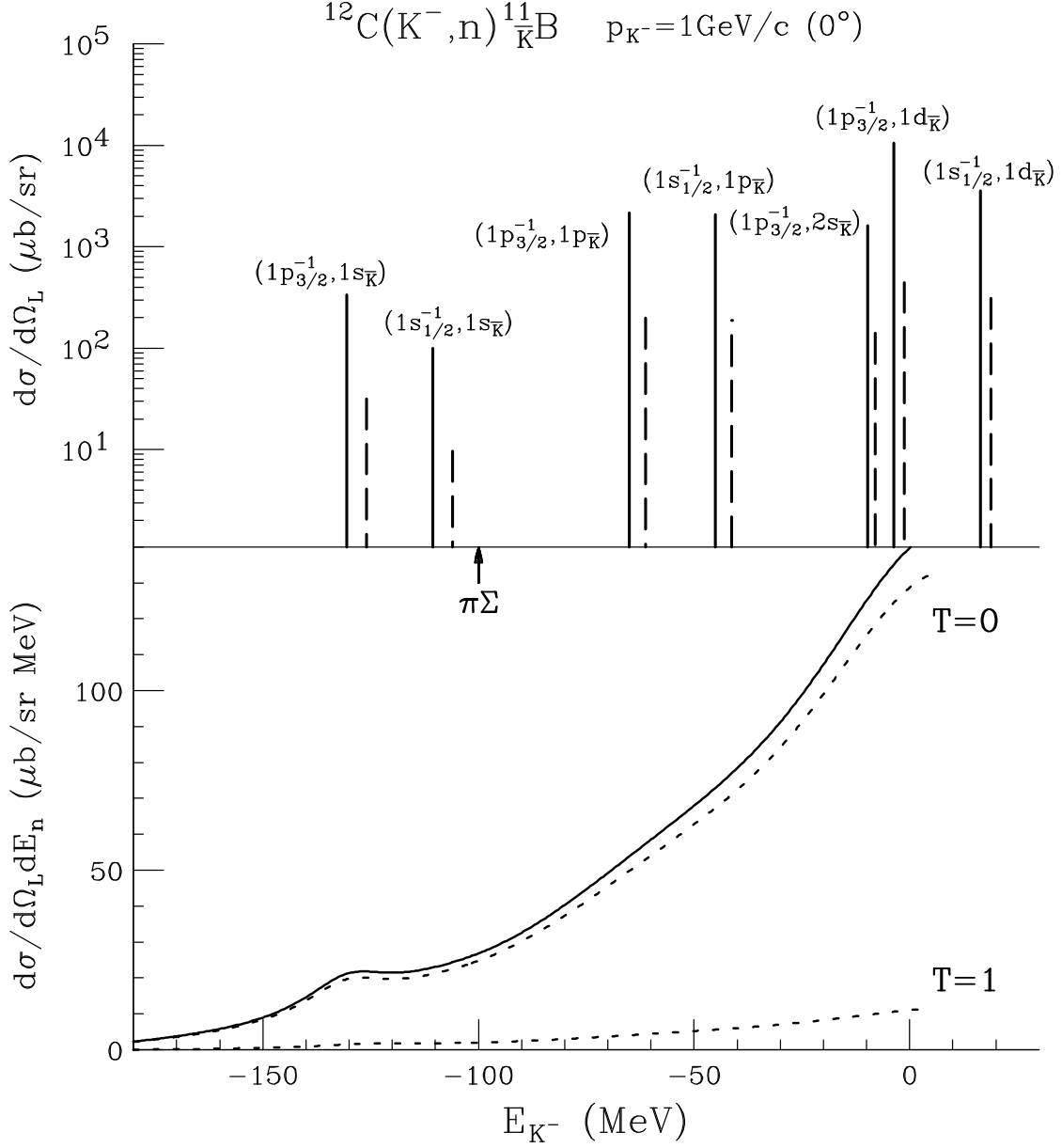}
  \caption{\label{fig:17}
  The integrated cross sections for $^{11}_{\bar{K}}$B 
  from $^{12}$C($K^-$, $n$) reactions at $p_{K^-}=$ 1.0 GeV/c and 
  $\theta_{\rm lab}=$ 0$^\circ$, 
  using the DD potential (top). 
  The solid and dashed lines denote the values of the 
  cross sections for $T=$ 0 and 1 states, respectively, omitting 
  the imaginary parts of the potential.
  The inclusive ($K^-$, $n$) spectrum for $^{11}_{\bar{K}}$B 
  from $^{12}$C($K^-$, $n$) reactions at $p_{K^-}=$ 1.0 GeV (bottom).
  The dotted curves denote the contributions from isospin $T=0$ and $T=1$ states. 
  See the caption of Fig.~\ref{fig:14}.
  }
  \end{center}
\end{figure}

In Fig.~\ref{fig:17}, we show the inclusive spectrum 
from $^{12}$C($K^-$, $n$) reactions at $p_{K^-}=$ 1.0 GeV/c,
which can be populate both $T=$ 1 and $T=$ 0 states of  
$^{11}_{\bar{K}}$B, as shown in Eq.~(\ref{eqn:e37}). 
Here we assumed that there appear the isospin good 
quantum number states in $^{11}_{\bar{K}}$B, using 
the DD potentials and omitting the Coulomb potential. 
Then, we input the Fermi-averaged cross sections for $\Delta T=$ 0 
and $\Delta T=$ 1, which are 19.6 mb/sr and 1.7 mb/sr, respectively, 
as shown in Fig.~\ref{fig:11}. 
We find that the shape and magnitude of the inclusive spectrum 
are almost determined by the contribution of the $T=0$ 
configuration. 
The cross sections of the $T=$ 0 states are almost 
10 times larger than those of 
the $T=1$ ones in the ($K^-$, $n$) reaction. 

The analysis of the ($K^-$, $N$) reactions for
the experimental data on light nuclear targets is expected 
to clarify the important information 
concerning the formation mechanism for $K^-$ nuclear bound states 
and the nature of the $\bar{K}$-nucleus potentials. 
Particularly, we believe that the comparison between 
the ($K^-$, $p$) and ($K^-$, $n$) spectra provides 
the isospin properties for the $K^- + N \to N + \bar{K}$ 
reactions and the $\bar{K}$-nucleus potentials, 
and the imaginary parts of the potentials, 
observing the spectrum of $\pi\Lambda$ decay processes,
\begin{equation}
[^{11}_{K^-}{\rm B}] \to \pi+\Lambda+{\rm nucleus}, 
\label{eqn:e55c}
\end{equation}
below the $\pi\Sigma$ threshold.

\Figuretable{FIG. 16}

\subsubsection{$^{28}{\rm Si}(K^-,\, p)$ reactions}

\begin{figure}[htb]
  \begin{center}
  \includegraphics[width=0.9\textwidth]{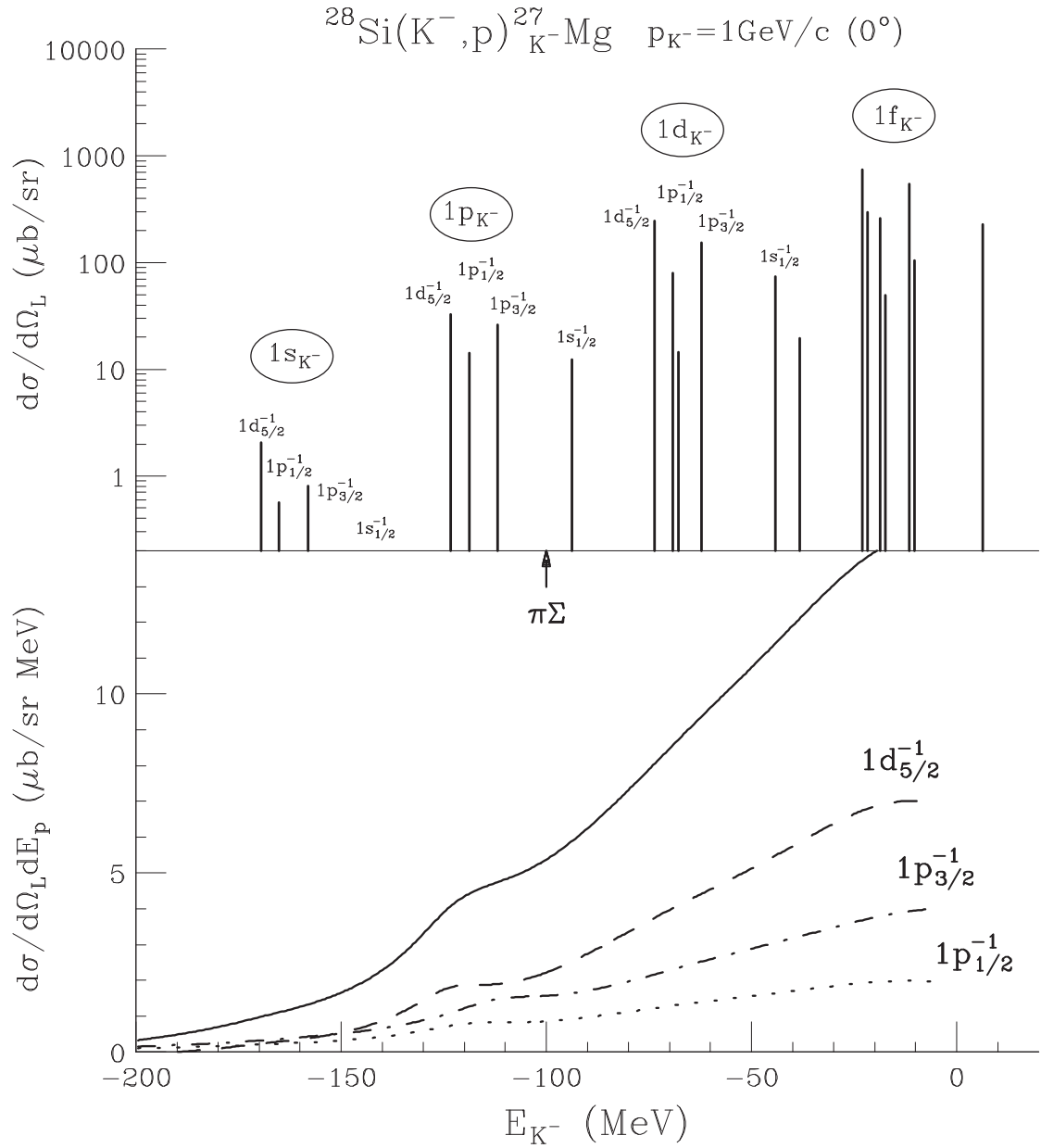}
  \caption{\label{fig:18}
  The integrated cross sections (top) and the corresponding proton spectrum (bottom) 
  for $^{27}_{K^-}$Mg from $^{28}$Si($K^-$, $p$) reactions at 
  $p_{K^-}=$ 1.0 GeV/c and $\theta_{\rm lab}=$ 0$^\circ$. 
  The dashed, dotted and dot-dashed curves denote the contributions from 
  $1d_{5 \over 2}$, $1p_{1 \over 2}$, $1p_{3 \over 2}$ proton-hole states, 
  respectively. 
  See the caption of Fig.~\ref{fig:14}.
  }
  \end{center}
\end{figure}

For a heavier $^{28}$Si target, 
we show the numerical results of the integrated lab cross 
sections for the forward ($K^-$, $p$) reactions 
at the incident $K^-$ lab momentum $p_{K^-}=$ 1.0 GeV/c,
as listed up in Table~\ref{tab:table5}. 
In Fig.~\ref{fig:18}, we display the integrated cross sections 
and the inclusive spectrum, together with the contributions of 
$1d_{5 \over 2}$, $1p_{1 \over 2}$ and $1p_{3 \over 2}$ 
proton-hole states.  
Note that the background and the continuum states for $K^-$ 
which arise above the $K^-$ threshold,  
are not taken into account.
We find that the cross sections for the ``spin-stretched'' states coupled to 
the $1d_{5 \over 2}$, $1p_{1 \over 2}$, $1p_{3 \over 2}$ 
and $1s_{1 \over 2}$ hole states are populated
selectively by the high momentum transfer, 
as seen clearly in Fig.~\ref{fig:18} (top).  
However, we confirm also that it is difficult to observe a
clear signal of the deeply-bound $K^-$ nuclear state such 
as $(1s)_{K^-}$, because several states above the $\pi\Sigma$ 
threshold have a broad width of $\sim$100 MeV,
smudging out the narrow signal.
Consequently, we believe that in order to avoid this 
difficulties at the present stage, 
one of the best ways might be to choose $s$-shell few-nucleons targets, 
$^3$H, $^3$He and $^4$He, 
as proposed by the forthcoming E15 experiments at J-PARC.

\Figuretable{TABLE V}

\Figuretable{FIG. 17}

\section{Deeply-bound antikaonic states on $s$-shell nuclei}
\label{sect:4}

We study the ($K^-$,$N$) reactions on the $s$-shell nuclear
target, considering the transitions 
$(1s_{1 \over 2})_N \to (1s)_{\bar{K}}$ 
to final $\bar{K}$ nuclear states of total spin $S$ and total 
isospin $T$
where all the nucleons are in $s$-shell.
These states give a testing ground for isospin properties of 
the production and decay processes in $\bar{K}$ nuclei, since 
the core-nucleus is often spin and isospin unsaturated.

The production amplitudes for the nuclear ($K^-$, $N$) reactions 
on the target $A$ in Eq.~(\ref{eqn:e22}),
${F}_{fi}=\langle \Psi_B |{\hat F}| \Psi_A \rangle$, 
are written in the $LS$-coupling scheme as 
\begin{eqnarray}
{F}_{fi}^{(K^-,\, n)}
&=& \bar{f}_{K^-n \to nK^-} {S}^{1/2}_{S_B,K^-}
    \langle K^-\otimes{^{(A-1)}{Z}} |{ \rho}_{fi} |A \rangle \nonumber\\
&&
+\bar{f}_{K^-p \to n\bar{K}^0} {S}^{1/2}_{S_B,{\bar{K}^0}}
    \langle \bar{K}^0\otimes{^{(A-1)}{\rm (Z-1)}} |{ \rho}_{fi} |A \rangle 
\label{eqn:e56}
\end{eqnarray}
for ($K^-$, $n$) reactions, and 
\begin{eqnarray}
{F}_{fi}^{(K^-,\, p)} 
&=& \bar{f}_{K^-p \to pK^-} {S}^{1/2}_{S_B,K^-}
    \langle K^-\otimes{^{(A-1)}{(Z-2)}} |{ \rho}_{fi} |A \rangle
\label{eqn:e57}
\end{eqnarray}
for ($K^-$, $p$) reactions. 
Here the spectroscopic amplitude ${S}^{1/2}_{S_B,\bar{K}}$ 
for the spin $S_B$ and charge $\bar{K}$ channel is given by 
\begin{equation}
{S}^{1/2}_{S_B,\bar{K}}
 =    \Bigl\langle 
        \bigl[X^{(C')}_{{S_{C'}},{T_{C'}}} 
        X^{(\bar{K})}_{0,{1 \over 2}}\bigr]_{S_B}
        X^{(N)}_{{1 \over 2},{1 \over 2}}\Big|
      \sum_{j=1}^{\rm A}{\hat{\cal O}}_{N_j}  
      \Big|
      X^{(\bar{K})}_{0,{1 \over 2}}
      {\cal A}\bigl[X^{(C)}_{{S_C},{T_C}} 
        X^{(N)}_{{1 \over 2},{1 \over 2}}
        \bigr]_{S_AT_A}
      \Bigr\rangle. 
\label{eqn:e58} 
\end{equation}
where 
$ X^{(\bar{K})}_{0,{1 \over 2}}$, $X^{(N)}_{{1 \over 2},{1 \over 2}}$
and  $X^{(C)}_{{S_C},{T_C}}$
($X^{(C')}_{{S_{C'}},{T_{C'}}}$) denote the spin-isospin 
functions $X_{S,T}$ for $\bar{K}$, $N$ and the core-nucleus in the 
target nucleus (the core-nucleus in the $\bar{K}$ nuclear states), respectively, 
and ${\cal A}$ is an antisymmetric operator for nucleons in the nucleus. 
The transition amplitude for $A \to B=[\bar{K}\otimes{^{(A-1)}{Z}}]$
is given as
\begin{eqnarray}
 \langle \bar{K}\otimes{^{(A-1)}{Z}} |{ \rho}_{fi} |A \rangle
 &=& 
 (-)^{J_B + L - {1 \over 2}}\, i^L 
  \sqrt{(2\ell_A+1) (2\ell_B+1) (2J_A+1) (2L+1)}\, \nonumber\\
 && \times 
    \left( \begin{array}{ccc}
                  \ell_B   &  L &   \ell_{A} \\
                  0   &  0        &  0      \\
    \end{array}  \right)  
    \left\{ \begin{array}{ccc}
                  \ell_{A}  &  J_{A}    &  {1 \over 2}       \\
                  J_B      &   \ell_B       &  L \\
    \end{array}  \right\} 
    \langle \Phi_{{C'}} | \Phi_{C} \rangle
    F_{\ell_B\ell_A L}(q),
\label{eqn:e59}
\end{eqnarray}
where the factor
$\langle \Phi_{C'} | \Phi_{C} \rangle$
denotes an overlapping between the core-nucleus for 
the initial and final states, 
taken into account the effects of the nuclear 
shrinkage (core-polarization) which are expected to be found 
in the $\bar{K}$-nuclear states.
The form factor $F_{\ell_B\ell_A L}(q)$ is 
given by
\begin{eqnarray}
 F_{\ell_B\ell_A L}(q)
 &=& \int_0^\infty r^2d{r}\, (\tilde{\varphi}_{n \ell_B}(r))^*
      \tilde{j}_{L0}(p_N, p_{K^-}, 0^\circ;r)\,
      \varphi^{(N)}_{\ell_A}(r),
\label{eqn:e60}
\end{eqnarray}
where $\varphi^{(N)}_{\ell_A}(r)$ is a relative wave function 
for the $N$-nucleus system, respectively, 
as a function of the relative coordinate 
between the nucleon and the core-nucleus, and 
$\tilde{\varphi}_{n \ell_B}(r)$ is a biorthogonal wave function 
for unstable $\bar{K}$-nucleus systems; 
$\tilde{j}_{L0}(p_N, p_{K^-},0^\circ;r)$
is a partial distorted-wave with the angular momentum transfer $L$ 
at $\theta_{\rm lab}=$ 0$^\circ$
from the incoming $K^-$ to the outgoing nucleon.
The factors of $M_C/M_B$ and $M_C/M_A$ 
in Eq.(\ref{eqn:e26}) 
are very important for the recoil effects in $s$-shell light nuclei.
Therefore, the production lab cross section for the ($K^-$,~$n$) or 
($K^-$,~$p$) reaction is given as
\begin{eqnarray}
 \left({{d \sigma_{fi} \over d \Omega_N}(0^\circ)}\right)_{\rm lab}
 &=& \alpha(0^\circ){1 \over 2J_A+1} 
 \sum_{M_A}\sum_{M_B, m_s} 
 {\rm Re}({F}_{fi}^{(K^-,\, N)}{F}_{fi}^{(K^-,\, N)\dagger}),
\label{eqn:e60a}
\end{eqnarray}
where $m_s$ denotes a $z$-component of the outgoing nucleon.

\subsection{$^4$He target}

Akaishi and Yamazaki \cite{Akaishi99,Akaishi02} suggested to look for 
a deeply-bound $\bar{K}\otimes [NNN]$ $T=$0, $S=$1/2 state
having a binding energy of over 100 MeV and a relative narrow width 
of $\varGamma \simeq$ 20 MeV, 
because the main decay channel $\bar{K}N \to \pi\Sigma$ is 
 closed. 
Wycech also pointed out that the width of such states could be 
as small as 20 MeV \cite{Wycech86}.
Surprisingly, Suzuki et al. \cite{Suzuki04} reported the experimental 
evidence of the tribaryon S$^{0}$(3115) by the 
$^4$He(stopped $K^-$, $p$) reaction (KEK-PS E471).
But recently it has been withdrawn (KEK-PS E549/570) \cite{Iwasaki06a}.
From a viewpoint of microscopic few-body dynamics, we believe
that the isospin nature of $[\bar{K}\otimes[NNN]_{T=1/2}]$ 
configuration fully reveals itself 
because the $^3$He and $^3$H core-nucleus are spin-isospin unsaturated.
Therefore, we consider the $^4$He($K^-$,$N$) 
reaction in order to see isospin properties of ($K^-$, $N$) reactions
and also $\bar{K}N$ interactions.

Reducing the spectroscopic factor $S_{S_B,\bar{K}}^{1/2}$ in 
Eq.~(\ref{eqn:e58}), we obtain the production amplitudes 
of Eq.~(\ref{eqn:e56});
\begin{eqnarray}
{F}_{fi}^{(K^-,\, n)} 
&=& \bar{f}_{K^-n \to nK^-} 
    \langle K^- \otimes{^3}{\rm He} |{ \rho}_{fi} |{^4}{\rm He} \,\rangle
- \bar{f}_{K^-p \to n\bar{K}^0} 
    \langle \bar{K}^0\otimes{^3}{\rm H} |{ \rho}_{fi} |{^4}{\rm He} \, \rangle, 
\nonumber
\label{eqn:e67}\\
{F}_{fi}^{(K^-,\, p)} 
&=& - \bar{f}_{K^-p \to pK^-}
    \langle K^- \otimes{^3}{\rm H} |{ \rho}_{fi} |{^4}{\rm He} \,\rangle.
\label{eqn:e68}
\end{eqnarray}
Substituting the spin-isospin functions in Eq.~(\ref{eqn:c4}) 
into Eq.~(\ref{eqn:e68}), thus, we can rewrite 
\begin{eqnarray}
{F}_{fi}^{(K^-,\, n)}  
&=  & 
\bar{f}_{\Delta T=0}
\langle {^{\,\, 3}_{\bar K}}{\rm H}\,;{T_B=0} |{ \rho}_{fi} | {^4}{\rm He} \rangle
+
\bar{f}_{\Delta T=1} 
\langle {^{\,\, 3}_{\bar K}}{\rm H}\,;{T_B=1} |{ \rho}_{fi} | {^4}{\rm He} \rangle, 
\nonumber
\label{eqn:e69}\\
{F}_{fi}^{(K^-,\, p)}  
&=  & 
-\sqrt{2} \, \bar{f}_{\Delta T=1}
\langle {^{\,\, 3}_{\bar K}}{\rm n}\,;{T_B=1} |{ \rho}_{fi} | {^4}{\rm He} \rangle,
\label{eqn:e70}
\end{eqnarray}
where $\bar{f}_{\Delta T=0}$ and $\bar{f}_{\Delta T=1}$
are the (normalized) Fermi-averaged isoscalar 
$\Delta T=$ 0 and isovector 
$\Delta T=$ 1 transition amplitudes for the $^4$He target, respectively;
\begin{eqnarray}
\bar{f}_{\Delta T=0} 
& =& 
(\bar{f}_{K^-n \to nK^-}+
\bar{f}_{K^-p \to n\bar{K}^0})/\sqrt{2}
=\bar{f}^{(0)}/\sqrt{2}, \nonumber\label{eqn:e71}\\
\bar{f}_{\Delta T=1} 
& =& 
(\bar{f}_{K^-n \to nK^-}
-\bar{f}_{K^-p \to n\bar{K}^0})/\sqrt{2}
=\bar{f}^{(1)}/\sqrt{2}
=\bar{f}_{K^-p \to pK^-}/\sqrt{2}.
\label{eqn:e72}
\end{eqnarray}
Here we can confirm the isospin relation of the cross sections 
in Eq.(\ref{eqn:e47}) and the dominance of the $T=$ 0 excitation 
by the ($K^-$, $n$) reaction.
In Table~\ref{tab:table6}, we show the relative production 
cross sections by the forward ($K^-$, $N$) amplitude for 
$\bar{K}\otimes[NNN]$ states 
with $T_C=$ 1/2 and $S_C=$ 1/2. 

\Figuretable{TABLE VI}

\subsection{$^3$He and $^3$H targets}

Now we focus on the three-body $\bar{K}NN$ system as a deeply-bound 
$\bar{K}$ nuclear state. This state is expected to be 
the lightest and the most important system for the $\bar{K}$ 
bound state \cite{Nogami63,Yamazaki07}, and it is 
populated from the ($K^-$, $N$) reaction on the $^3$He or $^3$H 
target.
For $^3$He($K^-$, $N$) reactions, 
the production amplitudes for the $\bar{K} \otimes [NN]$ $S=$ 0
states on charge $\bar{K}$ basis are written as 
\begin{eqnarray}
{F}_{fi}^{(K^-,\, n)} 
&=& {1 \over \sqrt{2}}\,
\bar{f}_{K^-p \to n\bar{K}^0} 
    \langle \bar{K}^0 \{pn\} |{ \rho}_{fi} |{^3}{\rm He} \, \rangle 
- \bar{f}_{K^-n \to nK^-} 
    \langle K^- \{pp\} |{ \rho}_{fi} |{^3}{\rm He} \,\rangle,
\label{eqn:e61}\nonumber\\
{F}_{fi}^{(K^-,\, p)} 
&=& {\sqrt{2}}\,
\bar{f}_{K^-p \to pK^-} 
    \langle K^-\{pn\} |{ \rho}_{fi} |{^3}{\rm He} \,\rangle.
\label{eqn:e62}
\end{eqnarray}
Substituting the spin-isospin functions listed up in Eq.~(\ref{eqn:c2})
into Eq.~(\ref{eqn:e62}), 
we rewrite these amplitudes as 
\begin{eqnarray}
{F}_{fi}^{(K^-,\, n)}  
&=  & 
\bar{f}_{\Delta T=0}
\langle {^{\,\, 2}_{\bar K}}{\rm H};{T_B=1/2} 
|{ \rho}_{fi} | {^3}{\rm He} \rangle
+ \bar{f}_{\Delta T=1} 
\langle {^{\,\, 2}_{\bar K}}{\rm H};{T_B=3/2}
|{ \rho}_{fi} |{^3}{\rm He} \rangle,  \nonumber
\label{eqn:e63}\\
{F}_{fi}^{(K^-,\, p)}  
&=  & 
{1 \over \sqrt{6}} \, \bar{f}_{\Delta T=0}
\langle {^{\,\, 2}_{\bar K}}{\rm n};{T_B=1/2} 
|{ \rho}_{fi} | {^3}{\rm He} \rangle
+
{1 \over \sqrt{3}} \, \bar{f}_{\Delta T=1}
\langle {^{\,\, 2}_{\bar K}}{\rm n};{T_B=3/2} 
|{ \rho}_{fi} | {^3}{\rm He} \rangle,
\label{eqn:e64}
\end{eqnarray}
where $\bar{f}_{\Delta T=0}$ and $\bar{f}_{\Delta T=1}$
are the Fermi-averaged (normalized) isoscalar 
$\Delta T=$ 0 and isovector 
$\Delta T=$ 1 transition amplitudes for the $^3$He target, respectively:  
\begin{eqnarray}
\bar{f}_{\Delta T=0} 
& =& 
\sqrt{2/3}\,
(\bar{f}_{K^-n \to nK^-}
+\bar{f}_{K^-p \to n\bar{K}^0}/2), \nonumber
\label{eqn:e65}\\
\bar{f}_{\Delta T=1} 
& =& 
-(\bar{f}_{K^-n \to nK^-}-\bar{f}_{K^-p \to n\bar{K}^0})/\sqrt{3}
=-\bar{f}_{K^-p \to pK^-}/\sqrt{3},
\label{eqn:e66}
\end{eqnarray}
involving the effects of the spectroscopic amplitude 
$S^{1/2}_{S_B,\bar{K}}$ due to spin-isospin nature 
on the unsaturated $^3$He target. 
In Table~\ref{tab:table7}, we show the relative formation 
cross sections of these states on $^3$He, 
where the core configuration is restricted by 
the Pauli principle 
to $\{S_C, T_C \}=\{0, 1 \}$ or $\{1, 0 \}$. 
For the $^3$H target, we show the relative 
formation cross sections at the forward ($K^-$,~$N$) reactions, 
as shown in Table~\ref{tab:table8}. It is noticed that 
the $^3{\rm H}(K^-,\, p)$ reaction can populate only the 
isospin $T_B=$ 1, $S=$ 1/2 states, in comparison with the case of 
the $^3{\rm He}(K^-,\, p)$ reaction.

\Figuretable{TABLE VII}
\Figuretable{TABLE VIII}

In a previous paper \cite{Koike07}, we examined 
the $^3$He(in-flight $K^-$,~$n$) 
reaction at $p_{K^-}$ = 1.0 GeV/c, $\theta_{\rm lab}=$ 0$^\circ$,  
with some simplified assumption. 
Only the $K^- + n \to n + K^-$ forward 
scattering has been considered, omitting the 
$K^- + p \to n + \bar{K}^0$ charge exchange process 
which can also contribute to the $[\bar{K}\otimes\{pp \}_{T=1}]_{T=1/2}$
formation through 
the coupling between $[{K^-}\otimes\{pp \}]$ and 
$[\bar{K}^0\otimes \{pn \}]$ channels.
Replacing the transition amplitude $\bar{f}_{K^-n \to nK^-}$ 
in Eq.~(\ref{eqn:e62})
by $-\bar{f}_{\Delta T=0}$, we can roughly estimate 
this contribution \cite{Koike07}; 
the cross section of the $K^-pp$ bound state is
enhanced by about 18 \% with the Fermi-averaged amplitudes.
It is noticed that the $\bar{K}NN$ state with $T$=1/2 dominates 
the ($K^-$,~$n$) reaction at $p_{K^-}$= 1.0 GeV/c
because 
$| \bar{f}_{\Delta T=0}|^2/|\bar{f}_{\Delta T=1}|^2 \simeq$ 14.
This nature justifies the assumption that 
we treat a single channel of $[\bar{K}\otimes \{NN \}_{T=1}]_{T=1/2}$ 
as $K^-pp$ restrictedly \cite{Koike07}.
We believe that a full coupled-channel calculation is needed
in order to get more quantitative results. 
Moreover, a choice of 
the parameters in the eikonal distorted waves also changes
the absolute value of the cross section, 
but the distortion effect is not significant for $^3$He. 
For the in-flight ($K^-$,~$N$) reaction, it would be not 
appropriate to use the decay rate measured by $K^-$ absorption 
at rest~\cite{Katz70},
considering that its value depends on atomic orbits where $K^-$ is 
absorbed through atomic cascade processes~\cite{Onaga89}. 
More theoretical and experimental considerations are needed.

\section{Summary and Conclusion}
\label{sect:6}

We have investigated theoretically the formation of deeply-bound 
antikaonic $K^-/\bar{K}^0$
nuclear states by the ($K^-$,~$N$) reaction, 
introducing the complex effective number in the DWIA.
We have discussed the isospin properties of the ($K^-$,~$N$) reaction 
on the basis of the Fermi-averaged elementary 
amplitudes of the $K^- + p \to p + K^-$,  $K^- + n \to n + K^-$
and $K^- + p \to n + \bar{K}^0$ processes, 
and the integrated cross sections for the nuclear ($K^-$,~$N$) reaction 
at the incident $K^-$ lab momentum $p_{K^-}$ = 1.0 GeV/c and 
$\theta_{\rm lab} = 0^{\circ}$,
concerning the kinematical condition.
The results are summarized as follows:

\begin{itemize}
\item[(1)] The deeply-bound $\bar{K}$ states with 
the isospin $T=0$ can be populated dominantly 
by the ($K^-$,~$n$) reaction on closed 
shell targets, e.g., $^{12}$C and $^{28}$Si, 
because of the isospin nature of the $K^-+ N \to N + \bar{K}$ 
amplitudes.

\item[(2)] The ($K^-$,~$N$) reaction differs kinematically 
from hypernuclear production of ($\pi^+$,~$K^+$) and ($K^-$,~$K^+$) 
reactions, so that the kinematical factors of 
$\alpha(0^\circ)$ and $\beta(0^\circ)$ are larger than 1, thus, 
the cross sections are enhanced.

\item[(3)] 
The integrated cross sections of 
deeply-bound $\bar{K}$ nuclear states 
for the ($K^-$,~$N$) reaction on nuclear 
$^{12}$C and $^{28}$Si targets, can be obtained fully by 
the complex effective nucleon number $N_{\rm eff}$; 
${\rm Re}N_{\rm eff}$ and ${\rm Arg}N_{\rm eff}$ 
enable to see the structure of the $\bar{K}$ 
unstable bound states.

\item[(4)] The deeply-bound $\bar{K}$ nuclear states 
for the ($K^-$,~$N$) reaction on $s$-shell nuclear targets, 
$^3$He, $^3$H and  $^4$He, have a strong isospin dependence of 
the cross sections due to the spin-isospin unsaturated nuclear 
core-states.

\end{itemize}

\noindent
In conclusion, 
the ($K^-$, $n$) reaction at $p_{K^-}=$ 1.0 GeV/c 
excites preferentially $\Delta T=$ 0 states 
in the deeply-bound $\bar{K}$ region, and 
is complementary to the ($K^-$,~$p$) reaction
which excites only isovector $\Delta T=$ 1 states.
The complex effective number approach provides
insight on the structure of the bound state spectrum.
Although the inclusive nucleon spectra calculated at 
$p_{K^-}=$ 1.0 GeV/c for $^{12}$C and $^{28}$Si targets
do not show distinct peak structure in the $\bar{K}$ 
bound region, they show substantial strength in this region,
thus promising to shed light on the depth of the 
$\bar{K}$ nuclear potential.
We advocate measuring nucleon ($K^-$, $N$) spectra on
s-shell targets, $^3$H, $^3$He and $^4$He, attempting to 
resolve the isospin structure of deeply-bound $\bar{K}$ 
nuclear states. This investigation is in progress.

\begin{acknowledgments}
We would like to thank Professor M. Iwasaki, 
Professor Y. Akaishi, 
Dr. Y. Hirabayashi and Dr. A. Umeya 
for many valuable discussions.
We are pleased to acknowledge Professor A. Gal for useful comments.
This work is supported by Grant-in-Aid for Scientific Research on
Priority Areas (No. 17070002 and No. 17070007). 

\end{acknowledgments}

\begin{appendix}

\section{Inclusive cross section for the $A({a},{b})$ reaction
and the kinematical factor $\beta$ in Eq.(\ref{eqn:e22})}
\label{sect:a1}

In order to see the formulation for the inclusive cross section 
of Eq.(\ref{eqn:e25}), 
we consider a nuclear two-body reaction 
\begin{equation}
   {a} + {A} \to  {b} + {B},
\label{eqn:a1}
\end{equation}
where $a$, $b$, $A$ and $B$ denote the incident (incoming), 
the detected (outgoing), the target and the residual particles, respectively. 
The differential cross section in the lab frame 
 can be expressed \cite{Sakurai67} as 
\begin{equation}
d\,^6\sigma_{fi}=\frac{(2\pi)^4}{v_a}
\delta(E_b+E_B-E_a-E_A)\delta({\bm p}_b+{\bm p}_B-{\bm p}_a-{\bm p}_A)
|T_{fi}|^2{d{\bm p}_b \over {(2\pi)^3}}{d{\bm p}_B \over {(2\pi)^3}},
\label{eqn:a2}
\end{equation}
where $v_a=p_a/E_a$, and the nuclear $T$-matrix for the transition $a +A \to b + B$ 
within the impulse approximation is defined by 
\begin{equation}
T_{fi} = \langle \Psi_{B}|\chi^{(-)*}_{b} 
\hat{\cal O} \chi^{(+)}_{a}| \Psi_{A} \rangle,
\label{eqn:a3}
\end{equation}
where $\chi^{(-)*}_{b}$ and $\chi^{(+)}_{a}$ are distorted waves 
of the outgoing $b$ and the incoming $a$, respectively. 
$\hat{\cal O}$ denotes a transition operator.
Integrating over ${d{\bm p}_B}$ in Eq.(\ref{eqn:a2}), 
the differential cross section can be written as 
\begin{equation}
d\,^3\sigma_{fi}=\frac{1}{(2\pi)^2 v_{a}}
\delta(E_f({\bm p}_b)-E_i)|T_{fi}|^2 p^2_bdp_b d\Omega_b,
\label{eqn:a4}
\end{equation}
where $E_f({\bm p}_b)=E_b({\bm p}_b)+E_B({\bm p}_a+{\bm p}_A-{\bm p}_b)$ and 
$E_i=E_a({\bm p}_a)+E_A({\bm p}_A)$. 
Summing over all the final states $f=\{b,B\}$, 
we obtain the differential cross section for the 
inclusive reaction within the DWIA factorized the two-body elementary 
$T$-matrix $a + N \to b + Y$ \cite{Morimatsu94};
\begin{eqnarray}
d\,^3\sigma
&=&\sum_f d\,^3\sigma_{fi}  \nonumber\\
&=& \frac{p_bE_b}{(2\pi)^2 v_{a}}
|\langle {\bm p}^{(0)}_Y {\bm p}^{(0)}_b|t
|{\bm p}_N{\bm p}_a \rangle |^2 dE_b d\Omega_b \nonumber\\
&& \times {1 \over 2J_A+1}\sum_{M_A}\sum_{b,B} \langle \Psi_A |
 \chi^{(+)*}_{a}\hat{\cal O}^\dagger \chi^{(-)}_{b}  |\Psi_B \rangle
\delta(E-E_B)\langle \Psi_B |
\chi^{(-)*}_{b} \hat{\cal O} \chi^{(+)}_{a}| \Psi_A \rangle \label{eqn:a5}\\
&=&{{p_b E_b} \over {(2\pi)^2}v_a}
|\langle {\bm p}^{(0)}_Y {\bm p}^{(0)}_b|t
|{\bm p}_N{\bm p}_a \rangle |^2  dE_b d\Omega_b \nonumber\\
& & \times (-){1 \over \pi}{\rm Im}
\left\langle \Psi_A \bigl|\chi^{(+)*}_{a}\hat{\cal O}^\dagger \chi^{(-)}_{b} 
{1 \over {E-H_B+i\epsilon}}
\chi^{(-)*}_{b} \hat{\cal O} \chi^{(+)}_{a}\bigr| \Psi_A \right\rangle, 
\label{eqn:a6}
\end{eqnarray}
where $\langle {\bm p}^{(0)}_Y {\bm p}^{(0)}_b|t
|{\bm p}_N{\bm p}_a \rangle$ denotes the $T$-matrix for 
the elementary $a + N \to b + Y$ reaction process in the lab frame, 
and ${\bm p}^{(0)}_Y$ and ${\bm p}^{(0)}_b$ denote the momenta of 
$Y$ and $b$, respectively,  
and $E= E_a({\bm p}_a)+E_A({\bm p}_A)- E_b({\bm p}_b)$.
Here we used  the relation
\begin{eqnarray}
\sum_B | \Psi_B \rangle \delta(E-E_B)\langle \Psi_B |
&=& (-){1 \over \pi}{\rm Im}\left[
{1 \over {E-H_B+i\epsilon}} \right],
\label{eqn:a7}
\end{eqnarray}
where $[{E-H_B+i\epsilon}]^{-1}$ is  Green's function 
for many-body final states $B$ including the $Y$+core-nucleus system.
Note that $dE_b= p_bdp_b/E_b$ is obtained from the energy-momentum 
relation for the detected $b$, $E_b=\sqrt{{\bm p}_b^2 + m_b^2}$, 
because the $\delta$-function in Eq.~(\ref{eqn:a7}) is held in Green's 
function.

In the case of the elementary process, $a + N \to b + Y$, 
the differential cross section at $\theta_{\rm lab}$ 
in the lab frame is written as 
\begin{equation}
\biggl({d\sigma \over d{\Omega_b}}\biggr)^{aN \to bY}_{\rm lab}
={{p^{(0)}_b E^{(0)}_b} \over {(2\pi)^2}v_a}
{ {p^{(0)}_b E^{(0)}_Y } \over 
{p^{(0)}_b E^{(0)}_Y  + 
E^{(0)}_b ({p^{(0)}_b - p_a \cos\theta_{\rm lab}})}}
|\langle {\bm p}^{(0)}_Y {\bm p}^{(0)}_b|t
|{\bm p}_N {\bm p}_a \rangle |^2,
\label{eqn:a8}
\end{equation}
where $E^{(0)}_b$ and $E^{(0)}_Y$ denote 
energies of the detected $b$ and the residual $Y$, respectively,
and the superscript $(0)$ refers to the kinematics for the 
two-body reaction on a nucleon ($N$) target. 
Substituting Eq.~(\ref{eqn:a8}) into Eq.~(\ref{eqn:a6}), 
we obtain the double-differential 
cross section for the inclusive $A$($a$, $b$) reaction as  
\begin{eqnarray}
\biggl({d^2\sigma \over {dE_b}d{\Omega_b}}\biggr)
&=& \beta 
\biggl( {d \sigma  \over d \Omega_b}\biggr)^{aN \to bY}_{\rm lab}
S(E),
\label{eqn:a9}
\end{eqnarray}
where the strength function is defined as
\begin{eqnarray}
S(E)
&=&
{1 \over 2J_A+1}\sum_{M_A}\sum_{b,B}
\Big|\langle \Psi_B|\chi^{(-)*}_{b} \hat{\cal O} \chi^{(+)}_{a}|\Psi_A \rangle
\Bigr|^2 \delta(E + E_b -E_i) \nonumber\\
&=& (-){1 \over \pi}{\rm Im} \left[ \sum_{\alpha' \alpha} 
 \int\,d{\bm r}' d{\bm r}
  F_{\alpha'}^\dagger({\bm r}') G_{\alpha' \alpha}(E;{\bm r}',{\bm r})
  F_{\alpha}({\bm r})\right]
\label{eqn:a10}
\end{eqnarray}
with the kinematical factor $\beta$ which is defined by
\begin{equation}
 \beta=\left(1+ {E^{(0)}_b \over E^{(0)}_Y}
        {{p^{(0)}_b - p_a \cos\theta_{\rm lab}} 
        \over p^{(0)}_b} \right){p_b E_b \over p^{(0)}_b E^{(0)}_b}.
\label{eqn:a11}
\end{equation}
One can find the expression of Eq.~(\ref{eqn:a9}), 
in several articles applying Green's function 
to hypernuclear physics and the related subjects 
\cite{Morimatsu94,Tadokoro95,Koike07}.
For the forward ($K^-$, $N$) reaction ($\theta_{\rm lab}=0^\circ$) 
in the lab frame, 
we replace $p_a$, $p_b$, $E_b$ and $E_Y$ by 
$p_{K^-}$, $p_N$, $E_N$ and $E_{\bar{K}}$, respectively, 
and we find that $q(0^\circ)=p_{K^-}-p^{(0)}_N$ 
is negative for the $K^-+N \to N+\bar{K}$ reaction.
Thus, the kinematical factor is written by 
\begin{eqnarray}   
\beta(0^\circ)
&=& \left(1 + {E_N^{(0)} \over E_{\bar{K}}^{(0)}}
              {p_N^{(0)}-p_{K^-} \over p^{(0)}_N} \right)
              {p_N \, E_N \over p_N^{(0)} E_N^{(0)}} \nonumber \\
&=&
\biggl(1- {v^{(0)}_{\bar{K}} \over v^{(0)}_N} \biggr)
{p_N E_N \over p^{(0)}_N E^{(0)}_N},
\label{eqn:a12}
\end{eqnarray}
where $v^{(0)}_N=p^{(0)}_N/E^{(0)}_N$ is the velocity of  
the detected nucleon, and 
$v^{(0)}_{\bar{K}}=q(0^\circ)/E^{(0)}_{\bar{K}}$ for 
the residual $\bar{K}$.
In the text, we represent the cross section of Eq.(\ref{eqn:e25}), 
as a function the energy-transfer $\omega$ instead of $E$.

\section{Integrated cross section for the formation of the 
$\bar{K}$ nuclear bound states}
\label{sect:a2}

In order to see the relation between the integrated cross section
of Eq.(\ref{eqn:e53})
and the inclusive cross section of Eq.(\ref{eqn:e25}),
we will perform explicitly the energy-integration of Eq.(\ref{eqn:e52}).
Here we assume a $Y$-nuclear bound state in the nucleus $B$, 
using an optical potential which gives unstable bound states with
complex eigenvalues 
$E_{n\ell}=\varepsilon_{n\ell}-i\varGamma_{n\ell}/2$
for simplicity.
We can expand Green's function in the $\bar{K}$-nucleus 
bound region as   
\begin{eqnarray}
G_{\ell}(E;{r}',{r})
&=&
\sum_n G_{n \ell}^{(pole)}(E;{r}',{r}) + G_{\ell}^{(bg)}(E;{r}',{r}),
\label{eqn:b1}
\end{eqnarray}
where the summation over $n$ includes all the pole of the $S$-matrix 
in the complex $k$-plane, and $G_{\ell}^{(bg)}(E;{r}',{r})$ indicates 
the background contribution. 
The pole contribution for a $(n \ell)$ unstable bound state can 
be expressed as 
\begin{eqnarray}
G_{n \ell}^{(pole)}(E;{r}',{r})
= {\varphi_{n \ell}(r')(\tilde{\varphi}_{n \ell}(r))^* 
\over E-E_{n \ell}+i\epsilon}
\label{eqn:b2}
\end{eqnarray}
with $\varphi_{n \ell}(r)$ denoting a radial wave function of 
the bound state, and $\tilde{\varphi}_{n \ell}$ is a 
biorthogonal one for $\varphi_{n \ell}$, as normalized by Eq.~(\ref{eqn:e56c}).
Thus, the contribution of the strength function can be written as 
\begin{eqnarray}
S^{(pole)}(E)
&=& (-){1 \over \pi}{\rm Im} \left[ 
 \int\,d{\bm r}' d{\bm r} F^\dagger({\bm r}') Y_{\ell}(\hat{\bm r}') 
 G_{n \ell}^{(pole)}(E;{r}',{r})
 Y_{\ell}^*(\hat{\bm r}) F({\bm r})
 \right] \nonumber\\
&=& (-){1 \over \pi}{\rm Im}
\left[{ N_{n \ell}^{(pole)} \over E-E_{n \ell}+i\epsilon}\right]
\nonumber \\
&=& {1 \over \pi}
\left\{{\varGamma_{n \ell}/2
\over {(E-\varepsilon_{n \ell})^2+\varGamma_{n \ell}^2/4}}
{\rm Re}N_{n \ell}^{(pole)}
-{{E-\varepsilon_{n\ell} 
\over {(E-\varepsilon_{n \ell})^2+\varGamma_{n \ell}^2/4}}
{\rm Im}N_{n \ell}^{(pole)}} 
\right\},
\label{eqn:b3}
\end{eqnarray}
where the strength of a pole is given by
\begin{eqnarray}
N_{n \ell}^{(pole)}
&=& 
    (2J_B+1)(2j_N+1)(2 \ell_{\bar{K}}+1)
    \left( \begin{array}{ccc}
     \ell_{\bar{K}} & j_N & J_B  \\
     0  & -{1 \over 2} & {1 \over 2}
    \end{array} \right)^2
    F({q})F^\dagger({q}), 
\label{eqn:b4}
\end{eqnarray}
where the form factor $F(q)$ is given as Eq.(\ref{eqn:e55}).
Here we assumed that the final states have the 
($j^{-1}_{N},n\ell_{\bar{K}})J_B$ configurations.
The integrated cross section can be evaluated by the energy integration;
\begin{eqnarray}
\biggl({d\sigma \over d{\Omega_b}}\biggr)
&=& \int {dE_b}\, 
\biggl({d^2\sigma \over dE_b d{\Omega_b}}\biggr) 
=
\beta \biggl( {d \sigma  \over d \Omega_b}\biggr)^{aN \to bY}_{\rm lab}
\int {dE_b}\, S^{(pole)}(E).
\label{eqn:b6}
\end{eqnarray}
Changing the $dE_b$-integration to $dE_B$-integration,
$dE_b=
|{{\partial E_b}/{\partial p_b}}|
|{{\partial E_B}/{\partial p_b}}|^{-1}
dE_B$,
with the momentum conservation, where
\begin{eqnarray}
\biggl|{{\partial E_b} \over {\partial p_b}}\biggr|
&=&{p_b \over E_b}, \qquad
\biggl|{{\partial E_B} \over {\partial p_b}}\biggr|=
{{{p_b E_B  + E_b ({p_b - p_a\cos{\theta_{\rm lab}}})}  
\over {E}_b E_B }},
\label{eqn:b7}
\end{eqnarray}
we can obtain 
\begin{eqnarray}
\int {dE_b}\, S^{(pole)}(E) 
&=&
(-){1 \over \pi}{\rm Im}
\int {dE_b}\, 
\left[{ N_{n \ell}^{(pole)}
 \over E-E_{n \ell}+i\epsilon}\right] \nonumber\\
&=&
\left(1 + \frac{E_b}{E_{B}}
\frac{p_{b}-p_a\cos{\theta_{\rm lab}}}{p_b} \right)^{-1}
(-){1 \over \pi}{\rm Im} 
\int {d{E}_B}\,
\left[{ N_{n \ell}^{(pole)}
 \over E-E_{n \ell}+i\epsilon}\right] \nonumber\\
&=&
\left(1 + \frac{E_b}{E_{B}}
\frac{p_{b}-p_a\cos{\theta_{\rm lab}}}{p_b} \right)^{-1}
{\rm Re}N_{n \ell}^{(pole)},
\label{eqn:b8}
\end{eqnarray}
where we substituted Eq.~(\ref{eqn:b3}) and used the relations
\begin{eqnarray}
\int_{-\infty}^{\infty}d{\bar E}
{ \varGamma/2 \over {({\bar E}-E')^2+\varGamma^2/4}}=\pi,
\quad 
\int_{-\infty}^{\infty}d{\bar E}
{ {\bar E}-E' \over {({\bar E}-E')^2+\varGamma^2/4}}=0.
\label{eqn:b9}
\end{eqnarray}
Therefore, the integrated cross section can be expressed as
\begin{eqnarray}
\biggl({d\sigma \over d{\Omega_b}}\biggr)
&=& 
\beta 
\left(1 + \frac{E_b}{E_{B}}\frac{p_{b}-p_a 
\cos{\theta_{\rm lab}}}{p_b} \right)^{-1}
\biggl( {d \sigma  \over d \Omega_b}\biggr)^{aN \to bY}_{\rm lab}
{\rm Re}N_{n \ell}^{(pole)} \nonumber \\
&=& \alpha 
\biggl( {d \sigma  \over d \Omega_b}\biggr)^{aN \to bY}_{\rm lab}
{\rm Re}N_{n \ell}^{(pole)}, 
\label{eqn:b10}
\end{eqnarray}
and the kinematical factor $\alpha$ \cite{Dover83} has a relation to 
$\beta$,
\begin{equation}
 \alpha = \beta
{\biggl(1+ {E_b \over E_B}{{p_b - p_a \cos\theta_{\rm lab}} 
       \over p_b} \biggr)^{-1}}, 
\label{eqn:b11}
\end{equation}
where ($\cdots$)$^{-1}$ is often called the recoil factor \cite{Frullani84}. 
Especially, we confirm that for the forward reaction;
\begin{eqnarray}
\alpha(0^\circ)
&=& 
{\biggl(1+{({p^{(0)}_b-p_a)/E^{(0)}_Y}
 \over p^{(0)}_b/{E^{(0)}_b}} \biggr)}
{p_b E_b \over p^{(0)}_b E^{(0)}_b}
 \bigg/
{\biggl(1+{({ p_b - p_a)/E_B}
 \over p_b/{E_b}} \biggr)} \nonumber\\
&=& 
\left({{1- {v^{(0)}_Y/v^{(0)}_b} }
 \over 
{1- {{v}_B/v_b}}}\right)
{p_b E_b \over p^{(0)}_b E^{(0)}_b},
\label{eqn:b12}
\end{eqnarray}
where $v^{(0)}_Y$ is a velocity of $Y$ for the nucleon 
target, and $v_B$ is a velocity of $B$ for the nucleus target.
The factor of Eq.~(\ref{eqn:b12}) is equivalent to that given in 
Eq.(2.10) by Dover et al.~\cite{Dover83}.

\section{Spin-isospin states for the $s$-shell $\bar{K}$-nuclear 
systems}
\label{sect:a3}

In the present paper, we use the isospin states for $\bar{K}N$ systems
having the relation between the isospin and charge bases 
as the following:
\begin{eqnarray}
& \left| 1, + 1 \right\rangle^{\bar{K}N} &
  = \qquad| \bar{K}^0p   \rangle, \nonumber\\
& \left| 1, \, 0 \, \right\rangle^{\bar{K}N} &
  = \sqrt{1 \over 2} | \bar{K}^0n \rangle 
    + \sqrt{1 \over 2} |  {K}^-p \rangle, \nonumber\\
& \left| 0, \, 0 \, \right\rangle^{\bar{K}N} &
  = \sqrt{1 \over 2} | \bar{K}^0n \rangle 
    - \sqrt{1 \over 2} |{K}^-p  \rangle, \nonumber\\
& \left| 1, -1 \right\rangle^{\bar{K}N} &
  = \qquad| {K}^-n \rangle,
\label{eqn:c1}
\end{eqnarray}
where $\left| I, I_z \right\rangle^{\bar{K}N}$
denotes an isospin-function for 
the $\bar{K}N$ channel 
with isospin $I$ and the $z$-projection $I_z$. 
Note that the phase definition of 
$\left| 0, \, 0 \, \right\rangle^{\bar{K}N}$
is an opposite sign of that given by Gopal et al.~\cite{Gopal77},
i.e., 
$\left| 0, \, 0 \, \right\rangle^{\bar{K}N}=
-\left| 0, \, 0 \, \right\rangle^{\bar{K}N}_{\rm Gopal}$.

For $\bar{K}NNN$ systems, we assume a configuration with 
$S_C = 1/2$ and $T_ C= 1/2$ in the $3N$ core-nucleus, $^3$He or $^3$H.
Thus we represent the spin-isospin states for the microscopic 
three-nucleon system, i.e., 
$|^3{\rm He}\, \rangle^{3N}$ for isospin $z$-projection 
$\tau_C =$ $+$1/2 and  
$|^3{\rm H}\, \rangle^{3N}$ for $\tau_C =$ $-$1/2.
Thus, the spin-isospin states with total spin $S=$ 1/2 in 
$[\bar{K}\otimes[NNN]_{T=1/2}]$ 
are written as
\begin{eqnarray}
& \left| ^3_{\bar{K}}{\rm He} \right\rangle_{T=1, \tau=+1} &
= \qquad | \bar{K}^0 \rangle
  \left| ^3{\rm He}\,
  \right\rangle^{3N}, \nonumber\\
& \left| ^3_{\bar{K}}{\rm H} \right\rangle_{T=1, \tau=0} &
= \sqrt{1 \over 2} | \bar{K}^0 \rangle
  | ^3{\rm H}\, \rangle^{3N}
 +\sqrt{1 \over 2} | {K}^- \rangle
  | ^3{\rm He}\,\rangle^{3N}, \nonumber\\
& \left| ^3_{\bar{K}}{\rm H} \right\rangle_{T=0, \tau=0} &
= \sqrt{1 \over 2} | \bar{K}^0 \rangle
  | ^3{\rm H}\, \rangle^{3N}
 -\sqrt{1 \over 2} | {K}^- \rangle
  | ^3{\rm He}\,\rangle^{3N}, \nonumber\\
& \left| ^3_{\bar{K}}{\rm n} \right\rangle_{T=1, \tau=-1} &
= \qquad | {K}^- \rangle
  \left| ^3{\rm H}\, \right\rangle^{3N}.
\label{eqn:c4}
\end{eqnarray}

For $\bar{K}NN$ systems, the spin-isospin functions 
with total spin $S=0$ and and isospin $T$ are written as 
\begin{eqnarray}
& \left| ^2_{\bar{K}}{\rm He} \right\rangle_{T=3/2, \tau=+3/2} &
= \qquad | \bar{K}^0\{ pp \}  \rangle, \nonumber\\
& | ^2_{\bar{K}}{\rm H} \rangle_{T=1/2, \tau=+1/2} &
= \sqrt{1 \over 3}| {\bar{K}}^0 \{ pn \} \rangle
  -\sqrt{2 \over 3}| {K}^- \{ pp \} \rangle, \nonumber\\
& | ^2_{\bar{K}}{\rm H} \rangle_{T=3/2, \tau=+1/2} &
= \sqrt{2 \over 3}| {\bar{K}}^0 \{ pn \} \rangle
  +\sqrt{1 \over 3}| {K}^- \{ pp \} \rangle, \nonumber\\
& | ^2_{\bar{K}}{\rm n} \rangle_{T=1/2, \tau=-1/2} &
= \sqrt{2 \over 3}| {\bar{K}}^0 \{ nn \} \rangle
  -\sqrt{1 \over 3}| {K}^- \{ pn \} \rangle, \nonumber\\
& | ^2_{\bar{K}}{\rm n} \rangle_{T=3/2, \tau=-1/2} &
= \sqrt{1 \over 3}| {\bar{K}}^0 \{ nn \} \rangle
  +\sqrt{2 \over 3}| {K}^- \{ pn \} \rangle, \nonumber\\
& | K^-nn \rangle_{T=3/2, \tau=-3/2} &
= \qquad | {K}^-\{ nn \}  \rangle,
\label{eqn:c2}
\end{eqnarray}
and for total spin $S=1$;
\begin{eqnarray}
& | ^2_{\bar{K}}{\rm H} \rangle_{T=1/2, \tau=+1/2} &
= | \bar{K}^0[ pn ]  \rangle, \nonumber\\
& | ^2_{\bar{K}}{\rm n} \rangle_{T=1/2, \tau=-1/2} &
= | {K}^-[ pn ]  \rangle,
\label{eqn:c3}
\end{eqnarray}
Here we denoted the two-nucleon configuration of 
$\{N_1, N_2\}=N_1N_2+N_2N_1$ with $^1$S$_0$, $T_{NN}$=1 
and $[N_1, N_2]=N_1N_2-N_2N_1$ with $^3$S$_1$, $T_{NN}$=0.

\end{appendix}


\clearpage

\begin{table*}
\caption{\label{tab:table1}
Summary of the isospin relation for $K^-+N \to N + \bar{K}$
lab amplitudes ${f}_{K^-N \to N\bar{K}}$ 
where $\bar{K}$ denotes the $\bar{K}^0$ or $K^-$, 
to $s$-channel $f_I$ and $t$-channel $f^{(t)}$ amplitudes 
which are labeled by $I$ and $t$, respectively.
$\Delta Z$ is a charge transfer defined by $\Delta Z= (i'_2)_z-(i_2)_z-1$
for the reaction 
${\bm i}_1 + {\bm i_2} \to {\bm i}_1' + {\bm i}_2'$ 
where $(i)_z$ is a $z$-component of the particle isospin $i$. 
}
\begin{ruledtabular}
\begin{tabular}{llrcc}
 Reaction  & $N \to {\bar K}$ & $\Delta Z$
                   & \multicolumn{2}{c}{${f}_{K^-N \to N\bar{K}}$}  \\
\cline{4-5}
           &                  & 
                   & $t$-channel & $s$-channel  \\
\tableline
 $K^-$+$n$$\to$$n$+$K^-$
                   & $n \to K^-$   &  $-1$
                   & $(f^{(0)}+f^{(1)})/2$
                   & $f_1$  \\
 $K^-$+$p$$\to$$n$+$\bar{K}^0$
                   & $p \to \bar{K}^0$   &  $-1$
                   & $(f^{(0)}-f^{(1)})/2$
                   & $(f_1+f_0)/2$ \\
 $K^-$+$p$$\to$$p$+$K^-$
                   & $p \to K^-$   &  $-2$
                   & $f^{(1)}$ 
                   & $(f_1-f_0)/2$ \\
\end{tabular}
\end{ruledtabular}
\end{table*}

\begin{table*}[bth]
\caption{
\label{tab:table2}
The $K^-$ binding energies and widths of the $K^-$ nuclear 
$(n \ell)_{K^-}$ bound states for 
$^{11}_{K^-}$Be = [$K^-$-$^{11}$B] and 
$^{27}_{K^-}$Mg = [$K^-$-$^{27}$Al]
with the $\bar{K}$-nucleus DD potentials \cite{Mares06}.
These values are estimated without (Real only) and with (Full) 
the imaginary parts which are multiplied by the phase space 
factor \cite{Mares06}, of the potentials. The Coulomb potentials
are taken into account. $k^{(pole)}_{n \ell}$ denotes 
a corresponding pole position of the bound state in the complex 
momentum plane.
}
\begin{ruledtabular}
\begin{tabular}{llcccccc}
       
       && \multicolumn{3}{l}{Real only} & \multicolumn{3}{l}{Full} \\
          \cline{3-5} \cline{6-8}
& States & $-B_{n \ell}$ & $k^{(pole)}_{n \ell}$ & rms
       & $-B_{n \ell}$ & $\varGamma_{n \ell}$  
       & $k^{(pole)}_{n \ell}$  \\
       && (MeV) & (fm$^{-1}$) &  (fm)  
       & (MeV) & (MeV) & (fm$^{-1}$)  \\
\hline
$^{11}_{K^-}$Be \\

&$(1s)_{K^-}$ &  $-$130 & +$i$1.82 & 1.3  & $-$130  & 25 & $-$0.09+$i$1.82   \\
&$(1p)_{K^-}$ &  $-$65  & +$i$1.28 & 1.8  & $-$63   & 91 & $-$0.43+$i$1.33   \\
&$(2s)_{K^-}$ &  $-$11  & +$i$0.52 & 3.1  & $+$2.5  & 76 & $-$0.72+$i$0.67   \\
&$(1d)_{K^-}$ &  $-$4.3 & +$i$0.33 & 2.6  & $+$6.6  & 97 & $-$0.84+$i$0.73   \\

\hline
$^{27}_{K^-}$Mg \\
&$(1s)_{K^-}$ &$-$169 & +$i$2.07 &  1.6 &$-$169 & 22 & $-$0.07+$i$2.07   \\
&$(1p)_{K^-}$ &$-$123 & +$i$1.77 &  2.0 &$-$123 & 25 & $-$0.09+$i$1.77   \\
&$(1d)_{K^-}$ &$-$74  & +$i$1.37 &  2.3 &$-$72  & 88 & $-$0.40+$i$1.41   \\
&$(2s)_{K^-}$ &$-$68  & +$i$1.32 &  2.4 &$-$66  & 91 & $-$0.42+$i$1.36   \\
&$(1f)_{K^-}$ &$-$23  & +$i$0.76 &  2.7 &$-$17  &114 & $-$0.73+$i$0.99   \\
&$(2p)_{K^-}$ &$-$21  & +$i$0.74 &  3.1 &$-$12  &100 & $-$0.70+$i$0.90   \\
                                                         
\end{tabular}                                            
\end{ruledtabular}
\end{table*}


\begin{table*}[bth]
\caption{
\label{tab:table3}
The integrated lab cross sections of ${K}^-$ nuclear 
bound states ${J^\pi}$ for $K^-$-${^{11}{\rm B}}$ by transitions 
$(n\ell_j)_{N} \to (nl)_{K^-}$ 
in the forward ($K^-$, $p$) reaction on the $^{12}$C target
at the incident $K^-$ lab momentum $p_{K^-}=$ 1.0 GeV/c. 
The Fermi-averaged cross section of 
$\langle d\sigma(0^\circ)/d\Omega \rangle^{K^-p \to p K^-}_{\rm lab}$= 3.5 mb/sr 
and distortion parameters 
$\bar{\sigma}^{\rm tot}_{NN}=\bar{\sigma}^{\rm tot}_{K^-N}=$ 40 mb and 
$\alpha_{NN}=\alpha_{K^-N}=$ 0 
are used in the DWIA.
}
\begin{ruledtabular}
\begin{tabular}{rcccccrrr}
Transition 
& $E_{K^-}$ & $\varGamma_{K^-}$ & $q(0^\circ)$ & $\alpha(0^\circ)$ & $J^{\pi}$ 
& ${\rm Re}N_{\rm eff}$ &  ${\rm Arg}N_{\rm eff}$ & ${d\sigma(0^\circ)/d\Omega}$ \\
& {(MeV)} & {(MeV)} & (MeV/c) &  &  & ($\times 10^{-2}$) & (deg.) & ($\mu$b/sr) \\ 
\hline
$(1p_{3 \over 2})_p\to (1s)_{K^-}$  &  $-130$ &  $25$   &  $-375$ &1.78 & ${3 \over 2}^{+}$  & 0.98  &     $0.08$ &    61 \\
                  $\to (1p)_{K^-}$  &  $-63$  &  $91$   &  $-296$ &1.62 & ${1 \over 2}^{-}$  & 0.42  &    $37.3$ &    24 \\
                                    &         &         &         &     & ${3 \over 2}^{-}$$\oplus$
                                                                           ${5 \over 2}^{-}$  & 6.44  &    $-12.1$ &  365 \\
                  $\to (2s)_{K^-}$  &  $+2.5$ &  $77$   &  $-217$ &1.47 & ${3 \over 2}^{+}$  & 11.66 &     $0.66$ &   601 \\
                  $\to (1d)_{K^-}$  &  $+6.6$ &  $97$   &  $-212$ &1.46 & ${1 \over 2}^{+}$$\oplus$
                                                                           ${3 \over 2}^{+}$  & 24.39 &   $-16.4$ &  1249 \\
                                    &         &         &         &     & ${5 \over 2}^{+}$$\oplus$
                                                                           ${7 \over 2}^{+}$  & 12.83 &   $-37.3$ &   657 \\
$(1s_{1 \over 2})_p \to (1s)_{K^-}$ & $-110$  &  $35$   &  $-375$ &1.77 & ${1 \over 2}^{-}$  & 0.29  &     $4.15$ &   18 \\
                   $\to (1p)_{K^-}$ &  $-43$  &  $101$  &  $-296$ &1.62 & ${1 \over 2}^{+}$$\oplus$
                                                                           ${3 \over 2}^{+}$  & 6.90  &    $-0.75$ &  391 \\
                   $\to (2s)_{K^-}$ &  $+23$  &  $87$   &  $-217$ &1.47 & ${1 \over 2}^{-}$  &$-0.08$&    $98.3$ &  $-4.2$ \\
                   $\to (1d)_{K^-}$ &  $+27$  &  $107$  &  $-212$ &1.46 & ${3 \over 2}^{-}$$\oplus$
                                                                 ${5 \over 2}^{-}$  & 13.99 &   $-14.0$ &   717 \\
\end{tabular}
\end{ruledtabular}
\end{table*}

\begin{table*}[bth]
\caption{
\label{tab:table4}
Comparison of the integrated cross sections 
$(d\sigma(0^\circ)/d\Omega)_{\rm lab}$ 
for the formation of ${K}^-$ nuclear 
$(1s)_{{K}^-}$ bound state, 
together with the momentum transfers $q(0^\circ)$, 
by the forward ($K^-$, $p$) reaction at the incident 
$K^-$ lab momentum $p_{K^-}=$ 1.0 GeV/c.
}
\begin{ruledtabular}
\begin{tabular}{lccclcclcclcc}
&& \multicolumn{2}{l}{Present} 
&& \multicolumn{2}{l}{Ref.~\cite{Cieply01}} 
&& \multicolumn{2}{l}{Ref.~\cite{Kishimoto99}} 
&& \multicolumn{2}{l}{Ref.~\cite{Yamagata05}} \\
\cline{3-4} \cline{6-7} \cline{9-10} \cline{12-13} 
Target
& Transition 
 & $d\sigma/d\Omega$ & $q(0^\circ)$ && $d\sigma/d\Omega$ & $q(0^\circ)$ 
&& $d\sigma/d\Omega$ & $q(0^\circ)$ && $d\sigma/d\Omega$ & $q(0^\circ)$ \\
&& ($\mu$b/sr) & (MeV/c) && ($\mu$b/sr) & (MeV/c) 
&& ($\mu$b/sr) & (MeV/c) && ($\mu$b/sr) & (MeV/c) \\
\hline
$^{12}$C
& $1p_{3/2} \to (1s)_{K^-}$ &  61      & $-$375 & 
                          &  47\footnotemark[1]      & $-$369 &
                          &  100-490\footnotemark[1] & $-$369 &
                          &  65\footnotemark[2]      & $-$379 \\
$^{28}$Si
& $1d_{5/2} \to (1s)_{K^-}$ &  2.1     & $-$428 & 
                          &  6.0\footnotemark[3]     & $-$404 &
                          &  35-180\footnotemark[3]  & $-$404 &
                          &  2.7\footnotemark[4]     & $-$458 
\end{tabular}
\end{ruledtabular}
\footnotetext[1]{$B_{K^-}=$ 122 MeV and harmonic oscillator wave functions were used.}
\footnotetext[2]{$B_{K^-}=$ 133 MeV for a deep potential was used 
at $T_{K^-}=$ 600 MeV ($p_{K^-}=$ 0.976 GeV/c).}
\footnotetext[3]{$B_{K^-}=$ 144 MeV and harmonic oscillator wave functions were used.}
\footnotetext[4]{$B_{K^-}=$ 191 MeV for a deep potential was used 
at $T_{K^-}=$ 600 MeV ($p_{K^-}=$ 0.976 GeV/c).}
\end{table*}

\begin{table*}[bth]
\caption{
\label{tab:table5}
The integrated lab cross sections of ${K}^-$ nuclear 
bound states ${J^\pi}$ for $K^-$-${^{27}{\rm Mg}}$ by transitions 
$(n\ell_j)_{N} \to (nl)_{\bar{K}}$ 
in the forward ($K^-$, $p$) reaction on the $^{28}$Si target
at $p_{K^-}=$ 1.0 GeV/c. 
We listed up only the deeply bound states of $E_{\bar{K}} <$ $-$50 MeV. 
The Fermi-averaged cross section of 
$\langle d\sigma(0^\circ)/d\Omega \rangle^{K^-p \to p K^-}_{\rm lab}$= 3.5 mb/sr 
and distortion parameters 
$\bar{\sigma}^{\rm tot}_{NN}=\bar{\sigma}^{\rm tot}_{K^-N}=$ 40 mb and 
$\alpha_{NN}=\alpha_{K^-N}=$ 0 
are used in the DWIA.
}
\begin{ruledtabular}
\begin{tabular}{rcccccrrr}
Transition 
& $E_{K^-}$ &$\varGamma_{K^-}$ &  $q(0^\circ)$ & $\alpha(0^\circ)$ & $J^{\pi}$ 
& ${\rm Re}N_{\rm eff}$ &  ${\rm Arg}N_{\rm eff}$ & ${d\sigma(0^\circ)/d\Omega}$ \\
& {(MeV)} & {(MeV)} & (MeV/c) &  &  & ($\times 10^{-2}$) & (deg.) & ($\mu$b/sr) \\ 
\hline
$(1d_{5 \over 2})_p\to (1s)_{K^-}$  &  $-169$ &  $22$   &  $-428$ &1.93 & ${5 \over 2}^{-}$  & 0.0031  &  $1.91$ &  2.1 \\
                  $\to (1p)_{K^-}$  &  $-123$ &  $25$   &  $-373$ &1.81 & ${1 \over 2}^{+}$$\oplus$
                                                                           ${3 \over 2}^{+}$  & $-0.0004$  & $-117.1$ & $-0.01$ \\
                                    &         &         &         &     & ${5 \over 2}^{+}$$\oplus$
                                                                           ${7 \over 2}^{+}$  & 0.519   & $-0.7$ & 32.9 \\
                  $\to (1d)_{K^-}$  &  $-72$  &  $88$   &  $-312$ &1.68 & ${1 \over 2}^{-}$  & 0.298   & $-18.5$ & 17.5 \\
                                    &         &         &         &     & ${3 \over 2}^{-}$$\oplus$
                                                                           ${5 \over 2}^{-}$  & 0.637   & $4.4$  & 37.6 \\
                                    &         &         &         &     & ${7 \over 2}^{-}$$\oplus$
                                                                           ${9 \over 2}^{-}$  & 3.079   & $-17.4$ & 181.5 \\
                  $\to (2s)_{K^-}$  &  $-65$  &  $92$   &  $-304$ &1.67 & ${5 \over 2}^{-}$  & 0.316 &   $17.6$ &  18.4 \\
$(1p_{1 \over 2})_p\to (1s)_{K^-}$  &  $-165$ &  $26$   &  $-428$ &1.93 & ${1 \over 2}^{+}$  & 0.0083  &  $5.0$ &   0.6 \\
                  $\to (1p)_{K^-}$  &  $-118$ &  $29$   &  $-373$ &1.81 & ${1 \over 2}^{-}$  & 0.029   &  $-3.1$ &  1.9 \\
                                    &         &         &         &     & ${3 \over 2}^{-}$ & 0.193   &  $1.5$ &  12.3 \\
                  $\to (1d)_{K^-}$  &  $-68$  &  $92$   &  $-312$ &1.68 & ${1 \over 2}^{+}$$\oplus$
                                                                           ${3 \over 2}^{+}$  & 0.187  &   $24.0$ &  11.0 \\
                                    &         &         &         &     & ${5 \over 2}^{+}$$\oplus$
                                                                           ${7 \over 2}^{+}$  & 1.207 &   $-5.7$ &  71.2 \\
                  $\to (2s)_{K^-}$  &  $-61$  &  $95$   &  $-304$ &1.67 & ${3 \over 2}^{+}$  & $-0.0115$ & $140.0$ & $-$0.67 \\
$(1p_{3 \over 2})_p\to (1s)_{K^-}$  &  $-158$ &  $28$   &  $-428$ &1.93 & ${3 \over 2}^{+}$  & 0.012  &    $5.8$ &   0.81 \\
                  $\to (1p)_{K^-}$  &  $-112$ &  $31$   &  $-373$ &1.81 & ${1 \over 2}^{-}$  & 0.061  &    $-2.8$ &   3.9 \\
                                    &         &         &         &     & ${3 \over 2}^{-}$$\oplus$
                                                                           ${5 \over 2}^{-}$  & 0.350  &    $1.6$ &   22.2 \\
                  $\to (1d)_{K^-}$  &  $-61$  &  $94$   &  $-312$ &1.68 & ${1 \over 2}^{+}$$\oplus$
                                                                           ${3 \over 2}^{+}$  & 0.320 &   $26.1$ &  18.9 \\
                                    &         &         &         &     & ${5 \over 2}^{+}$$\oplus$
                                                                           ${7 \over 2}^{+}$  & 2.403 &   $-6.1$ &  141.7 \\
                  $\to (2s)_{K^-}$  &  $-54$  &  $97$   &  $-304$ &1.67 & ${3 \over 2}^{+}$  & $-0.025$ & $147.1$ &  $-1.45$ \\
$(1s_{1 \over 2})_p \to (1s)_{K^-}$ & $-140$  & $32$    &  $-428$ &1.93 & ${1 \over 2}^{-}$  & 0.0004  &  $16.4$ &  $0.03$ \\
                   $\to (1p)_{K^-}$ &  $-94$  &  $35$   &  $-373$ &1.81 & ${1 \over 2}^{+}$$\oplus$
                                                                           ${3 \over 2}^{+}$  & 0.195  &    $4.7$ &  12.4 \\
\end{tabular}                                 
\end{ruledtabular}
\end{table*}

\clearpage

\begin{table*}[bth]
\caption{
\label{tab:table6}
Relative production 
cross sections of deeply-bound $\bar{K}\otimes [NNN]$ states 
on $^4$He. $T$ and $S$ denote the 
isospin and spin of the $\bar{K}$ nuclear states, respectively. 
}
\begin{ruledtabular}
\begin{tabular}{cccc}
  $T$  &  $S$  &  $\sigma(K^-, n)$ & $\sigma(K^-, p)$ \\
\hline
  $0$ & ${1 \over 2}$    
   &  $|f_{K^-n \to n K^-}+f_{K^-p \to n {\bar K}^0}|^2$ 
   &  0  \\
  $1$ & ${1 \over 2}$  
   &  $|f_{K^-n \to n K^-}-f_{K^-p \to n {\bar K}^0}|^2$
   &  $2|f_{K^-p \to p K^-}|^2$  \\
\end{tabular}
\end{ruledtabular}
\end{table*}

\begin{table*}[bth]
\caption{
\label{tab:table7}
Relative production 
cross sections of deeply-bound $\bar{K}\otimes [NN]$ states 
on $^3$He. $T$ and $S$ denote the 
isospin and spin of the $\bar{K}$ nuclear states, respectively. 
$T_C$ and $S_C$ are the isospin and 
spin of the $NN$ core-nucleus states, 
respectively.
}
\begin{ruledtabular}
\begin{tabular}{cccc}
  $T$ ($T_C$) &  $S$ ($S_C$)  
  &  $\sigma(K^-, n)$ & $\sigma(K^-, p)$ \\
\hline
  ${1 \over 2}$ (0) & $1$ (1)  
   &  ${3 \over 2}|f_{K^-p \to n {\bar K}^0}|^2$ 
   &  ${3 \over 2}|f_{K^-p \to p K^-}|^2$ \\
  ${1 \over 2}$ (1) & $0$ (0) 
   &  ${2 \over 3}|f_{K^-n \to n K^-}+{1 \over 2}f_{K^-p \to n {\bar K}^0}|^2$ 
   &  ${1 \over 6}|f_{K^-p \to p K^-}|^2$  \\
  ${3 \over 2}$ (1) & $0$ (0) 
   &  ${1 \over 3}|f_{K^-n \to n K^-}-f_{K^-p \to n {\bar K}^0}|^2$ 
   &  ${1 \over 3}|f_{K^-p \to p K^-}|^2$  \\
\end{tabular}
\end{ruledtabular}
\end{table*}

\begin{table*}[bth]
\caption{
\label{tab:table8}
Relative production 
cross sections of deeply-bound $\bar{K}\otimes [NN]$ states 
on $^3$H. $T$ and $S$ denote the 
isospin and spin of the $\bar{K}$ nuclear states, respectively. 
$T_C$ and $S_C$ are the isospin and spin of the $NN$ core-nucleus states, 
respectively.
}
\begin{ruledtabular}
\begin{tabular}{cccc}
  $T$ ($T_C$) &  $S$ ($S_C$) 
                    &  $\sigma(K^-, n)$ & $\sigma(K^-, p)$ \\
\hline
  ${1 \over 2}$ (0) & $1$ (1)  
   &  ${3 \over 2}|f_{K^-n \to n K^-}|^2$ 
   &  0  \\
  ${1 \over 2}$ (1) & $0$ (0)  
   &  ${2 \over 3}|f_{K^-p \to n {\bar K}^0}+{1 \over 2}f_{K^-n \to n K^-}|^2$ 
   &  0  \\
  ${3 \over 2}$ (1) & $0$ (0) 
   &  ${1 \over 3}|f_{K^-p \to n {\bar K}^0}-f_{K^-n \to n K^-}|^2$ 
   &  $|f_{K^-p \to p K^-}|^2$  \\
\end{tabular}
\end{ruledtabular}
\end{table*}


\clearpage

\end{document}